\documentclass[]{aa}

\usepackage{graphicx}
\usepackage{natbib}
\usepackage{subfigure}

\def\gtrsim{\mathrel{\hbox{\rlap{\hbox{\lower4pt\hbox{$\sim$}}}\hbox{$>$}}}}
\def\lesssim{\mathrel{\hbox{\rlap{\hbox{\lower4pt\hbox{$\sim$}}}\hbox{$<$}}}}

\bibpunct{(}{)}{;}{a}{}{,} 

\begin{document}

\title{The contribution of very massive high-redshift 
SWIRE galaxies to the stellar mass function}


\author{Stefano Berta\inst{1}\fnmsep\inst{2}\fnmsep\inst{3}\fnmsep\thanks{SB was supported by the
	Ing. Aldo Gini Foundation}
        \and
        Carol~J. Lonsdale\inst{2}\fnmsep\inst{4}
	\and
	Maria Polletta\inst{2,5}
	\and
	Richard~S. Savage\inst{6}
	\and
	Alberto Franceschini\inst{1}
	\and
	Helen Buttery\inst{7}
	\and
	Andrea Cimatti\inst{8}
	\and
	Joao Dias\inst{7}
	\and
	Chiara Feruglio\inst{9}
	\and
	Fabrizio Fiore\inst{9}
	\and
	Enrico~V. Held\inst{10}
	\and
	Fabio La Franca\inst{11}
	\and
	Roberto Maiolino\inst{9}
	\and
	Alessandro Marconi\inst{12}
	\and
	Israel Matute\inst{7}
	\and
	Seb~J. Oliver\inst{6}
	\and
	Elena Ricciardelli\inst{1}
	\and
	Stefano Rubele\inst{1}
	\and
	Nicola Sacchi\inst{11}
	\and
	David Shupe\inst{13}
	\and
	Jason Surace\inst{13}
}

\offprints{Stefano Berta, \email{berta@mpe.mpg.de, ste\_atreb@yahoo.it}}

\institute{Dipartimento di Astronomia, Universit\`a di Padova, Vicolo dell'Osservatorio 3, 
35122 Padova, Italy.
\and 
Center for Astrophysics and Space Sciences, University of California, 
San Diego, 9500 Gilman Dr., La Jolla, CA 92093-0424, USA.
\and
Max-Planck-Institut f\"{u}r Extraterrestrische Physik (MPE),
Postfach 1312, 85741 Garching, Germany.
\and
Infrared Processing \& Analysis Center, California Institute of Technology
100-22, Pasadena, CA 91125, USA. 
\and
Institut d'Astrophysique de Paris, 98bis bld Arago, 75014 Paris, France.
\and
Astronomy Centre, CPES, University of Sussex, Falmer, Brighton BN19QJ, UK. 
\and
INAF -- Osservatorio Astronomico di Arcetri, Largo E. Fermi 5, 
50125 Firenze, Italy. 
\and 
Dipartimento di Astronomia, Universit\`a di Bologna, Via Ranzani 1,
40127 Bologna, Italy.
\and 
INAF -- Osservatorio Astronomico di Roma, Via Frascati 33, I-00044 
Monteporzio Catone, Italy. 
\and
INAF -- Osservatorio Astronomico di Padova, Vicolo dell'Osservatorio 5,
35122 Padova, Italy.
\and
Dipartimento di Fisica, Universit\`a\ degli Studi `Roma Tre', Via della Vasca Navale 84, I-00146 Roma, Italy. 
\and
Dipartimento di Astronomia e Scienza dello Spazio, Universit\`a\ di  
Firenze, Largo E. Fermi 2, 50125 Firenze, Italy.
\and
Spitzer Science Center, California Institute for Technology, 220-6, Pasadena, CA 91125, USA.
}

\date{Received 16 March 2007; accepted 24 September 2007}

\titlerunning{IR-peakers mass function}
\authorrunning{Berta S., et al. }

 
  \abstract
   {In the last couple of years a population of very massive ($M_\star>10^{11}$ M$_\odot$),
   high-redshift ($z\ge2$) galaxies has been identified, but its role in galaxy 
   evolution has not yet been fully understood.}
   {It is necessary to perform a systematic study of high-redshift massive galaxies,
   in order to determine the shape of the very massive tail of the stellar mass
   function and determine the epoch of their assembly.}
   {We 
   selected high-$z$ massive galaxies at 5.8$\mu$m, in the SWIRE
   ELAIS-S1 field (1 deg$^2$). Galaxies with the 1.6$\mu$m stellar peak redshifted
   into the IRAC bands ($z\simeq1-3$, called ``IR-peakers'') were identified. 
   Stellar masses were derived by means of spectro-photometric fitting and used to compute the 
   stellar mass function (MF) at $z=1-2$ and $2-3$. A parametric fit to the MF was 
   performed, based on a Bayesian formalism, and the stellar mass density of 
   massive galaxies above $z=2$ determined.}
   {We present the first systematic study of the very-massive tail of the galaxy
   stellar mass function at high redshift.
   A total of 326 sources were selected. The majority of these 
   galaxies have stellar masses in excess of $10^{11}$ M$_\odot$ and lie at $z>1.5$.
   The availability of mid-IR data turned out to be
   a valuable tool to constrain the contribution of young stars to galaxy SEDs, and thus 
   their $M_\star/L$ ratio. The influence of near-IR data and of the chosen 
   stellar library on the SED fitting are also discussed.
   The $z=2-3$ stellar mass function between $10^{11}$ and $\sim10^{12}$ M$_\odot$
   is probed with unprecedented detail.
   A significant evolution is found not only for galaxies with $M\sim10^{11}$ M$_\odot$, 
   but also in the highest mass bins considered.
   The comoving number density of these galaxies was lower by more than a factor of 10 at $z=2-3$, 
   with respect to the local estimate.
   SWIRE 5.8$\mu$m peakers  
   more massive than $1.6 \times10^{11}$ M$_\odot$ provide 30$-$50\% of 
   the total stellar mass density in galaxies at $z=2-3$.  
   }
   {}


   \maketitle



\section{Introduction}\label{sect:intro}

Tracing the formation of galaxies and understanding the epoch {\em when} the bulk 
of their baryonic mass was assembled represents one of the major problems of
modern cosmology, particularly controversial when dealing with massive
($M_{\textrm{stars}} > 10^{11}$ M$_\odot$) objects.

The assembly of massive galaxies is one of the critical
questions in the cosmic evolutionary scenario. The uniform properties 
of local early-type galaxies and of the fundamental plane have inspired 
the so called ``monolithic collapse'' scenario  
\citep{eggen1962,chiosi2002}, in which galaxies formed in the remote 
past through huge events of star formation and subsequently evolved 
passively across cosmic time.
On the other hand, in the more recent ``hierarchical'' scenario 
\citep{white1978,kauffmann1996,kauffmann1998,somerville1999},
massive galaxies assemble by
mergers of lower-mass units, with the most massive objects
being born in the latest stages of evolution, at $z\le1$.

The availability of several powerful tracers of star formation
(e.g. UV continuum, optical recombination lines, far-IR emission, sub-mm light)
has favored the popularity 
of studies of the comoving star formation density \citep{madau1996,lilly1996} 
in galaxies at various redshifts.
It is now well determined that the Universe experienced an epoch of 
enhanced star formation in the past, peaking at $z\simeq1-2$, with a 
subsequent decline of at least one order of magnitude to the present time
\citep[e.g.][ among others]{hopkins2004,rudnick2003,flores1999,madau1996}.

An alternative approach 
consists of studying the mass already assembled in galaxies, instead of the 
amount of stars being formed. The integral of the past star formation 
density provides the stellar mass density at a given epoch: a complementary
constraint on cosmic galaxy evolution.

Thus, the build up of the stellar mass across cosmic time has become 
one of the major topics in observational cosmology, and has overtaken the 
classic Madau-Lilly diagram as the central tool for studying galaxy evolution.
The large observational effort dedicated to this subject has shown 
that the global stellar mass density increases from early epochs to the 
low-redshift Universe \citep[e.g.][]{brinchmann2000,dickinson2003,fontana2003,
fontana2004,fontana2006,rudnick2003,rudnick2006,drory2005}. Very deep surveys 
have been exploited to describe the shape of the stellar mass function at 
high redshift \citep{fontana2006,drory2005,gwyn2005}, but a 
clear picture on the role of very massive galaxies has not yet emerged.

Several pieces of evidence exist that fully formed massive
galaxies were already in place at redshift $z\sim 2-3$.

A substantial population of luminous red galaxies at redshifts $z>2$ 
(known as ``distant red galaxies'', DRGs) was found in the 
Faint InfraRed Extragalactic Survey \citep[FIRES,][]{franx2003}.
Based on near-IR  spectroscopy \citep{foerster2004,vandokkum2004},
these galaxies turned out to be massive ($M=1-5\times10^{11}$ M$_\odot$),
evolved (ages of $1-2.5$ Gyr) systems, probably descendants 
of galaxies which started forming at redshift $z>4$. 
Based on FIRES data, \citet{rudnick2003}
inferred that DRGs contribute $\sim50$\% of the global stellar mass density
at $z=2-3$.

\citet{dickinson2003} exploited Hubble Deep Field North NICMOS data to
derive the stellar masses of galaxies up to $z=3$. The study of the global
stellar mass density highlighted that $50-75$\% of the present-day stellar mass was 
already in place at $z\simeq1$, while only $3-14$\% had been already assembled at 
$z\simeq2.7$.

Direct determinations of the galaxy mass function based on near-IR deep 
imaging by the Spitzer Space Telescope indicate 
that the number of massive galaxies does 
not significantly evolve up to at least $z\simeq1$ \citep{franceschini2006,fontana2006,bundy2005}.

Using data from the Gemini Deep Deep Survey \citep[GDDS,][]{abraham2004}, 
\citet{glazebrook2004} identified a population of red ($[I-K]>4$ Vega mag.)
galaxies at $z\simeq2$ with stellar masses in excess of $10^{11}$ M$_\odot$. 
These objects contribute roughly 30\% of the total stellar mass density of the Universe 
at that epoch. \citet{mccarthy2004} estimated the age of red galaxies at 
$z=1.3-2.2$ in the GDDS, deriving a median age of 1$-$3 Gyr, and a 
star formation history dominated by very powerful bursts (300$-$500 
M$_\odot$ yr$^{-1}$). These massive galaxies must have undergone
a rapid formation process at $z>1$.

Exploiting data from the K20 survey 
\citep{cimatti2002c,cimatti2002b,cimatti2002a}, 
\citet{daddi2004} identified few luminous $K$-band selected galaxies at $1.7 < z < 2.3$ 
with stellar masses  $M \simeq 10^{11}-5\times 10^{11}\ [M_\odot]$.
Combining deep K20 spectroscopy and HST-ACS imaging, \citet{cimatti2004} 
discovered four old, fully assembled spheroidal galaxies at $1.6 < z < 1.9$: 
the most distant such objects currently known. The stellar mass of these 
galaxies turned out to be in the range $1-3\times10^{11}$ M$_\odot$.
\citet{fontana2004} studied K20 galaxies as well, 
showing that massive ($M>10^{11}$ M$_\odot$) galaxies are easily found 
up to $z\simeq2$. These authors also report on the stellar mass function:
only mild evolution ($\sim$2$-$30\%) is detected to $z=1$, but 
only $\sim35$\% of the $z=0$ stellar mass locked up
in massive objects was assembled by $z=2$.

At even higher redshifts, \citet{rigopoulou2006} have recently studied a population 
of $z\sim3$ Lyman-break galaxies (LBGs) with stellar masses in excess of $10^{11}$ M$_\odot$.
\citet{mclure2006} identified nine LBGs 
at $z\ge5$ in the UKIDSS \citep{lawrence2006} survey, over an area of
0.6 deg$^2$. A stacking analysis suggests that the typical 
stellar mass of these sources is $>5 \times 10^{10}$ M$_\odot$.
\citet{mobasher2005} analyzed the properties of J-dropouts in the Hubble Ultra
Deep Field \citep[HUDF,][]{beckwith2006}, exploiting Spitzer photometry. 
They
identified a $z\sim6.5$ candidate that was interpreted as a 
post-starburst galaxy with a surprisingly high stellar
mass of $5.7 \times 10^{11}$ M$_\odot$ (but see, for example, Yan et al. \citeyear{yan2004}
for a different interpretation).

\citet{drory2005} studied the stellar mass function of galaxies in the 
FORS Deep Field \citep[FDF,][]{heidt2003} and GOODS/CDFS \citep{giavalisco2004} field,
over a total area of 90 arcmin$^2$ and found that the total stellar mass density at $z=1$
is 50\% of the local value. At $z=2$, 25\% of the local mass density was already assembled,
and at $z=3$ and $z=5$, at least 15\% and 5\% of stellar mass, respectively, was already in place.
Massive ($M>10^{11}$ M$_\odot$) galaxies existed over the whole redshift range probed,
up to $z=5$. The number density of these massive galaxies evolves very similarly to 
galaxies with $M>10^{10}$  M$_\odot$, decreasing by 0.4 dex to $z=1$, 
0.6 dex to $z=2$, and 1 dex to $z=4$. 

By analyzing the properties of $K$-selected galaxies in 
the $\sim131$ arcmin$^2$ of the GOODS-CDFS survey,
\citet{caputi2006} found that the vast majority ($85-90$\%) of 
local $M > 2.5 \times 10^{11}$ M$_\odot$ galaxies appears to be already in place 
at $z\sim1$. These authors also infer that roughly $65-70$\% of these galaxies 
assembled at $z=1-3$ by means of obscured, intense bursts of star formation, 
while the remaining could be in place at even higher redshifts ($z= 3-4$).

The observational challenge is that large volumes are needed to find representative samples 
of such rare very massive galaxies at high redshift.
The Spitzer Wide-area InfraRed Extragalactic survey 
\citep[SWIRE,][]{lonsdale2003,lonsdale2004} observed $\sim$49 deg$^2$ in the seven Spitzer 
channels, and is therefore ideal to 
find rare objects. Its volume is large enough to detect 
$\sim$85 DM haloes of mass $> 10^{14} M_\odot$ in the 
$2<z<3$ redshift range \citep{jenkins2001,mo2002}.
These haloes are predicted to host the most luminous ($L_{bol}>10^{12}\ L_\odot$)
and most massive ($M_\star>\textrm{several}\ 10^{11}\ M_\odot$)
galaxies ever to exist.

The Infrared Array Camera \citep[IRAC,][]{fazio2004}, 
onboard Spitzer \citep{werner2004}, samples the restframe 
near-IR light of distant galaxies.
A near-IR selection not only directly probes the low-mass stars dominating 
the baryonic mass of a galaxy, but also is minimally affected by dust 
extinction. Therefore Spitzer is best suited 
to the study of the stellar content of galaxies up to $z=3$.
Moreover Spitzer allows detection of galaxies that would be missed by restframe UV 
selection, for example distant red galaxies \citep[e.g.][]{daddi2004}.

We take advantage of the shape of near-IR spectral energy distribution (SEDs)
of galaxies to identify high-redshift objects on the basis of IRAC colors.
Our selection is based on the detection of the 1.6$\mu$m stellar peak in galaxies 
\citep{sawicki2002,simpson1999}, redshifted to the IRAC domain.
It is also worth noting that galaxies selected in this way benefit from a negative 
k-correction in the IRAC bands, because the slope of a galaxy's SED 
is negative redward of the 1.6$\mu$m peak.
In this way we performed a systematic search for $M\gtrsim10^{11}$ M$_\odot$
galaxies at $z>1$. 

The analysis is carried out in the central square degree
of the ELAIS-S1 SWIRE field, where optical, near-IR ($J$, $K_s$) photometry
and optical spectroscopy are available (Berta et al., \citeyear{berta2006},
Dias et al., in prep., La Franca et al., in prep.).
This area, and the sampled volume, are bigger than any other previously explored for studying 
very massive galaxies at $z=1-3$, which were limited to very deep, 
pencil-beam surveys \citep[e.g.][]{dickinson2003,drory2005,gwyn2005,fontana2006}.

This paper is structured as follows. Section {\bf 2} presents the data available in the 
ELAIS-S1 field; in Sect. {\bf 3} we present our selection criterion; then Sect. {\bf 4}
discusses the photometric estimate of redshifts.
Section {\bf 5} deals with the estimate of the stellar mass in galaxies, and 
presents a very detailed analysis of the influence of mid-IR and near-IR constraints 
on it. 
Experiments with different stellar libraries and IMFs are also discussed.
Section {\bf 6} presents the stellar mass function of our galaxies, including 
completeness correction and a parametric fit based on a Bayesian formalism.
Finally, Sects. {\bf 7} and {\bf 8} discuss results and draw our conclusions.

Throughout this work, we adopt a standard $H_0=71$ $[$km s$^{-1}$ Mpc$^{-1}]$, 
$\Omega_m=0.27$, $\Omega_\Lambda=0.73$ cosmology, unless otherwise stated.


\section{Available Data}\label{sect:data}

The ELAIS-S1 field ($\textrm{RA} = 00^{h}38^{m}30^{s}$,
$\textrm{Dec} = -44^\circ00'00''$, J2000.0) represents the minimum of 
the Milky Way 100 $\mu$m cirrus emission \citep{schlegel1998}
in the southern hemisphere, with an average emissivity of 0.38 MJy/sterad.

\subsection{Spitzer SWIRE data}

This area was targeted by the Spitzer Space Telescope \citep{werner2004}, 
as part of the {\em Spitzer Wide area Infra-Red Extragalactic} survey 
\citep[SWIRE,][]{lonsdale2003,lonsdale2004}. A total of $\sim7$ deg$^2$
were observed with the IRAC \citep{fazio2004} and MIPS \citep{rieke2004}
cameras.

Data processing is described by \citet{surace2004}, Surace et al. (in prep.),
Shupe et al. (in prep.) and Afonso Luis et al. (in prep.).
It consists of Basic Calibrated Data (BCD) by the SSC
pipeline plus post-processing aimed at artifact removal, mosaicking and
source extraction.
Mosaicking was performed with the SSC routine MOPEX, and
source extraction with SExtractor \citep{bertin1996}.  
For unresolved sources (i.e. the case examined here),
IRAC fluxes were extracted through a 1.9${\arcsec}$ diameter 
aperture and corrected 
to total fluxes following SSC prescriptions; MIPS fluxes 
were extracted by means of PRF fitting 
(see Surace et al., and MIPS Data Handbook 2006).

The resulting 5$\sigma$ depths are 4.1, 8.5, 43, 48, 400 $\mu$Jy
at 3.6, 4.5, 5.8, 8.0, and 24 $\mu$m respectively. 
Observations at 70 and 160 $\mu$m
lead to 26 and 166 mJy 5$\sigma$ limits (Surace et al., Shupe et al., 
Afonso Luis et al., in prep.).

For more details on Spitzer data reduction, calibration and catalog extraction see
the SWIRE delivery documentation 
\citep[][ and following releases\footnote{available 
at the Spitzer Science Center Legacy Program web page, 
http://ssc.spitzer.caltech.edu/legacy/}]{surace2004}.

\subsection{Optical ESIS imaging}

The ELAIS-S1 area is the target of extensive optical follow up
carried out with ESO telescopes: the {\em ESO-Spitzer 
Imaging extragalactic Survey} \citep[ESIS,][]{berta2006}.
The optical ancillary data cover $\sim 5$ deg$^2$ in the $B$, $V$, $R$ bands, observed 
with the Wide Field Imager (Baade et al. \citeyear{baade1999}, on the 2.2m ESO-MPI telescope
in La Silla) and in the $I$, $z$ bands, obtained with VIMOS (Le F{\` e}vre et al. 
\citeyear{lefevre2002}, on the VLT).
The $B$, $V$, $R$ observations and results over the central 1.5 deg$^2$ area were 
presented in \citet{berta2006}, while the VIMOS data are in an advanced 
reduction stage (Berta et al., in prep.). 

The reduction of $B$, $V$, $R$ data was performed by using IRAF standard tasks 
and self-built routines. 
Sky flat-field frames were acquired during each night and applied to images
obtained on the same date. In order to obtain a uniform background and photometric 
zeropoint across the field, super-sky-flat frames were built and applied for each 
observing night, taking care of masking very bright sources. 

Astrometric calibration was performed on observations of \citet{stone1999} astrometric fields.
In order to accurately map coordinates, a TNX algorithm was chosen, 
combining a gnomonic projection and non-linear polynomial distortions.
The r.m.s coordinate difference between the ESIS and the GSC 2.2 catalogs turned out to be 
$\sim0.1$ arcsec in RA and Dec. The relative astrometric accuracy between different 
ESIS pointings is better than 0.05 arcsec r.m.s.

Photometric calibration was obtained by observing \citet{landolt1992} standard fields, 
in order to calibrate photometric zeropoints. 
Since observations were spread across several years, particular care has been taken into
accounting for zeropoint differences between the numerous science frames.
Moreover, $BVR$ color-curves were built, and magnitudes were transformed to the Johnson-Cousins
standard photometric system. 
Catalogs were extracted by using SExtractor \citep{bertin1996}, 
Kron total 
magnitudes are adopted for extended sources, while aperture magnitudes, 
corrected using the observed PSF, are used for unresolved objects.
The final catalog reaches 
95\% completeness at $B,V\simeq25$ and $R\simeq24.5$.

We defer to \citet{berta2006} for a further description of these data and their analysis.

\subsection{Near-IR $J$ and $K_s$ imaging}

The square degree at the center of the ELAIS-S1 area includes 
$J$ and $K_s$ observations carried out with the 
SOFI \citep{moorwood1998} camera on the NTT, during four different
periods in 2002 and 2003 (Dias et al., in prep.).

Pre-reduction, sky-subtraction and mosaicking were carried out 
in the standard way, using the IRAF environment.
The astrometric mapping was calibrated by using the ESIS $R$ band 
images \citep{berta2006}, reaching a 0.22$^{\arcsec}$ and 0.16$^{\arcsec}$
r.m.s. uncertainty for RA and Dec, respectively.
Photometric calibration was computed using aperture 
photometry of point-like objects.
The average $J$ and $K_s$ magnitude difference 
for the sources in common with the 2MASS catalog 
is $\sim0.02$ mag in both bands.

Catalog extraction was performed using SExtractor \citep{bertin1996}.
In this work we make use of total magnitudes. 
The resulting catalog is 95\% complete to $J\sim19.8$ and  
$Ks=18.73$ (Vega).

The optical, near-IR and SWIRE/Spitzer catalogs were matched with a 
simple closest-neighbor algorithm, adopting a 1.0 arcsec matching 
radius \citep[see][]{berta2006}.

\subsection{Other imaging data}

The central area benefits from X-ray observation by the XMM-Newton telescope 
\citep{puccetti2006}. The whole ELAIS-S1 field was observed at 1.4 GHz 
with the ATCA radio telescope by \citet{gruppioni1999} down to a 80 $\mu$Jy r.m.s. and 
by Middelberg et al. (submitted) down to 30 $\mu$Jy r.m.s.
Finally, the ultraviolet Galaxy Evolution Explorer \citep[GALEX,][]{martin2005} Deep
Imaging Survey (DIS) included the ELAIS-S1 field.


\subsection{Available spectroscopic data}\label{sect:spec_data}

The ELAIS-S1 area has been spectroscopically surveyed by different projects, 
in particular focused on the central field.

\citet{lafranca2004} performed optical spectroscopy 
of ISOCAM 15$\mu$m counterparts, with the 2dF/AAT,
ESO-Danish 1.5m, ESO 3.6m and NTT telescopes, over the spectral 
range between 4000$-$9000 \AA.

During 2004 and 2005, the XMM-NIR area was the target
of 5000$-$9500 \AA\ low resolution spectroscopy, carried out with 
VIMOS-VLT. The primary targets of this survey are X-ray sources, $K$-selected 
galaxies ($K_s < 18.5$), and 24 $\mu$m SWIRE objects.
Further spectroscopy of optically bright ($R < 21$)
X-ray and mid-IR sources was obtained at the 3.6m/ESO
telescope, during Fall 2005. These observations will be presented in future
works by La Franca and collaborators.

Overall, at the present time, $\sim$1250 spectroscopic redshifts are available in the ELAIS-S1
SWIRE area.


\section{Selection criteria}\label{sect:selection}

In order to select high redshift galaxies dominated by stellar emission 
in the restframe near-IR, we have exploited the ``IR-peak'' technique, 
described in Lonsdale et al. (in prep.) and \citet{berta2007}.

The near-IR emission of a galaxy is characterized by a peak centered at 1.6$\mu$m, 
due to the combination of the Planck spectral peak of low-mass stars (dominated by type M),
a minimum in the H$^-$ opacity in stellar atmospheres and molecular absorptions
in the spectra of cold stars. 
The Spitzer IRAC camera was partly designed to detect this feature in 
high redshift galaxies \citep{sawicki2002,simpson1999}.
The peak is fully
sampled by the IRAC photometric bands when it falls in the 4.5 or 5.8 $\mu$m channel, 
i.e. when at least one band lies shortward and one longward of the peak. 
We thus focus our analysis on 4.5$\mu$m-peakers and 5.8$\mu$m-peakers (objects 
with SEDs peaking in the 4.5$\mu$m and 5.8$\mu$m channels).
The 1.6 $\mu$m peak is redshifted to the 4.5 and 5.8 $\mu$m channels 
for sources in the redshift range $z=1.5-3$.

We focus this paper on the ELAIS-S1 sub-area centered at $\textrm{RA}=00^{h}34^{m}48^{s}$ 
$\textrm{Dec}=-43^\circ30'38''$, 
where all BVR, $J$, $K_s$, IRAC and MIPS photometric data are available, for a 
total area of 1 deg$^2$. 
This area includes 48101 SWIRE sources, and 
29859 SWIRE+ESIS matches.
In this area, 2848 objects are detected at 24 $\mu$m by the 
MIPS camera, while the remaining Spitzer sources are detected by IRAC only.

We first apply a 5.8 $\mu$m flux cut at the 3$\sigma$ SWIRE depth (25.8 $\mu$Jy), 
in order to favor sources detected in the IRAC bands, which sample the restframe 
near-IR emission and hence the stellar mass (as dominated by low mass stars).
This sub-sample consists of 5546 objects (over one square degree).

\begin{figure*}[!ht]
\centering
\includegraphics[width=0.39\textwidth]{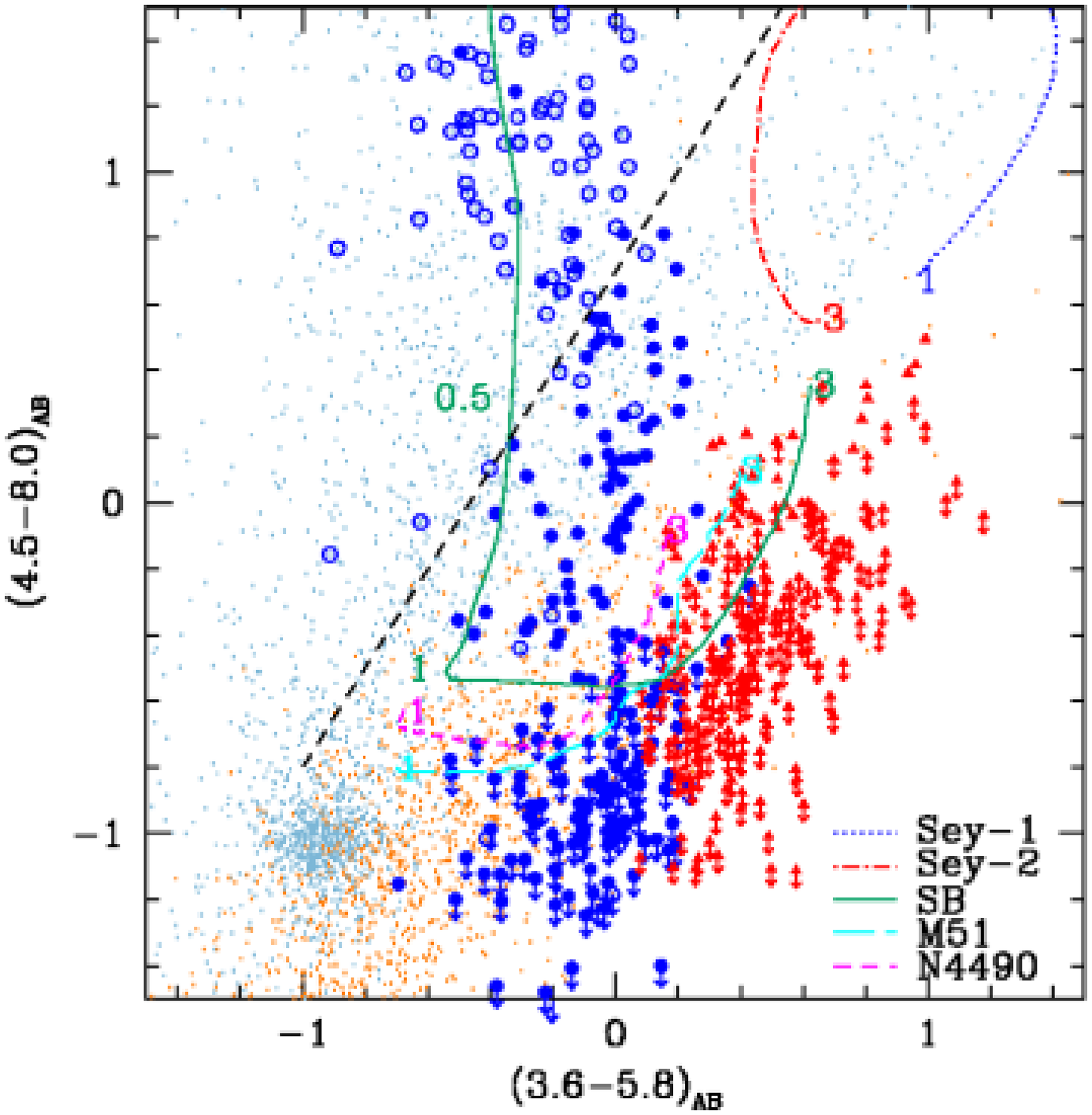}
\includegraphics[width=0.39\textwidth]{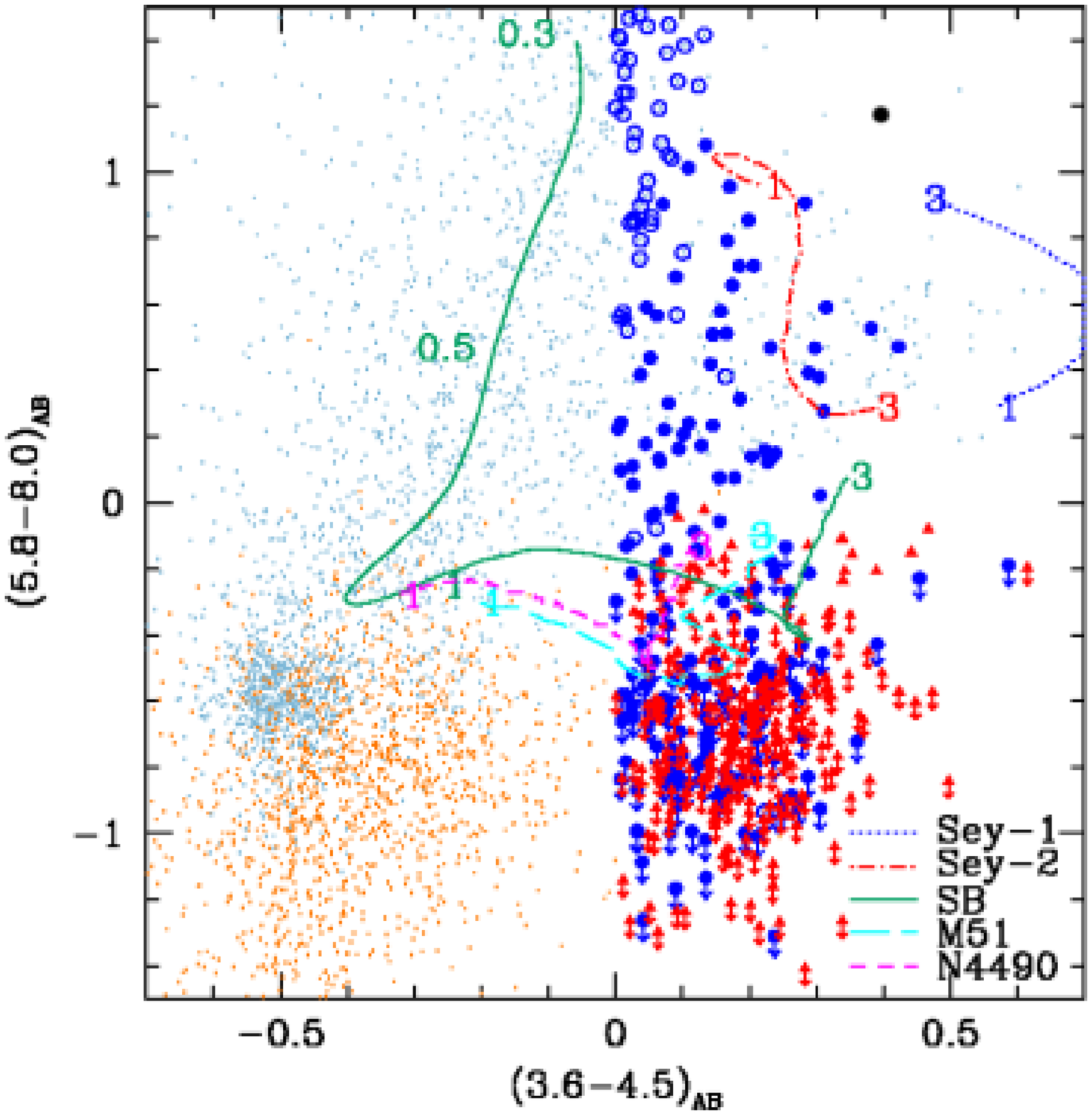}
\includegraphics[width=0.39\textwidth]{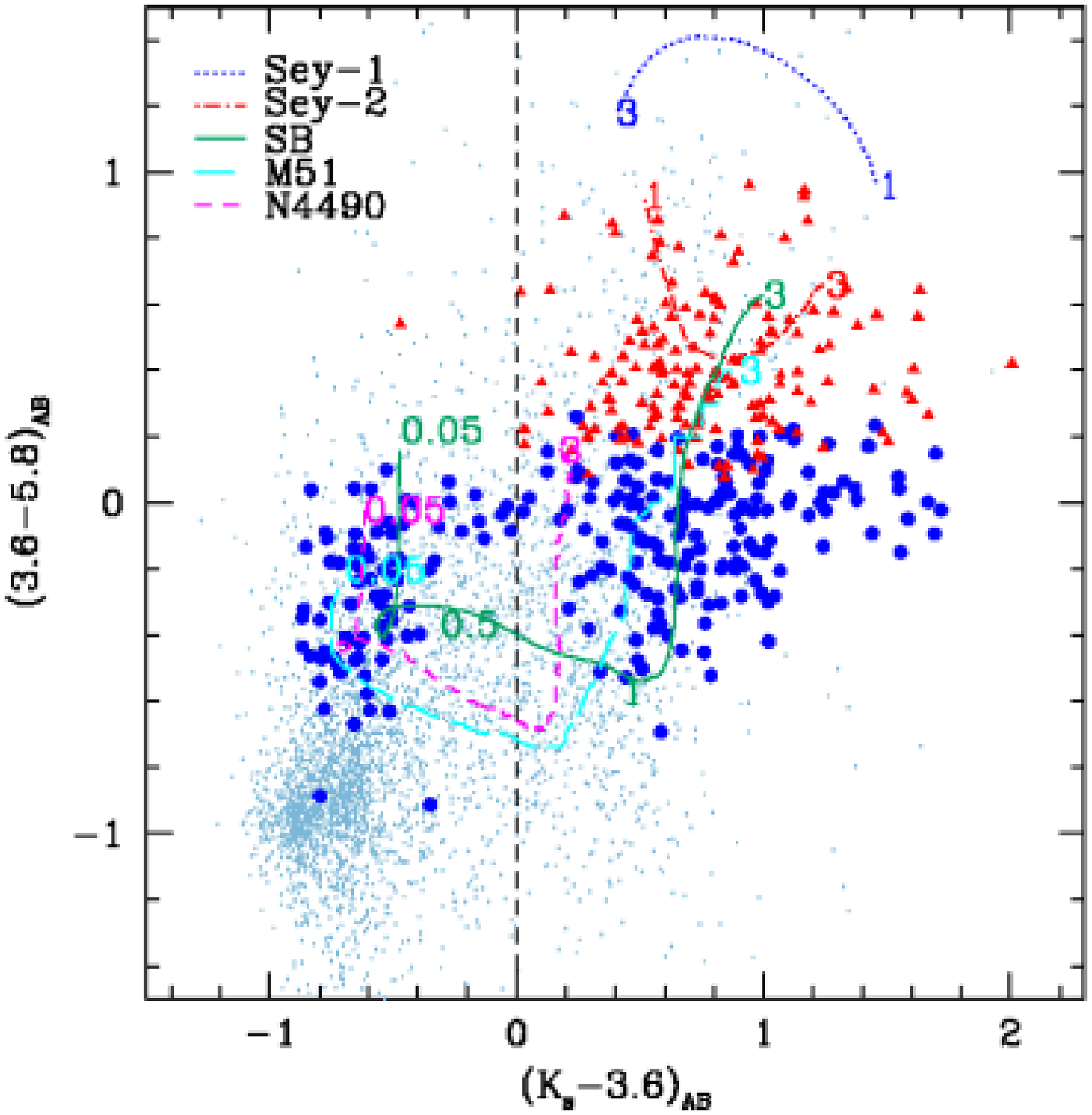}\\
\includegraphics[width=0.39\textwidth]{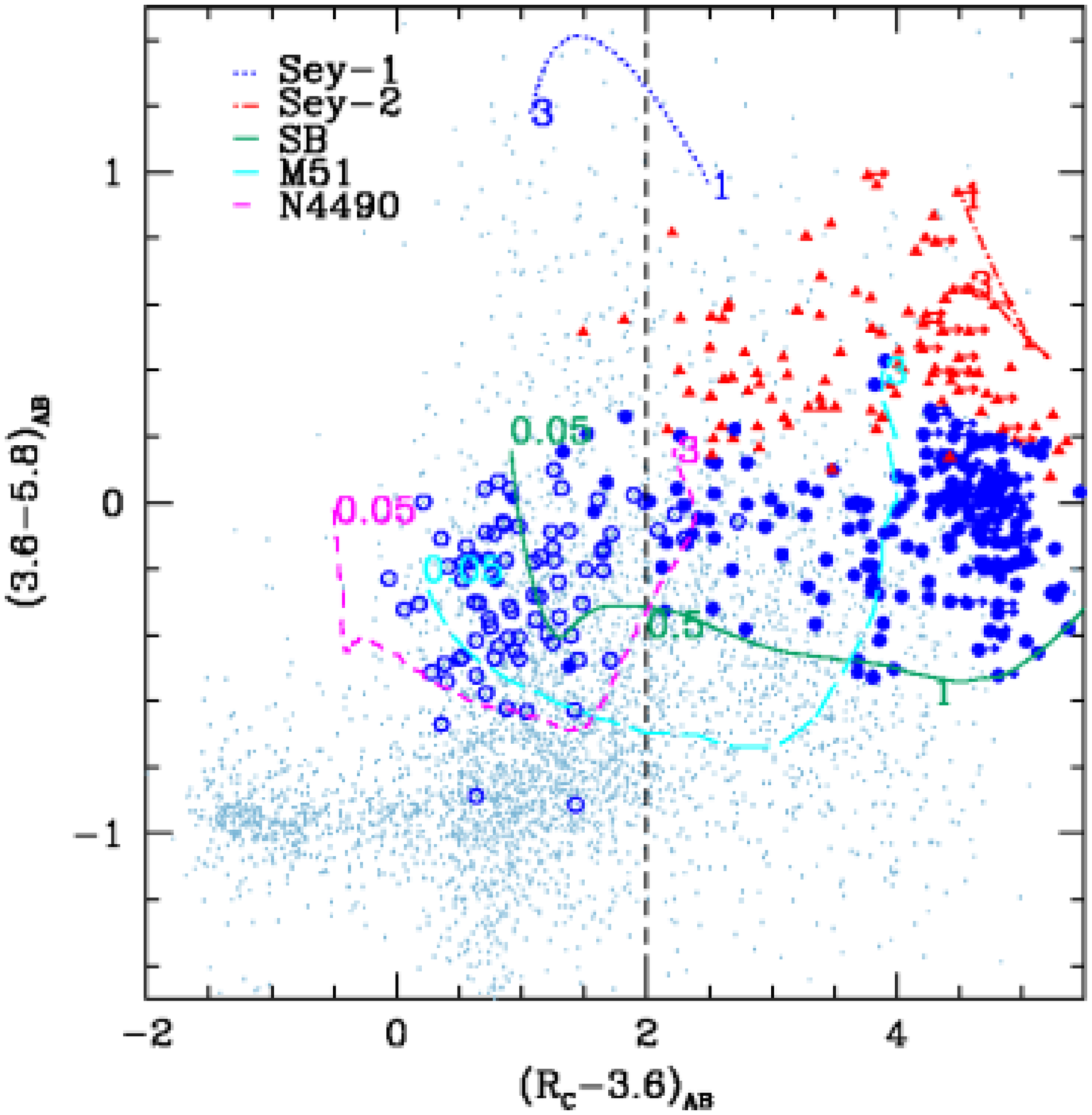}
\includegraphics[width=0.39\textwidth]{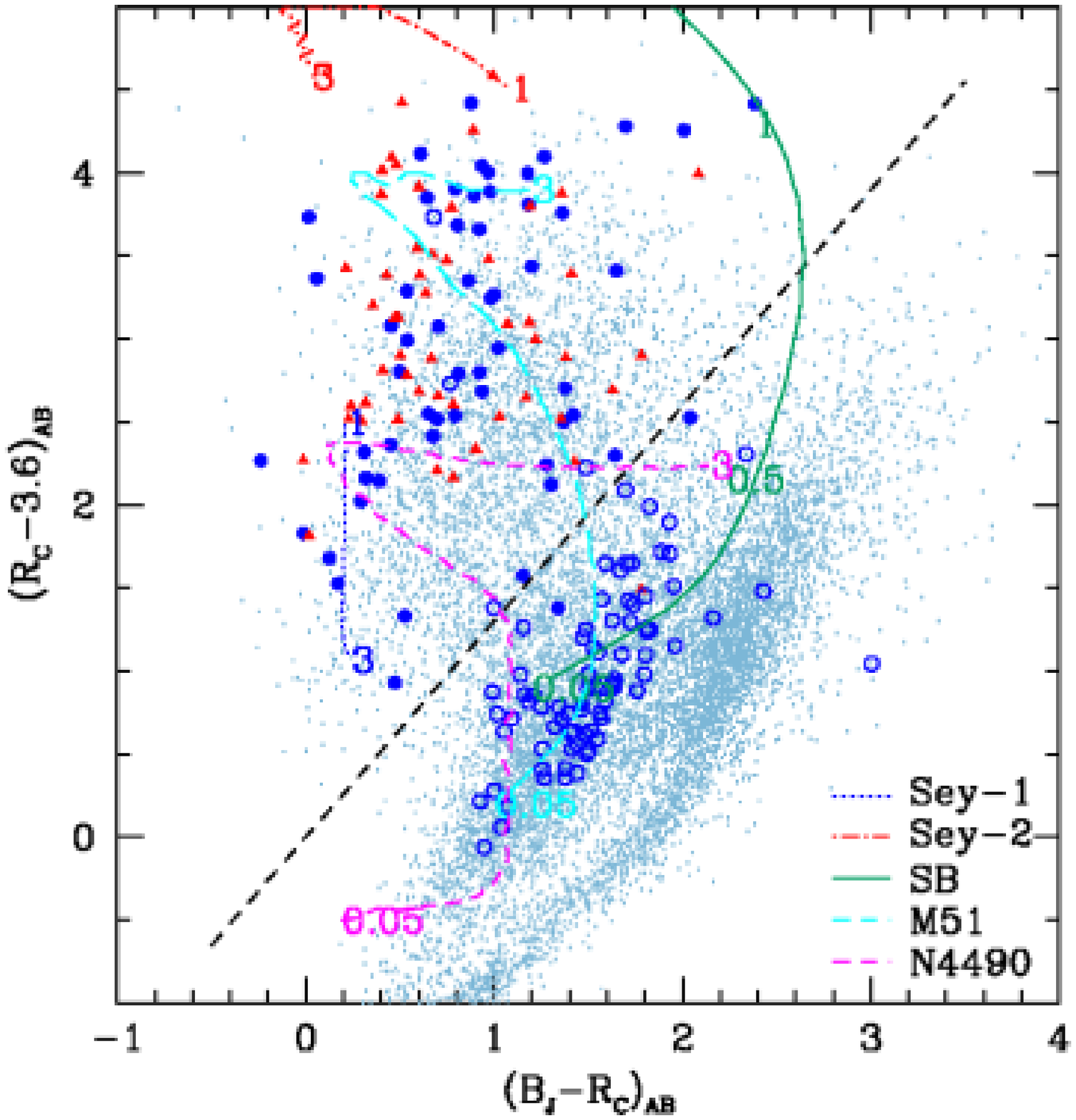}
\caption{Selection of the IR-peak galaxies. The two {\em top} panels show 
the position of 4.5$\mu$m-peakers (solid circles) and 5.8$\mu$m-peakers (triangles)
in the IRAC color space \citep{lacy2004,stern2005}. 
The {\em middle} plot highlights the use of $K_s$ band data to recognize 
low-redshift interlopers aliasing 4.5$\mu$m-peaker colors. The open circles in the other 4 
panels belong to sources with $(K_s-3.6)<0$ [AB mag]. Finally, the two {\em bottom} panels 
present additional information based on optical-IRAC colors. Small dots represent the 
general galaxy population in the SWIRE ELAIS-S1 field. Template tracks 
are overlaid for a Sy-1, Sy-2, starburst (M82), spiral (M51) and blue spiral (NGC4490) 
galaxies. The covered redshift range is reported aside each track.}
\label{fig:selection}
\end{figure*}

\subsection{IR-peak galaxies}

The IR-peak selection is well exemplified in the IRAC color space 
\citep{lacy2004,stern2005}, as shown in the top panels of Fig. 
\ref{fig:selection}. Here red triangles represent 5.8$\mu$m-peakers: formally
sources showing $S_\nu(3.6)<S_\nu(4.5)<S_\nu(5.8)>S_\nu(8.0)$.
Blue circles (either filled or not) are 4.5$\mu$m-peak objects, characterized by
$S_\nu(3.6)<S_\nu(4.5)>S_\nu(5.8)$. The difference 
between filled and open symbols is defined later in Section \ref{sect:use_JK}. We include in the 
plots also objects not detected in the 8.0 $\mu$m band, but for which the 
8.0 $\mu$m 3$\sigma$ upper limit is consistent with the 
definition of the IR-peak.
The small dots represent the general galaxy population detected in the central 
ELAIS-S1 SWIRE field, light blue having a 4 band detection, and orange having 
a 8.0 $\mu$m upper limit only.

We overplot a set of template tracks as a function of redshift:
a Seyfert-1 (dotted lines, Mrk 231, Fritz et al., \citeyear{fritz2006}), 
a Seyfert-2 (dot-dashed, IRAS 19254-7245, Berta et al., \citeyear{berta2003}),
a starburst galaxy (solid line, M82, Silva et al., \citeyear{silva1998}, enhanced with 
observed PAHs by F{\"o}rster-Schreiber et al. \citeyear{foerster2001}), a spiral 
(long-dashed, M51, Silva et al., \citeyear{silva1998}) and an optically-blue spiral 
(short-dashed, NGC4490, Silva et al., \citeyear{silva1998}) galaxy. 
The tracks are limited to the range $z=1-3$, for the sake of clarity, 
apart for the starburst one, which is extended down to $z\simeq0$.

\subsection{Possible aliases}

IR-peak-like colors could in principle be produced not only by the 
1.6 $\mu$m feature redshifted in the IRAC domain, but also by a strong 
3.3 $\mu$m PAH feature at lower redshift, as is demonstrated by 
Lonsdale et al. (in prep.) and \citet{berta2007}. In this case, the 
real 1.6 $\mu$m peak lies shortward of the IRAC channels, resulting in 
a blue NIR-IRAC observed color.

As far as 5.8$\mu$m-peakers are concerned, the requirement that 
$S_\nu(3.6)<S_\nu(4.5)$ automatically avoids low-redshift interlopers,
because it forces rejection of those sources for which the 1.6 $\mu$m peak 
falls at wavelengths shorter than the IRAC 3.6 $\mu$m channel.

\subsection{Use of near-IR and optical photometry}\label{sect:use_JK}

On the other hand, in the case of 4.5$\mu$m-peakers, only one IRAC band 
samples the SED on the blue side of the observed peak and there is no 
trivial way to avoid aliases with low-redshift objects with a bright 3.3 $\mu$m
PAH. In this case the $K_s$ data are of fundamental importance in 
breaking the degeneracy and ruling out interlopers.
The middle panel in Fig. \ref{fig:selection} involving the $K_s$ 
magnitude shows how near-IR data can effectively break this degeneracy and 
help in selecting $z=1-3$ objects only. The redshift tracks of starburst, spiral and blue 
spiral galaxies suggest that objects with $(K_s-3.6)_{AB}<0$ 
lie at $z<1$. Therefore we adopt this value in order to disentangle low- and high-$z$
sources. 

In the two top panels of Fig. \ref{fig:selection} (discussed above), 
open circles represent objects with $(K_s-3.6)_{AB}<0$, while 
filled circles lie above this threshold.
In the top left diagram \citep{lacy2004}, these low-redshift interlopers 
seem to be separated from the high-redshift objects and we empirically define 
the transition line: 
\begin{equation}\label{eq:lacy_transition}
(4.5-8.0)_{AB}=1.5\times (3.6-5.8)_{AB}+0.7\textrm{.}
\end{equation}
We then use this line to distinguish between low- and high-$z$ candidates among those 
sources that are not detected in the near-IR ($J$, $K_s$) survey.

\begin{figure*}[!ht]
\centering
\includegraphics[width=0.45\textwidth]{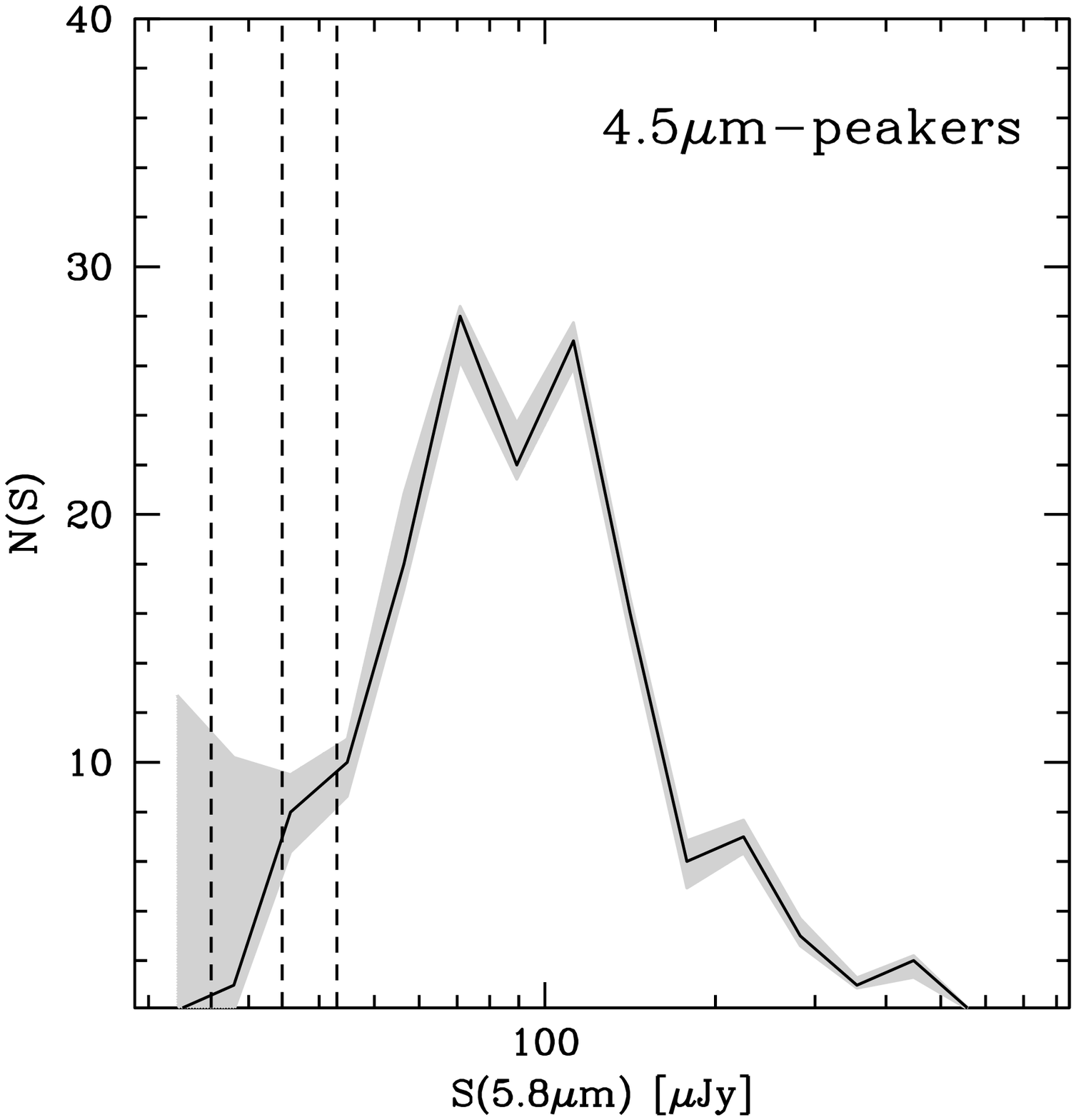}
\includegraphics[width=0.45\textwidth]{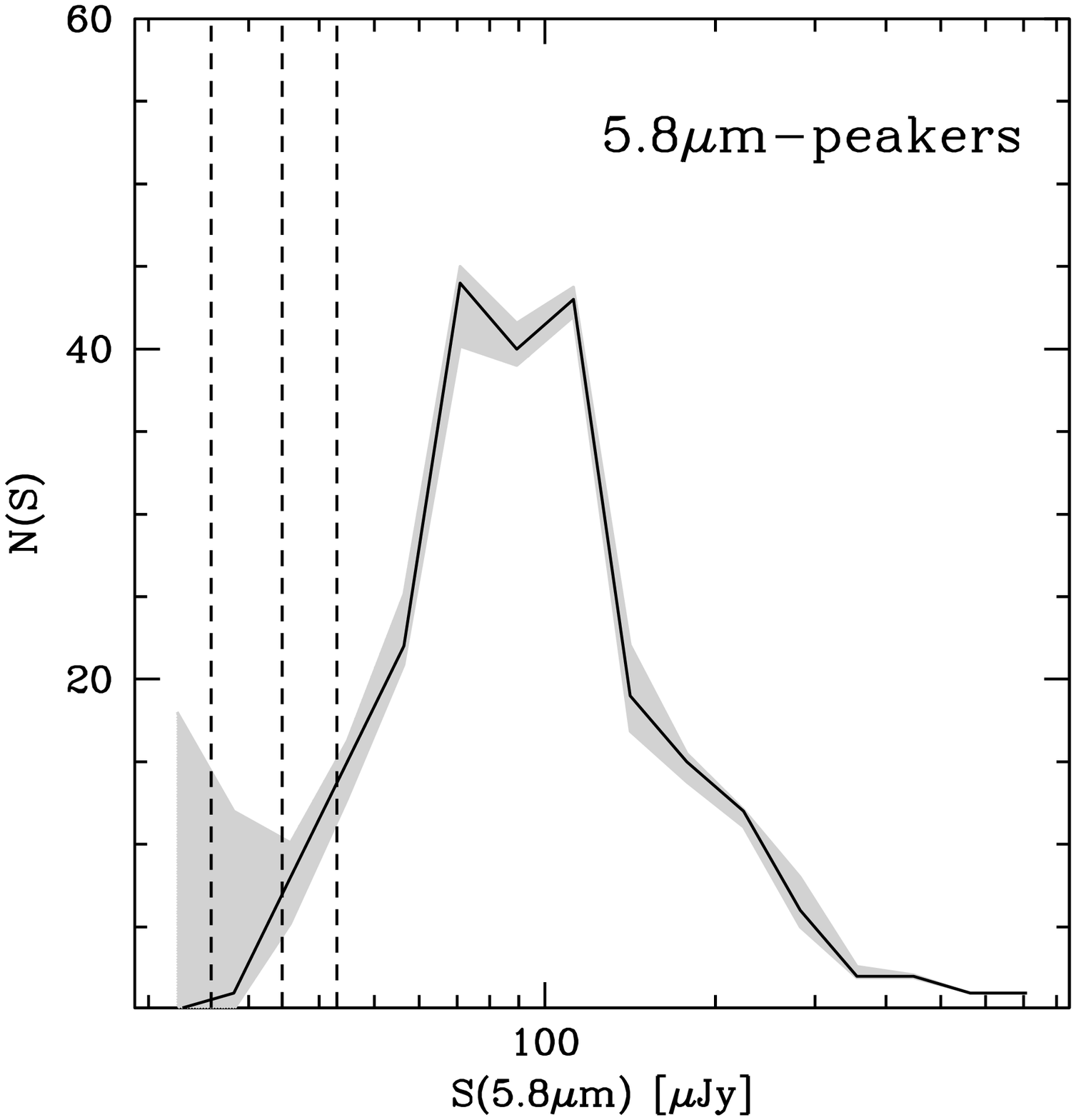}
\caption{Effect of sky-noise on IR-peakers selection. Random 
fluctuations of IRAC and K$_s$ fluxes can cause objects to 
be scattered in and out of the sample. Solid lines are 
the actual distribution of IR-peakers. Shaded areas are the result of 
simulations and represent the variation in the number of peakers due to random 
fluctuations of their colors. The three vertical lines refer to the 3, 4, 5 $\sigma$ 
thresholds at 5.8$\mu$m in the SWIRE ELAIS-S1 survey.}
\label{fig:noise_simulation}
\end{figure*}

It is also interesting to test whether the optical-IRAC color space shows any 
segregation of low-$z$ versus high-$z$ objects (as defined on the basis 
of the IR-peak technique and $K_s$ band photometry), in order to 
define alternative selection criteria to be used when no near-IR data are available.
The bottom plots in Fig. \ref{fig:selection} show a couple of diagrams 
involving the available optical BVR photometry from the ESIS survey 
\citep{berta2006}. The majority of IR-peakers with $(K_s-3.6)_{AB}<0$
lie in the $(R_C-3.6)_{AB}<2$ sub-space and below the line 
\begin{equation}\label{eq:opt_IRAC_transition}
(R_C-3.6)_{AB}=1.3\times (B_J-R_C)_{AB}\textrm{.}
\end{equation}
In the end, combining the IR-peak criterion and the near-IR and optical constraints,
our sample contains 149 and 231 sources peaking at 4.5$\mu$m and 5.8$\mu$m 
respectively, over one square degree. Among these, 149 (i.e. $\sim$39\%) have an 
optical detection (at least one band), 295 ($\sim$78\%) are detected in either 
$J$ or $K_s$, and 109 ($\sim$29\%) have a 24$\mu$m counterpart.
Finally, $\sim$77\% of the IR-peakers have no 8.0$\mu$m detection.
Table \ref{tab:select_numbers} summarizes these numbers.

\begin{table}
\centering
\begin{tabular}{l r}
\hline
\hline
Description  & Number\\
\hline
SWIRE sources & 48101 \\
SWIRE + opt. sources & 29859 \\
24$\mu$m sources & 2848 \\
$S(5.8)\ge25.8$ $\mu$Jy (3$\sigma$) & 5546 \\
\hline
IR-peakers & 380 \\
IR-peakers 4$\sigma$ & 326 \\
\hline
4.5$\mu$m-peakers & 149 \\
4.5$\mu$m-p. + 24$\mu$m & 44 \\
4.5$\mu$m-p. + optical & 69 \\
4.5$\mu$m-p. + NIR & 140 \\
4.5$\mu$m-peakers 4$\sigma$ & 123 \\
\hline
5.8$\mu$m-peakers & 231 \\
5.8$\mu$m-p. + 24$\mu$m & 65 \\
5.8$\mu$m-p. + optical & 80 \\
5.8$\mu$m-p. + NIR & 155 \\
5.8$\mu$m-peakers 4$\sigma$ & 203 \\
\hline
\end{tabular}
\caption{Summary of IR-peaker selection in the central square degree of the 
SWIRE ELAIS-S1 field. The descriptions ``+ optical'' and ``+ NIR'' refer to
sources with at least one optical or JK detection.}
\label{tab:select_numbers}
\end{table}

\subsection{Effect of sky noise}\label{sect:skynoise}

Random fluctuations of fluxes within the 
photometric uncertainties in the bands defining 
the IR-peaker selection can cause non-peaker
objects to 
be scattered into the sample and actual peakers 
to fall out of it.

It is worth noting that, since the 3.6$\mu$m and 4.5$\mu$m 
detection thresholds are much fainter than at 5.8$\mu$m (see Section \ref{sect:data}),
our sources are detected above $\sim$15$\sigma$ in the two bluest IRAC
channels, and sky-noise affects their fluxes by a factor smaller that $\sim$5\%.

On the other hand, random fluctuations in the 5.8$\mu$m and 8.0$\mu$m bands 
can significantly modify the colors of these objects and scatter them in and out
of the sample.

In order to quantify this effect, we have performed a bootstrap simulation
of IR-peaker selection, starting from the multi-wavelength SWIRE catalog 
in the ELAIS-S1 field (48101 sources). The procedure applies a random 
fluctuation to the K$_s$-to-8.0$\mu$ fluxes of all SWIRE sources, 
with a dispersion given by the sky-noise associated to each source.

This process was looped 10000 times and a new IR-peaker selection was performed at each step.
Figure \ref{fig:noise_simulation} reports the result of 
this simulation, for 4.5$\mu$m-peakers (left panel) 
and 5.8$\mu$m-peakers (right). Solid lines represent  
the actual distribution of IR-peakers, while 
shaded areas are the result of simulations.
Vertical dashed lines represent the 3, 4, 5 
$\sigma$ detection thresholds in the SWIRE ELAIS-S1 5.8$\mu$m survey.

This analysis shows that, at the 3 $\sigma$ detection level
random noise fluctuations significantly affect the selection
of IR-peakers, and tend to diverge.
On the other hand at the 4 and 5 $\sigma$ flux levels the contamination 
decreases to $\sim20$\% and $\sim15$\% respectively.

Therefore the subsequent analysis will be carried out 
only for those IR-peakers detected above the 4 $\sigma$ level 
(34.4 $\mu$Jy) in the 5.8 $\mu$m band. Thus the sample of IR-peakers reduces
to 326 objects, 123 of which are 4.5$\mu$m-peakers and the remaining 203
are 5.8$\mu$m-peakers. The electronic Table associate to this work 
reports the main data of the final selected sample.


\section{Photometric redshifts}\label{sect:photo_z}

The spectroscopic survey in ELAIS-S1 has targeted mainly 
X-ray sources and generally low-redshift ($z\le1$) galaxies.
Unfortunately, the few sources detected in the IR-peaker redshift range are 
classified as AGNs, on the basis of optical spectroscopy and 
show power-law IRAC colors or bright 8.0$\mu$m excesses due to 
torus warm dust. These objects do not belong to the IR-peaker
sample.

As a consequence, we need to rely mostly on photometric redshift.
Nevertheless, recent spectroscopic analyses of subsamples of IR-peakers in other 
SWIRE fields (Lockman Hole, ELAIS-N1, ELAIS-N2) have confirmed that 
the adopted selection effectively identifies galaxies at $z\simeq1.5-3.0$.
\citet{weedman2006} performed IRS \citep{houck2004} mid-IR spectroscopy 
of IR-peakers detected at 24$\mu$m and found redshifts between $z=1.6-2.0$ 
in 90\% of the examined cases.
\citet{berta2007} obtained Keck UV-optical (restframe) spectroscopy 
of IR-peakers, confirming the photometric selection between $z=1.3$ and 2.5.

We have derived photometric redshifts by using the {\tt Hyper-z} code \citep{bolzonella2000}.
We adopt a semi-empirical template library including GRASIL \citep{silva1998}
models of spiral and elliptical galaxies,
M82 and Arp220 templates (Silva et al.) upgraded with observed PAH mid-IR features
\citep{foerster2001,charmandaris1999}, 
and a ULIRG template \citep[IRAS 19254-7245,][]{berta2003}.

The {\tt Hyper-z} performance has been optimized for galaxies: we 
have excluded from the analysis all sources with a type-1 AGN spectroscopic 
classification and all those showing a power-law
like IRAC SED. In this way AGN-dominated 
objects are avoided. This class is particularly delicate to 
fit with a template-based procedure and photometric redshifts are 
typically mis-interpreted in 50\% of the cases (see for example Berta et al. \citeyear{berta2007}), 
because sharp features are missing in their SEDs.
Moreover, the IR-peak criterion automatically selects the sample against power-law AGNs.

\begin{figure}[!ht]
\centering
\includegraphics[width=0.49\textwidth]{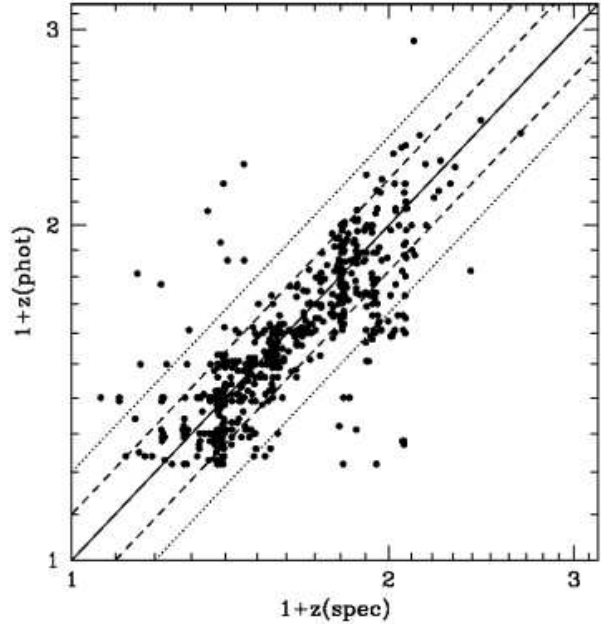}
\caption{Comparison of photometric and spectroscopic redshifts for all 
galaxies with available spectroscopy in the ELAIS-S1 field. 
Objects classified as type-1 AGNs on the 
basis of spectroscopy or having a power-law like IRAC SED have been excluded
from this analysis. Dashed and dotted lines represent $\pm10$\% and $\pm20$\%
uncertainty levels.}
\label{fig:zphot_zspec}
\end{figure}

The estimate of photometric redshifts was performed including  
the optical, $J$, $K_s$ and IRAC ($3.6-8$ $\mu$m) data in the $\chi^2$ computation, but 
ignoring the MIPS 24$\mu$m flux. In fact, including the 24$\mu$m data 
turned out to produce a higher degree of degeneracy and aliasing. 

The choice of templates in the library and the allowed extinction have been 
tested on galaxy populations in the ELAIS-S1 field, taking advantage 
of the available spectroscopic redshifts.
Figure \ref{fig:zphot_zspec} reports the comparison of spectroscopic
and photometric redshifts, as obtained with {\tt Hyper-z}, 
The dashed and dotted lines represent $\pm$10\% and $\pm$20\% uncertainties. 
The difference between the two estimates has an r.m.s. of 0.19 and  
s.i.q.r.\footnote{s.i.q.r. = semi inter-quartile range, defined as $(q_3 - q_1)/2$, where $q_1$ and $q_3$
are the first and third quartiles. The first quartile is the number below which 25\%
of the data are found and the third
quartile is the value above which 25\% of the data are found.} of 0.076.

\begin{figure}[!ht]
\centering
\rotatebox{-90}{\includegraphics[height=0.49\textwidth]{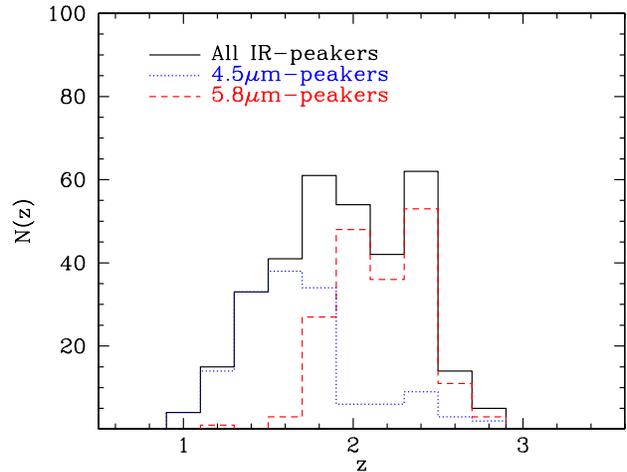}}
\caption{Distribution of photometric redshifts of IR-peakers in ELAIS-S1, 
as derived with the code {\tt Hyper-z} \citep{bolzonella2000}.}
\label{fig:zdistr_IR_peakers}
\end{figure}

The main outliers are sources with few photometric data available, 
especially those with a detection in the two more sensitive IRAC channels (3.6$\mu$m and 4.5$\mu$m) and 
one optical band (typically R) only.
The photometric code calibrated in this way was then used to derive redshifts 
for the general IR-peak population selected as described above.

The redshift distribution of IR-peakers is shown in Fig. 
\ref{fig:zdistr_IR_peakers}, where 
4.5$\mu$m-peakers are represented by the dotted line and 5.8$\mu$m-peakers 
by the dashed one.


\section{The estimate of stellar mass}\label{sect:fit}

The estimate of the stellar mass of the galaxies in
our sample has been obtained with the code {\tt Sim-Phot-Spec} (SPS), developed in 
Padova \citep{berta2004,poggianti2001}.

This code performs mixed stellar population (MSP) spectro-photometric synthesis.
Several phases in the life of a simple stellar population (SSP) are combined 
together, adopting a different star formation rate (SFR) for each age. 
Each SSP phase is meant to represent a formation episode of average constant
SFR, over a suitable time period $\Delta t$.
The effective number of SSP ages involved in the fit depends on the 
redshift of the given source, in order not to exceed the age of the 
Universe at that redshift. The maximum number of SSP phases 
considered for our sample is 6, corresponding to a redshift $z=1$, and 
decreases to 4 if $z=3$. Moreover, during the minimization phase, 
the code checks for SSPs that do not effectively contribute to the emitted
light, either in the optical and IR, and deletes them. In the end the number 
of SSPs contributing to the final fit is typically 3-4.

Age-selective extinction is applied, assuming that the oldest stars have 
abandoned the dusty medium long ago.
Keeping in mind that disc populations are on average affected by a moderate $A_V$ 
\citep[$<1$ mag, e.g.][]{kennicutt1992}, the maximum allowed absorption for 
stars older than 1 Gyr is $A_V=0.3-1.0$ magnitudes. For younger populations
the color excess gradually increases, but is limited to $A_V\le5$.

The best fit is found by $\chi^2$ minimization of the differences between 
the observed photometric data and the synthetic SED, 
including photometric uncertainties in all bands. 
Photometric redshifts, as computed with {\tt Hyper-z} 
\citep[][ see Sect. \ref{sect:photo_z}]{bolzonella2000}
are adopted. The SPS code takes into full account the uncertainty associated 
with the photometric redshift estimate, while exploring parameter space.
This uncertainty is of the order of $0.1-0.2$, depending on how many 
photometric bands are available. When the given source is not 
detected in one (or more) bands, upper limits are used.

For each galaxy in the sample, the SPS code builds a large number of models 
($\sim10^5$), exploring the parameter space by means of the ``Adaptive Simulated 
Annealing'' algorithm \citep[ASA,][]{ingber2001,ingber1989,berta2004}. 
First the parameters are varied with a coarse resolution, and then the 
algorithm focuses on the found minima.
In order to avoid ``freezing'' in local minima, 
re-annealing is applied and the code literally ``jumps'' out of the
minimum in order to explore a different region of the parameter space 
and find the absolute best fit.
In this way, the possible fluctuations of colors, within 
the measured photometric errors, are accounted for and are automatically
propagated to the output stellar mass uncertainty.

The SPS code is optimized to derive the assembled stellar mass
in galaxies and the associated uncertainty due to degeneracies in star 
formation history (SFH) space \citep[see][]{berta2004}. 
The $\chi^2$ of each model considered 
during the minimization (i.e. each combination of parameter values)
is recorded and 1, 2, 3$\sigma$ contours are computed.
In this way, the resulting uncertainty on the stellar mass estimate 
accounts also for minimal- and maximal-mass models, derived from the 
projection of the parameters space into ``young'' and ``old'' sub-spaces.
The resulting best fit masses and 3 $\sigma$ mass ranges are reported 
in an electronic Table.

The adopted SSP library is based on the Padova 
evolutionary sequences of stellar models \citep{fagotto1994a,fagotto1994b,bressan1993}
and isochrones \citep{bertelli1994}, and was computed
by assuming a solar metallicity and a Salpeter initial mass function (IMF) between 0.15 and 120 
M$_\odot$. The SSP spectra were built with the \citet{pickles1998} spectral library, 
and extended to the near-IR with the \citet{kurucz1993} atmosphere models.
Nebular features were added through the ionization code CLOUDY \citep{ferland1990}.
The spectra thus obtained provide a reliable description of simple stellar generations up
to $\sim 5\ \mu$m (restframe). Beyond this wavelength, dust 
emission is no longer negligible.

We assume that the total energy absorbed in the UV-optical
domain is processed by dust in the thick molecular clouds embedding 
young stars and re-emitted in the mid- and far-IR (8-1000 $\mu$m) in the form 
of a starburst template.
By convolving the template with the 24$\mu$m passband,
the MIPS 24$\mu$m flux expected for the given model is computed.
Finally, this synthetic flux is compared to the observed 24$\mu$m data and 
included in the $\chi^2$ computation. 
Mid-IR photometry is mainly sensitive to the power of the ongoing
starburst, as well as to the amount of dust obscuring it, therefore
it provides a valuable constraint on the amount of dust 
and the strength of young stellar populations \citep[see also][]{berta2004}.

\begin{figure*}[!ht]
\centering
\includegraphics[width=0.47\textwidth]{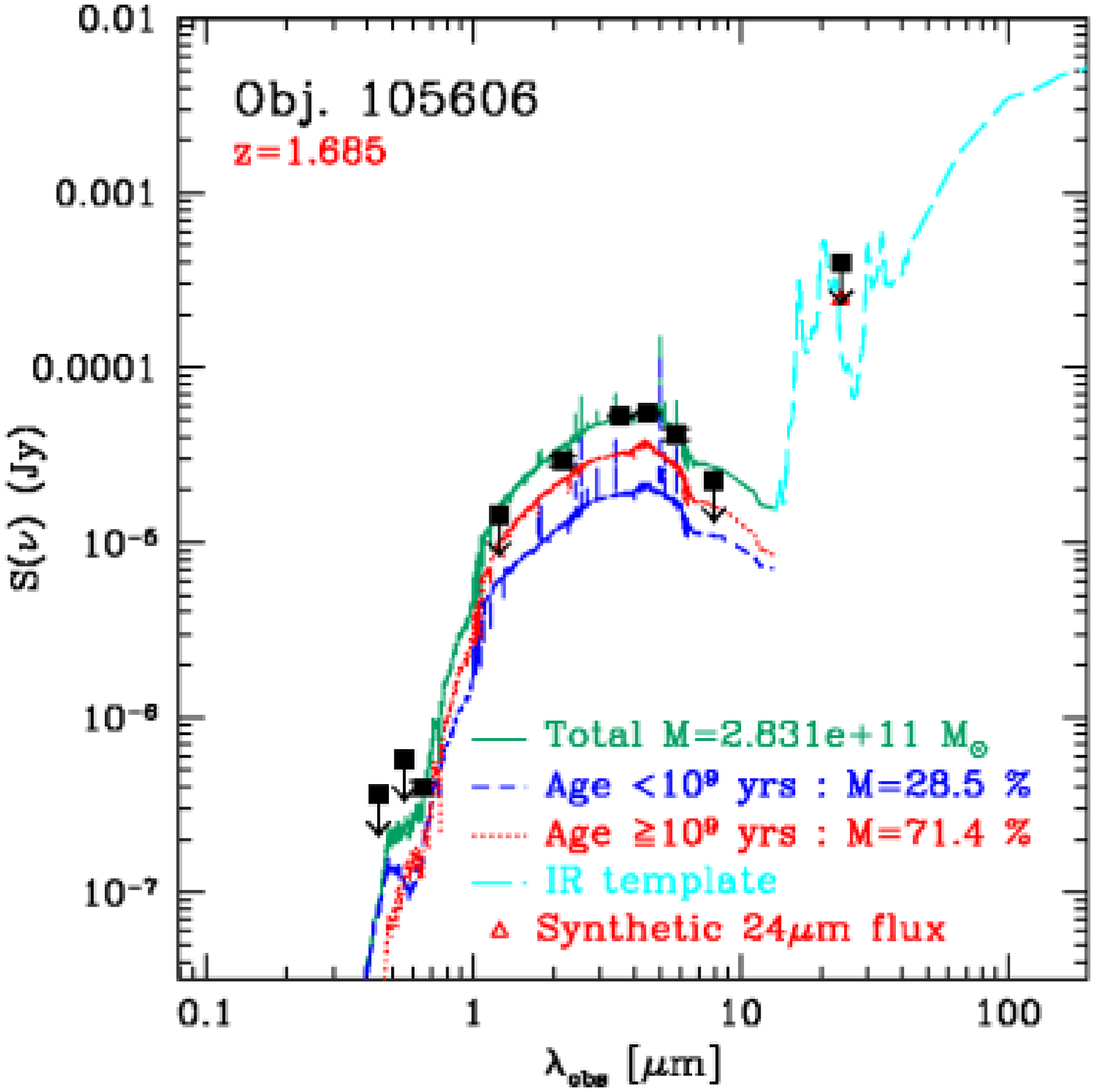}
\includegraphics[width=0.47\textwidth]{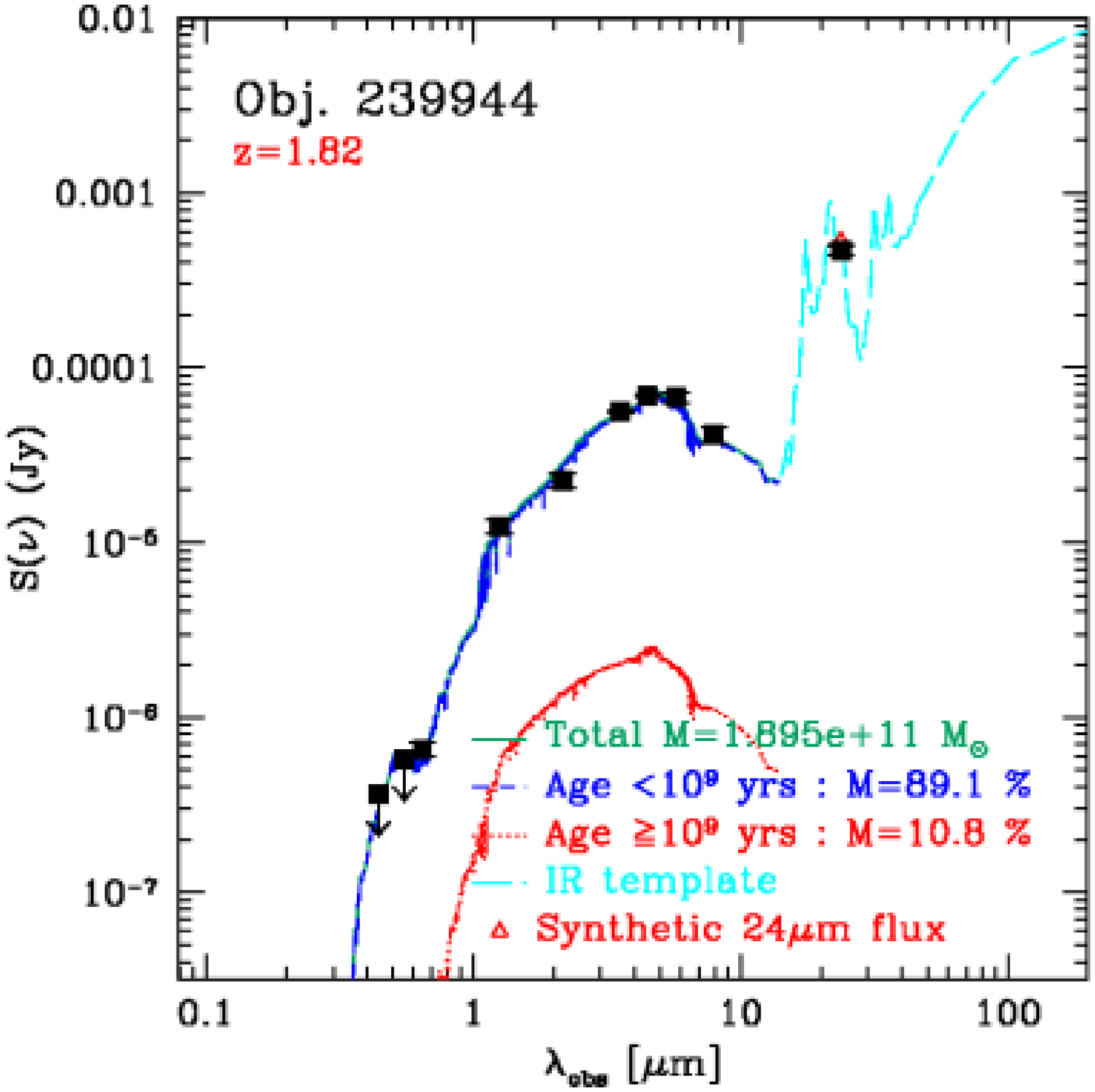}
\includegraphics[width=0.47\textwidth]{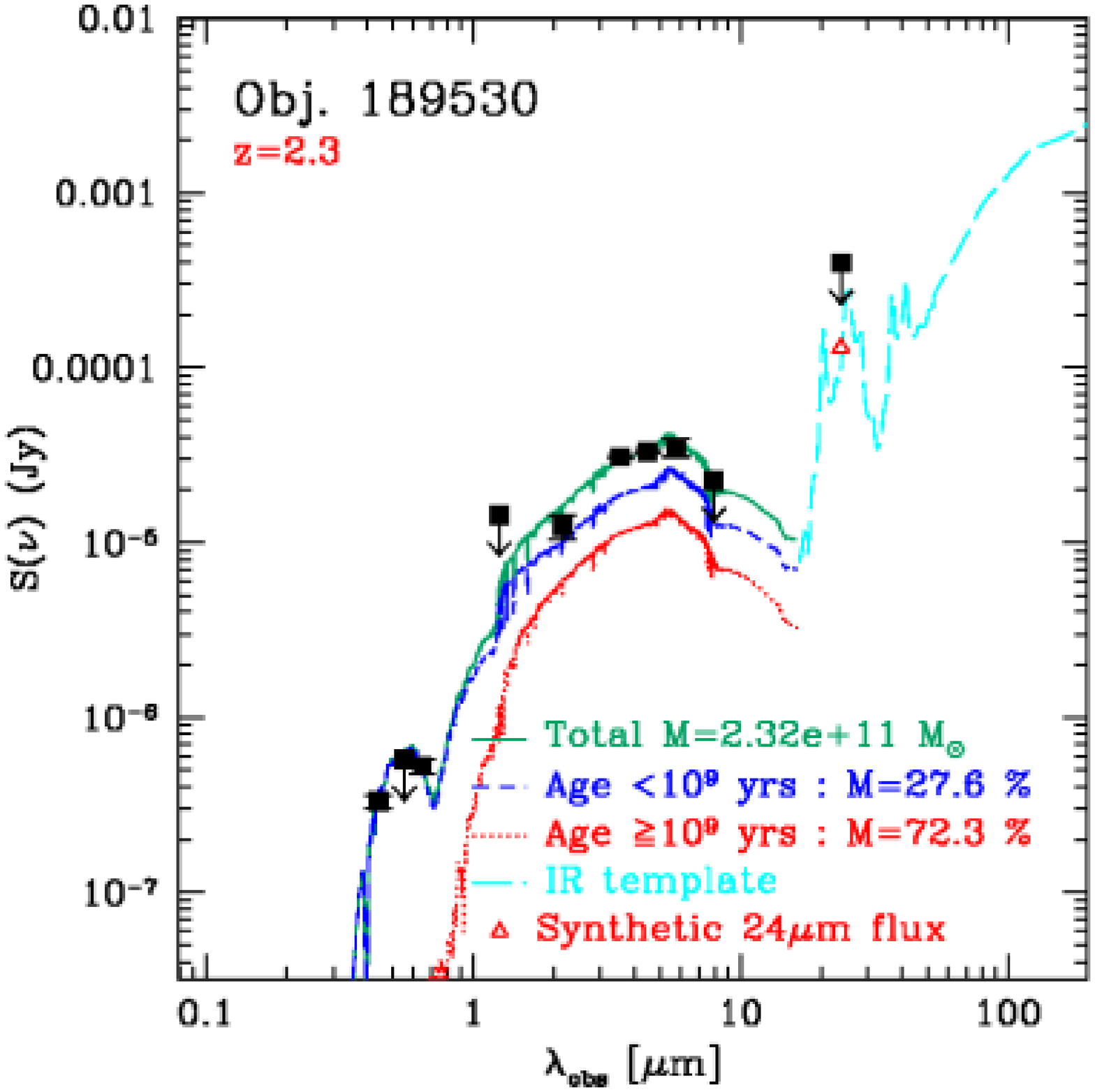}
\includegraphics[width=0.47\textwidth]{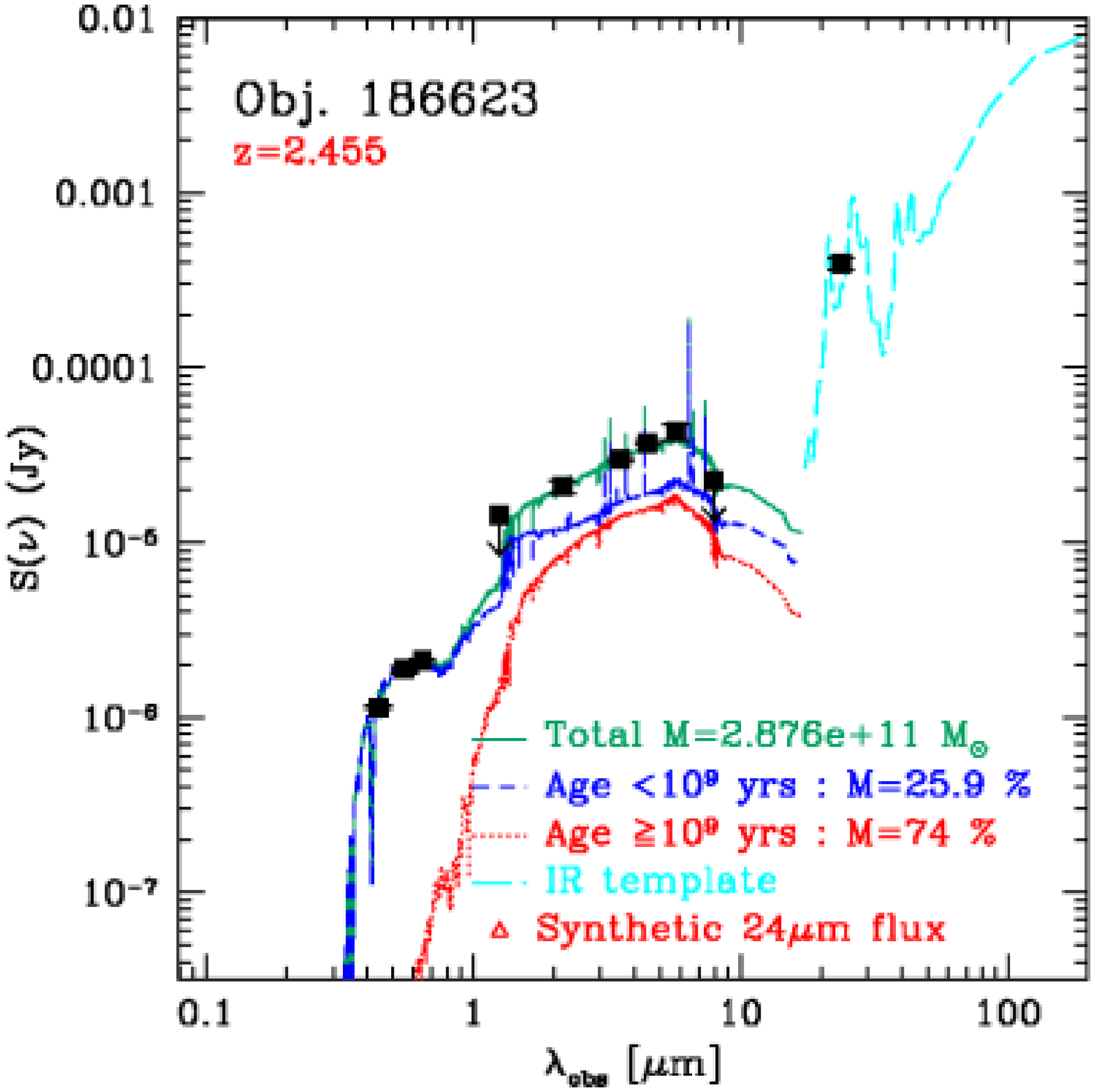}
\caption{Examples of SED fits of IR-peak sources. The {\em top} panels 
belong to 4.5$\mu$m peakers, the {\em bottom} ones show two 5.8$\mu$m-peak galaxies.
The two sources on the {\em left} are not detected at 24$\mu$m, and only an upper limit
is available, while those on the {\em right} have an actual 24$\mu$m 
flux above 250 $\mu$Jy (i.e. the 3$\sigma$ SWIRE limit in ELAIS-S1).
The dashed line represents the contribution to the best fit model by young and intermediate-age 
($<$1 Gyr) stars, while the dotted line is the light emitted by older stars
(age $\ge$1 Gyr). The solid line is the total best fit emission up to 5$\mu$m (restframe).
The long-dashed line longward of 5$\mu$m (restframe) is the adopted starburst template
re-processing the UV-optical light absorbed by dust to the IR.}
\label{fig:fits}
\end{figure*}

\begin{figure*}[!ht]
\centering
\rotatebox{-90}{
\includegraphics[height=0.49\textwidth]{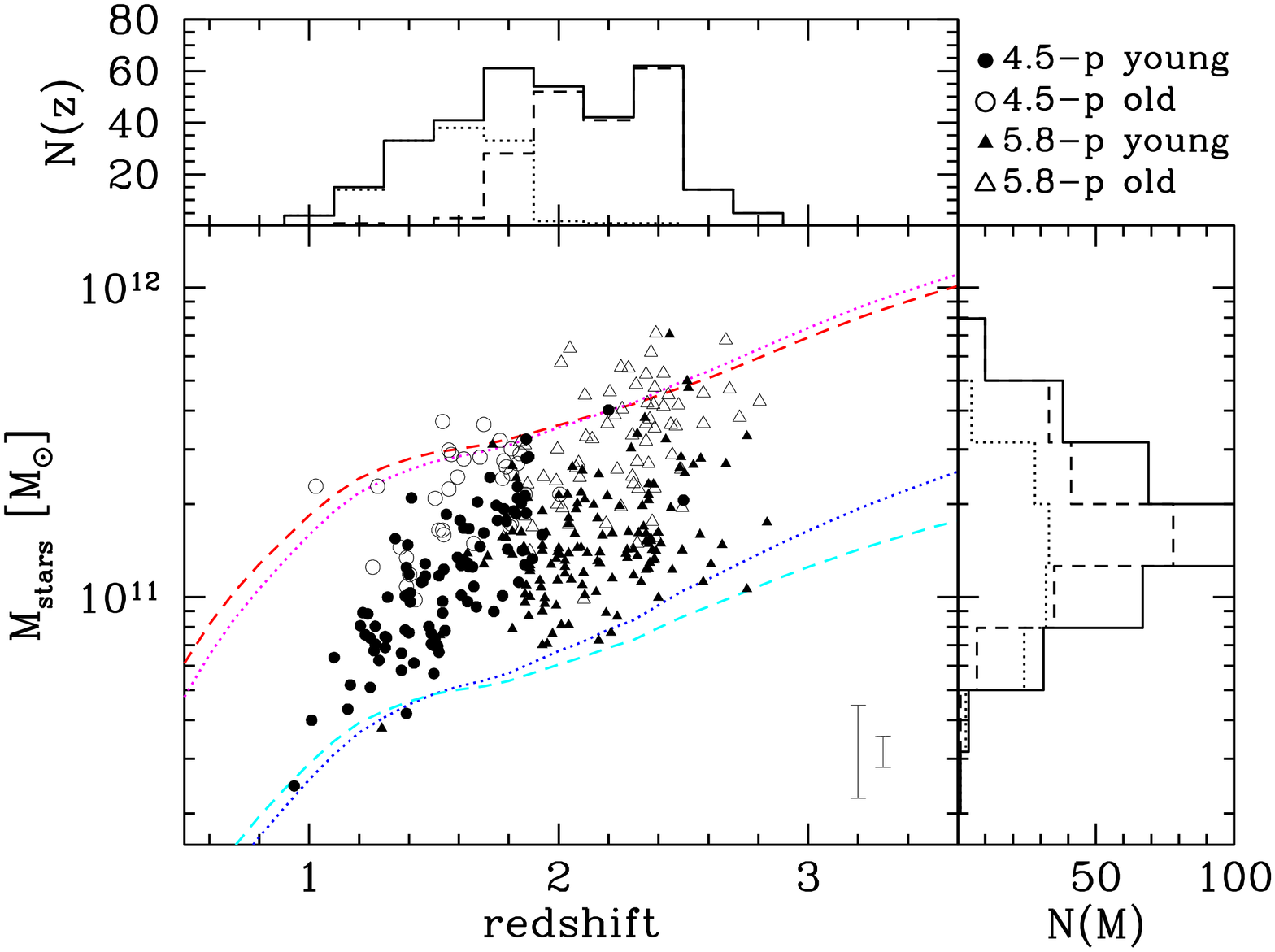}	
}
\rotatebox{-90}{
\includegraphics[height=0.49\textwidth]{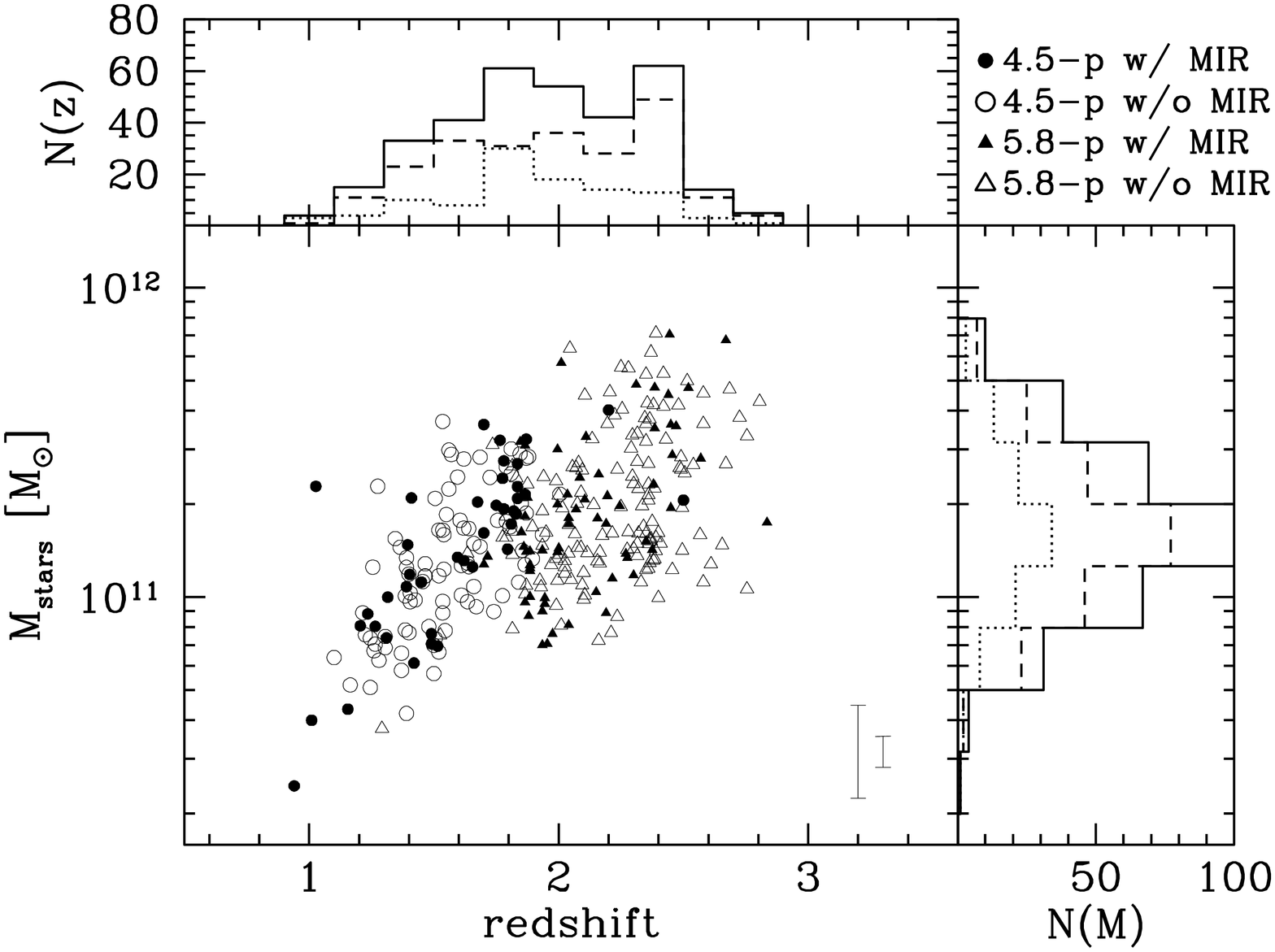}	
}
\caption{Distribution of stellar masses as a function of redshift. Circles and 
triangles represent 4.5$\mu$m- and 5.8$\mu$m-peakers respectively.
{\em Left} panel: filled and open symbols belong to sources dominated (best fit) 
by young or old stellar populations. The mass thresholds expected 
for the maximal and minimal $M_\star/L$ ratios in the sample are overplotted as
dashed (5.8$\mu$m-peakers) and dotted (4.5$\mu$m-peakers) lines.
Histograms are plotted for 
4.5$\mu$m-peakers (dotted) and 5.8$\mu$m-peakers (dashed). The solid histogram 
is the total distribution of sources. {\em Right} panel: filled and open symbols belong to
sources with or without a 24$\mu$m detection. Histograms have the same 
meaning, dotted ones representing objects detected by MIPS and dashed ones objects not detected.}
\label{fig:mass_z_normal_fit}
\end{figure*}

Typically an M82 template is adopted, but a different choice would not 
significantly affect the resulting stellar mass estimate, as demonstrated 
in \citet{berta2004}.
Increasing observational evidence exists that high-$z$
IR-peakers detected in the mid-IR resemble the M82 prototype.
Based on Spitzer mid-IR IRS spectroscopy, \citet{weedman2006} found that
$z\simeq1.9$ IR-peak galaxies  
are dominated by bright PAH features and lack deep silicate 10$\mu$m absorption.
\citet{rowanrobinson2005} studied and classified the SEDs of SWIRE sources over 6.5 deg$^2$
in the ELAIS-N1 fields, finding that M82-like starbursts are 3 times 
more numerous than colder Arp220-like objects. 
Finally, millimeter (250 GHz) observations of SWIRE 24$\mu$m-selected 
IR luminous galaxies, performed with the MAMBO bolometer on the IRAM/30m telescope
(Lonsdale et al., in prep.)
showed that the 1.2mm/24$\mu$m flux ratio of these sources resembles that 
of M82, lower that for an Arp220-like population.

Figure \ref{fig:fits} shows  a few examples of SED fits of IR-peak sources.
Overplotted on the observed fluxes is the best fit solution of the MSP synthesis:
the dashed line refers to young-intermediate (age $<1$ Gyr) stellar populations, and the dotted line
is the contribution from old (age $\ge1$ Gyr) stars. The solid line provides the 
total emitted light in the optical and near-IR wavelength range. 
Longward of 5$\mu$m (restframe), the long-dashed line is the starburst  
template used to model the IR emission from dust, heated by young stars.
The open triangle represents the 24$\mu$m flux predicted by the model, which 
often is beneath the observed data-point.

Figure \ref{fig:mass_z_normal_fit} reports the distribution of stellar 
masses as a function of redshift. 
Circles belong to 4.5$\mu$m-peakers, while triangles represent 5.8$\mu$m-peak 
sources. 

In the left panel, open symbols indicate objects whose best fit synthetic SEDs are
dominated by old (age $\ge10^9$ yr) stars, while for filled circles the bulk
of the luminosity is provided by younger populations.
The dashed and dotted histograms belong to 5.8$\mu$m- and 4.5$\mu$m-peakers.
Objects with a resulting lower mass are dominated by younger stellar
populations, as expected on the basis of the dependence of the $M_\star/L$ ratio on 
age. The effect of Malmquist bias is evident not only from the trend seen 
in the $M_\star$ {\em vs.} $z$ plot, but also from the distribution 
of sources in the mass space: 4.5$\mu$m-peak galaxies have masses
lower than 5.8$\mu$m-peakers, on average.

Typical error bars are shown: uncertainties on the stellar mass can be as high 
as 0.3 $dex$ for sources that lack two ($B$ and $V$) or all optical bands
 and --- in the worst cases --- also near-IR photometric data. 
This situation is represented by the big error bar. When good multi-wavelength
coverage is available, the uncertainty on the stellar mass reduces significantly (small 
error bar).

The mass thresholds expected at the 4 $\sigma$ flux limit of $S(5.8\mu\textrm{m})=34.4$ 
$\mu$Jy, as derived from the maximal and minimal $M_\star/L$ ratios
in the SED fitting results, are overlaid on the data. Dotted lines belong to 4.5$\mu$m-peakers and 
dashed lines to 5.8$\mu$m-peakers. See Sect. \ref{sect:completeness} for 
a discussion on the completeness of the sample.

The right panel of Fig. \ref{fig:mass_z_normal_fit} shows the same plot, but coded on the basis of 
24$\mu$m detections. Filled and open symbols refer to sources detected or not
detected in the MIPS 24$\mu$m channel, respectively, at the 3$\sigma$ level. 
The dashed and dotted 
histograms have the same meaning. 

\subsection{Effect of 24 $\mu$m data on the mass}\label{sect:fit_no24}

\begin{figure*}[!ht]
\centering
\includegraphics[width=0.6\textwidth]{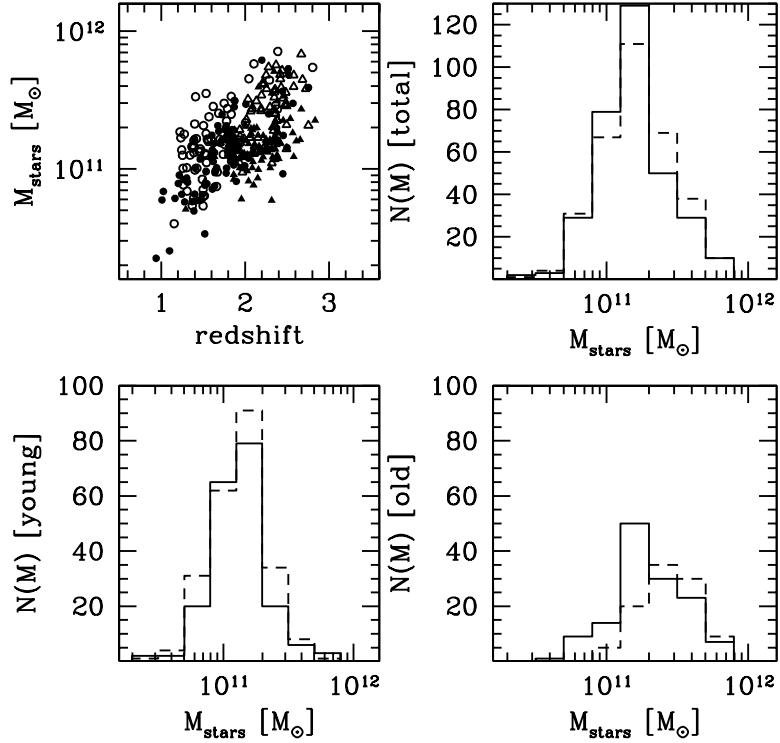}
\caption{Results of SED fitting, ignoring the 24$\mu$m 
fluxes (solid histogram), and comparison to the fit including 24$\mu$m 
(dashed, see Fig. \ref{fig:mass_z_normal_fit}). In the {\em top left} panel, filled 
and open symbols represent sources dominated (best fit) by young or old stellar populations,
respectively.
The mass distributions for these two sub-classes are shown in the two {\em bottom} 
panels.}
\label{fig:fit_no24}
\end{figure*}

The availability of mid-IR observations provides a valuable tool for 
constraining the amount of dust and the luminosity of young stellar populations
in the synthesized models, thus reducing the uncertainty on the derived stellar 
mass \citep[e.g.][]{berta2004}.

In order to understand the effect of the 24$\mu$m flux on the mass, i.e.
how it effectively influences the star formation history of the best fit 
solution, we have performed a fitting run ignoring the available 24$\mu$m
photometry.

The results are shown in Fig. \ref{fig:fit_no24}. The top-left panel shows the distribution 
of stellar masses as a function of redshift (symbols are the same as in Fig.
\ref{fig:mass_z_normal_fit}), while the top-right plot compares the 
new results (solid line, obtained excluding the 24 $\mu$m flux from the 
minimization) to the standard fit mass distribution (dashed histogram). The two bottom panels 
present the same comparison, for sources dominated by young and old stellar populations respectively
(more that 50\% of the total mass being assembled in stars younger or older than 1 Gyr).
 
If no mid-IR detection is available, the amount of dust and young stars 
are free to vary with no control, and very extinguished ($A_V\ge4$)
young populations, not visible in the optical domain, may exist.
As a consequence, the amount of young stellar mass is found to be larger, the total stellar mass 
inferred from SED fitting smaller on average, and the mass spread slightly wider.
The availability of MIPS data provides a valuable  
constraint on the recent history of star formation of galaxies and 
helps to avoid a significant number of SED fitting solutions that could not 
be a priori ruled out.

\subsection{Effect of near-IR (JK) data on the fit}\label{sect:fit_noJK}

The near-IR $J$ and $K_s$ data turned out to be very useful at the 
selection stage, in particular to exclude low-redshift interlopers with 
IRAC colors similar to 4.5$\mu$m-peakers.

\begin{figure*}[!ht]
\centering
\rotatebox{-90}{
\includegraphics[height=0.7\textwidth]{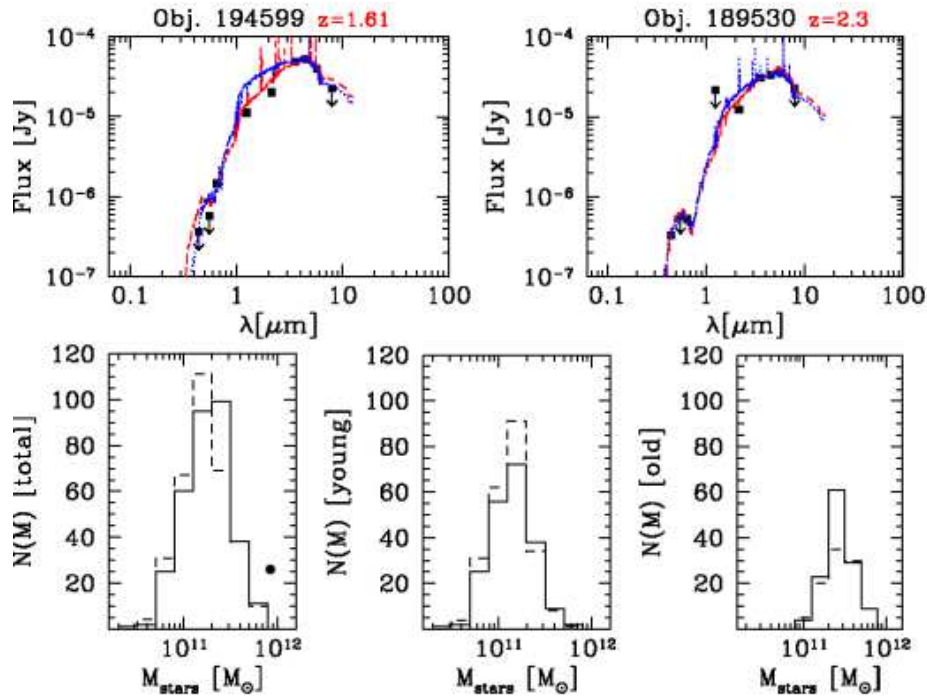}
}
\caption{Comparison of SED fitting with and without the $J$, $K_s$ data.
The {\em top} panels show an example of 4.5$\mu$m- ({\em left}) and 5.8$\mu$m-peaker
({\em right}). The dotted line shows the best fit obtained ignoring the 
real $J$, $K_s$ data, while the dashed line is the standard fit.
The three {\em bottom} panels show the distribution of masses for all sources
and young/old-dominated galaxies, for the standard fit (dashed histograms) and 
the fit obtained ignoring the $J$, $K_s$ data (solid lines).}
\label{fig:fit_noJK}
\end{figure*}

It is worthwhile exploring how observations at these wavelengths (1-2 $\mu$m)
constrain the estimate of stellar masses, by means of SED fitting.
We have therefore performed another fitting run, without taking into account 
the $J$, $K_s$ photometry. Figure \ref{fig:fit_noJK} shows a couple of 
SEDs and the comparison to masses from the standard fit.

The top panels include the SEDs of a 4.5$\mu$m-peaker (left) and a 5.8$\mu$m-peak
galaxy (right). The dashed line represents the fit obtained with the 
standard technique, i.e. accounting for all the available photometric data, 
while the dotted lines are the best fit models obtained 
by excluding the near-IR ($J$ and $K_s$) fluxes from the minimization.

In some cases, the $J$, $K_s$ data turn out to be useful
to constrain the age of the dominant stellar populations 
hosted by IR-peakers, since they sample the 
depth of the D4000 break, when combined with optical data. Nevertheless, 
the $J$, $K_s$ constraint turns out to only be effective, in this respect, 
for a small fraction of cases ($\le15$\%). The top panels in Fig. \ref{fig:fit_noJK}
show two examples among those for which these data are most useful.

It is worth noting that the advantage of having near-IR data is larger for 
4.5$\mu$m-peakers than for objects that peak at longer wavelength (in the 5.8
$\mu$m channel, i.e. at higher redshift). For the latter, the SED slope blueward of the 
1.6$\mu$m peak is, in fact, already defined by the $3.6-4.5-5.8$ $\mu$m colors
well enough to reasonably constrain the D4000 break, and therefore the best fit solutions
obtained with or without $J$, $K_s$ do not differ too much. 

Several authors \citep[e.g.][]{berta2004} had shown that the 
introduction of IRAC data in SED fitting provides tighter constraints on 
the stellar mass, reducing its uncertainty by a factor as high as 5, for 
$z>2$ sources.
In addition, \citet{wuyts2006} infer that $JHK$ band data can reduce 
the uncertainty on the stellar mass of blue galaxies with 
$(U-V)_{\textrm{rest}}<1$, while IRAC photometry is really effective only 
for redder sources, having a $R-$IRAC color
significantly redder than 1; this is the case for our IR-peakers.

Overall, the distributions of stellar masses, as obtained with and without 
the near-IR $J$ and $K_s$ data (Fig. \ref{fig:fit_noJK}), are not significantly different, 
confirming the result that these data do not play a fundamental role in deriving the
stellar mass of IR-peakers. On the other hand, as we have already pointed out, 
near-IR magnitudes are critical to avoid low-$z$ aliases. We also expect
these data to be more effective in constraining the age and $M_\star/L$ ratio of galaxies
in other redshift bins, e.g. when the blue side of the 1.6$\mu$m feature 
is not fully sampled by IRAC, $z\le1.5$.

\subsection{Choosing the IMF}

The choice of an initial mass function (IMF) different from 
Salpeter's can significantly affect the estimate of the stellar mass.

The \citet{salpeter1955} IMF has a constant $\alpha=1.35$ slope throughout 
the whole mass range considered, but 
in fact it was never measured down to 0.15 M$_\odot$ by Salpeter.

More recent determinations of the IMF showed that a flatter slope is 
needed at low masses ($M\le1$ M$_\odot$), in order to reproduce the observed Galactic data
\citep[e.g.]{miller1979,kennicutt1983,kroupa1993,kroupa2001,chabrier2003}
Consequently, assuming Salpeter's IMF results in including too many
low-mass stars (which dominate the stellar mass budget) in the galaxy modeling.
Introducing a drop-off at $M\le1$ M$_\odot$ leads to lower values of the stellar
mass of the analyzed galaxies.

At the bright end, \citet{elmegreen2006} pointed out that above 1 M$_\odot$ the IMF slope
is not steeper than Salpeter's; \citet{miller1979} and the other IMFs with a drop at
the highest masses were based on galactic disk measurements
which cannot be safely used to trace the high-mass end of the IMF,
because of the complicated SFH of the Galaxy. 

We study here the effect of a different choice in the IMF, by performing 
a new fitting run with the \citet{chabrier2003} IMF instead of the
classic \citet{salpeter1955} one.

The \citet{chabrier2003} IMF is described 
by a power-law for $M>1$ M$_\odot$ and a lognormal form
below this limit.
\citet{chabrier2003} and \citet{kroupa2001} are very similar to each other, 
but here we prefer to adopt the former because it is physically motivated and
provides a better fit to counts of low-mass stars and brown dwarfs in
the Galactic disc (see for
example Chabrier \citeyear{chabrier2001,chabrier2002,chabrier2003b} and also 
Bruzual \& Charlot \citeyear{bruzual2003}). 

\begin{figure*}[!ht]
\centering
\rotatebox{-90}{
\includegraphics[height=0.7\textwidth]{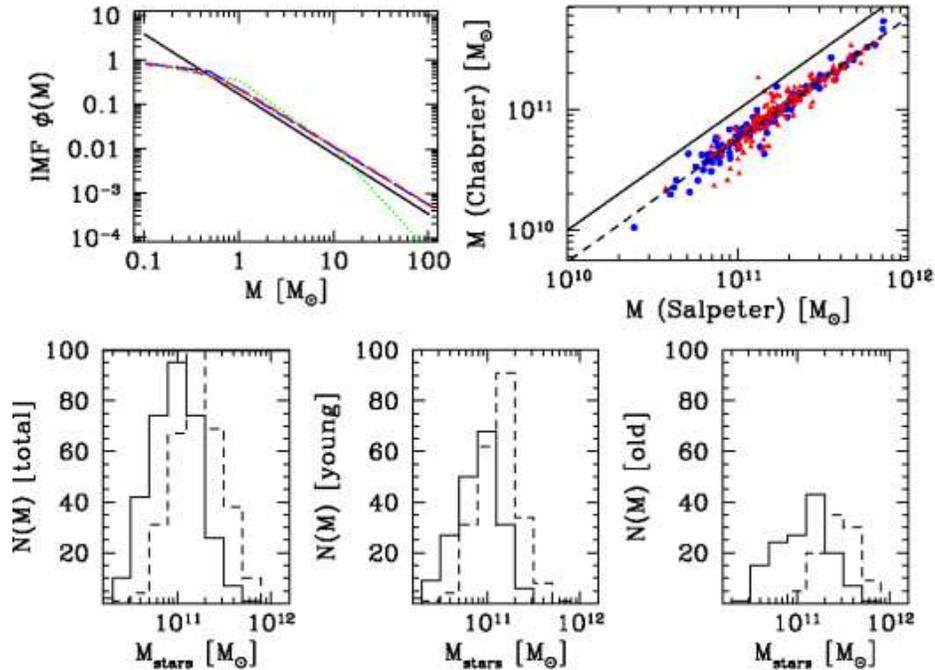}
}
\caption{Comparison of IR-peakers stellar masses as derived with a \citet{salpeter1955}
and a \citet{chabrier2003} IMFs. The {\em top left} panel shows different 
derivations of the IMF: Salpeter (\citeyear{salpeter1955}, solid line), 
Miller \& Scalo (\citeyear{miller1979}, dotted), Kroupa (\citeyear{kroupa2001}, dot-dashed) 
and Chabrier (\citeyear{chabrier2003}, dashed, almost superimposed over Kroupa's). The {\em top-right} panel shows 
the direct comparison of stellar masses in the two examined cases, for 4.5$\mu$m-peakers (circles) and 
5.8$\mu$m-peakers (triangles). The {\em bottom} panels show the stellar mass
distribution for the Salpeter (dashed) and Chabrier (solid) fits. The
three panels report the results for all sources and for objects dominated by
young-intermediate (age $<10^9$ yr) or old ($\ge10^9$ yr) SSPs.}
\label{fig:imf}
\end{figure*}

The results of this analysis are reported in Fig. \ref{fig:imf}, where
we show the shape of different IMF's (top left panel), and the direct comparison
between stellar masses as obtained with the Salpeter and Chabrier IMFs (top
right panel). In the bottom panels, the comparison of the stellar mass distribution 
for the two cases is shown. Solid histograms represent stellar masses 
derived with the Chabrier IMF and dashed ones with Salpeter's.

The diagrams show that the \citet{chabrier2003} IMF leads to systematically 
lower masses than the \citet{salpeter1955}, as expected. The difference 
between the two turns out to be $\sim$0.3 dex, represented by the dashed 
line in the top right panel of Fig. \ref{fig:imf}. The solid line sets the 
1:1 ratio.

The stellar mass distribution is roughly rigidly shifted to lower masses (bottom
panels). 
As far as the splitting into sources dominated by young-intermediate (age $<10^9$ yr) 
and old stars ($\ge10^9$ yr), a few objects migrate from the former to the
latter sub-samples, but the overall distributions are maintained.

Despite the fact that the choice of a \citet{chabrier2003} or \citet{kroupa2001} 
IMF restricts the inclusion of too many low-mass stars, 
the majority of literature studies on the 
stellar mass function of galaxies are based on the \citet{salpeter1955}
description of the IMF
\citep[e.g][]{fontana2006,fontana2004,drory2005,drory2004,franceschini2006,rudnick2006,dickinson2003,
papovich2001,shapley2001,sawicki1998}. 
For this reason, we will derive the stellar mass function of IR-peakers from 
the Salpeter-based SED fitting, keeping in mind that a different choice 
\citep[e.g.][]{chabrier2003} would produce a shift of the mass function 
to lower masses (e.g. by about 0.3 dex).

\subsection{Models with TP-AGB phase}\label{sect:fit_Maraston}

Recently, \citet{maraston2006} have performed stellar population synthesis
of $z=1.4-2.7$ galaxies from the GOODS Spitzer survey \citep{dickinson2003},
adopting the \citet{maraston2005} library and studying the differences
in the derived parameters, with respect to the Padova 
\citep[e.g.][]{fagotto1994a} models.

\begin{figure}[!ht]
\centering
\includegraphics[height=0.49\textwidth]{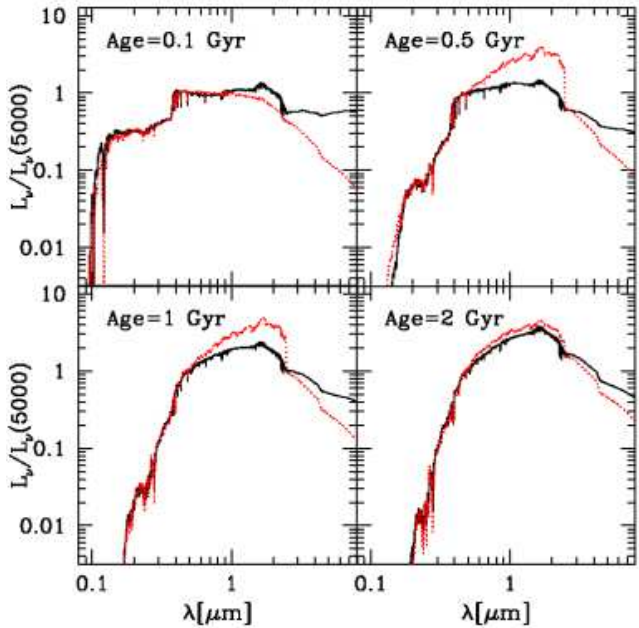}
\caption{Comparison of SSPs from the Padova 1994 \citep[e.g.][ solid lines]{fagotto1994a}
and the \citet{maraston2005} library (dotted), for a Salpeter IMF and solar 
metallicity. The critical age range, when the TP-AGB phase is active, is shown.
The SEDs are normalized at 5000 \AA.}
\label{fig:padova_maraston}
\end{figure}

The \citet{maraston2005} library is based on the Frascati tracks 
\citep[e.g.][]{cassisi1997}. The main difference between the 
two is that the Padova tracks include a certain
amount of convective overshooting on the main sequence (MS), whereas the
Frascati tracks do not; moreover the temperature distribution of the red giant 
branch (RGB) phase is shifted to cooler temperatures in the Padova tracks 
for $Z\ge Z_\odot$.
As a consequence, the MS lifetime is longer for the Padova tracks and the 
RGB phase is delayed, with respect to the Frascati models.

In any case, the key difference of the \citet{maraston2005} approach is the 
way the thermally-pulsing asymptotic giant branch (TP-AGB) phase is 
included in the evolution of stellar populations, i.e. by means of 
a semi-empirical fuel consumptions table, in contrast to
``a posteriori'' recipes used in isochrone synthesis. TP-AGB stars dominate the 
near-IR (e.g. $K$ band) luminosity for SSP ages between 0.3 and 2 Gyr (for 
a Salpeter IMF and solar metallicity), while main-sequence stars dominate 
in the optical (e.g. $V$ band).

At the transition between the early-AGB branch 
and the TP-AGB, i.e. when the TP-AGB phase begins, 
the near-IR luminosity of stars significantly increases, and
therefore the near-IR mass-to-light ratio ($M_\star/L_K$) of the SSP
drops suddenly by a factor $3-5$. In models without this 
phase, $M_\star/L$ monotonically increases, 
independent of wavelength.
At ages between 0.8 and 1 Gyr, the models based on the 
Padova tracks have $M_\star/L_K$ larger than the \citet{maraston2005}
ones, as a consequence of the different treatment of the TP-AGB phase. At older ages, 
$M_\star/L_K$ is smaller for the Padova 1994 case, because of the cooler RGB phase.

\begin{figure*}[!ht]
\centering
\rotatebox{-90}{
\includegraphics[height=0.7\textwidth]{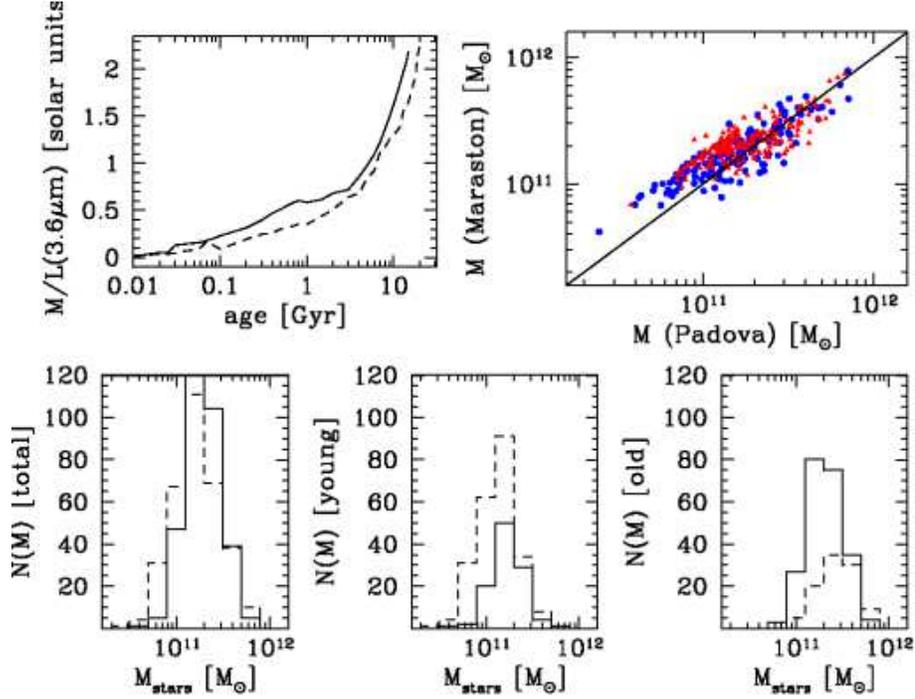}
}
\caption{Comparison of stellar mass distributions as obtained with the \citet{maraston2005} 
SSP library (solid lines) and the 
Padova 1994 \citep[e.g.][]{fagotto1994a} models (dashed). The {\em top left} panel
reports the restframe mass-to-light ratio in the 3.6$\mu$m IRAC band, as computed
for the two SSP libraries, a Salpeter IMF and solar metallicity. The {\em top right}
panel shows the direct comparison of stellar masses, derived in the two cases, for 4.5$\mu$m-peakers
(circles) and 5.8$\mu$m-peakers (triangles).
The {\em bottom} panels 
show the comparison of the stellar mass distribution for all sources and 
objects dominated by young/old stars (in the best fit).}
\label{fig:fit_maraston}
\end{figure*}

\citet{maraston2006} model the observed SEDs of high-$z$ galaxies, 
adopting different star formation histories (SFH): an instantaneous burst, 
a exponentially declining star formation (SF), a prolonged burst, and a constant SF.
As a result, they find that the 
\citet{maraston2005} library leads to systematically younger best fit solutions
for these systems, and hence to lower stellar masses, with respect to 
the Padova 1994 library. The IRAC Spitzer data 
turn out to be very useful in constraining the mass, and the \citet{maraston2005}
models provide a better fit to the observed photometry.

With this in mind, we have performed a similar analysis on our 
sample of IR-peakers, adapting the \citet{maraston2005} SSP library 
to the SPS code and running it in the same exact way as before.
Figure \ref{fig:padova_maraston} compares the SEDs of the Padova and 
Maraston SSPs, in the age range in which the TP-AGB phase is active.

The results of this further analysis are reported in Fig. \ref{fig:fit_maraston},
where the stellar masses derived with the Maraston library (solid histogram) are compared 
to those obtained earlier with our ``standard'' fit (dashed).
The top-left panel reports the $M_\star/L_{3.6}$ ratio for the SSPs in the 
two libraries and the top-right plot shows the direct comparison of stellar 
masses as derived in the two cases.
As far as the low-mass end of the distribution is concerned, 
stellar masses based on the Maraston library turn out to be higher than in the 
Padova-94 case. Conversely, the Padova-94 fit provides slightly 
lower masses at the bright end. 

The free-form MSP technique adopted for fitting the IR-peakers
produces a different result, compared to that of \citet{maraston2006}.
The new library of models produces best fit solutions characterized by 
completely different SFH's, with respect to the previous ones, based on the 
Padova tracks. The relative contribution of young and old stars in the best fits
changes significantly, with the fraction of objects dominated by old populations 
increasing in the Maraston case. The two bottom plots show the $M_\star$ distribution
for sources dominated (in mass) by old and young stars: not only the balance 
between old and young SSPs changes, but also the distributions peak at lower masses 
(for old-dominated sources) and higher masses (for young-dominated ones), than in 
the Padova-94 case.
On the other hand, the overall stellar mass distribution (top right panel)
is not significantly changed, apart for a shift of $M<10^{11}$ M$_\odot$ galaxies
to higher masses.

\subsection{Possible AGN contributions}\label{sect:poss_agn}

First of all, the sample of IR-peakers has been checked against
bright X-ray sources, thanks to the available XMM-Newton 
survey in the area \citep{puccetti2006}. Ten sources turn out to host 
an AGN X-ray component, seven 4.5$\mu$m-peakers and three 5.8$\mu$m-peakers.
The former are characterized by a bright 8.0$\mu$m excess, 
likely due to the presence of torus warm dust in the mid-IR SED
(see also Berta et al. \citeyear{berta2007}, Lonsdale et al. 2007, for an extensive description 
of SED shapes). 
The three 5.8$\mu$m-peakers emitting in the X-rays show a smooth stellar peak, 
diluted by the AGN dust. These sources have been excluded form further analyses.

The remaining sources show a sharp IR-peak, and, 
in performing the SED analysis, we have assumed that only stellar
emission contributes to the observed SEDs, while no AGN component is present.
In fact, the warm dust in the AGN torus would emit at IRAC-MIPS
wavelengths, producing a power-law SED or 
diluting the 1.6$\mu$m (restframe) stellar peak.
Hence a sharp stellar peak 
shifted to the 4.5 or 5.8 $\mu$m bands should identify
sources whose near-IR emission is dominated by stars.
Nevertheless, the presence of a 
possible AGN component cannot be fully excluded.  

\citet{berta2007} have analyzed UV-optical (restframe) 
spectra of IR-peakers observed with the Keck-I 10m telescope, selected 
in order to show no evidence of AGN components on IRAC color-color
diagrams \citep[e.g.][]{stern2005,lacy2004}.
Because of instrument limitations, the sample was limited to the brightest
IR-peakers in the SWIRE survey, with $r^\prime\lesssim 24.5$ (Vega).

These authors find evidence for AGN emission lines in 62\% of the detected 
IR-peakers, two thirds belonging to the type-1 and one third to the type-2 classes.
The spectroscopic redshift of these sources lies between $z=1.3-2.0$, 
including 5.8$\mu$m-peakers. 
In order to explain the presence of the 1.6$\mu$m peak in the 4.5 and 5.8$\mu$m
channels, a multi-component (stars and AGN) fit to the SEDs of these sources was performed.
The AGN component partially dilutes the 1.6$\mu$m peak, 
and --- more importantly --- modifies 
its shape, resulting in an apparent shift to longer wavelengths.

At optical restframe wavelengths, the AGN emission is overwhelmed by stars in 
the host galaxy, while it
re-emerges in the IRAC domain, especially in the two longest
wavelength channels (5.8 and 8.0 $\mu$m): 
torus dust is not negligible for these sources, and thus might affect the 1.6$\mu$m-peak 
selection.
All the sources hosting an AGN were detected at 24 $\mu$m above 250 $\mu$Jy,
have moderate 24$\mu$m excess ($[3.6-24]=1.5-2.5$ in 
AB units) and show a wide range of optical-IRAC colors $(R_C-3.6)_{AB}=2-4$.
AGN features were identified only in sources above $r^\prime\lesssim 23.8$,
corresponding to $V_J\simeq23.6$ (AB) for a gray source 
\citep{fukugita1996}.

If an AGN component were present, the estimate of the stellar mass
would be affected in two different ways, operating in the same direction,
i.e. decreasing the actual mass, with respect to that measured by ignoring the AGN.
Firstly, the torus warm dust emission would result in a lower redshift 
than expected for IR-peakers, and the 
luminosity (hence the mass) would be consequently lower. Moreover
the torus would contribute to the observed IRAC fluxes, therefore  
the light in stars (hence the mass) would be even smaller.

We can not a priori rule out the presence of an AGN component;
nevertheless, it is worth noting that only 48 sources (i.e. $\sim$12.5\% 
of the IR-peaker sample in ELAIS-S1) are brighter than $V_J=23.6$ [AB]. 
Among these, only 23 are also detected at 24$\mu$m, corresponding to $\sim$6\%
of the sample, and finally only $\sim$3.5\% (13 objects) show a moderate 
$(3.6-24)$ color.

We will not discuss this problem further, since the fraction of 
sources that might be affected is very small.


\section{The mass function}\label{sect:mass_function}

The results obtained from spectro-photometric synthesis have been 
exploited to derive the contribution of $z\simeq1-3$ IR-peakers to the
galaxy global stellar mass density, $\rho^\star$.

The derived stellar masses are used to build the stellar 
mass function of IR-peakers in two ways: with the $V_a$ technique and 
applying the STY \citep{sandage1979} method. The latter uses
a parametric function, $\Phi\left(M,\textrm{param}\right)$ and 
a Monte Carlo Markov Chain algorithm adopting a Bayesian 
formalism.

\subsection{The accessible volume}\label{sect:access_V}

\begin{figure*}[!ht]
\centering
\includegraphics[height=0.49\textwidth]{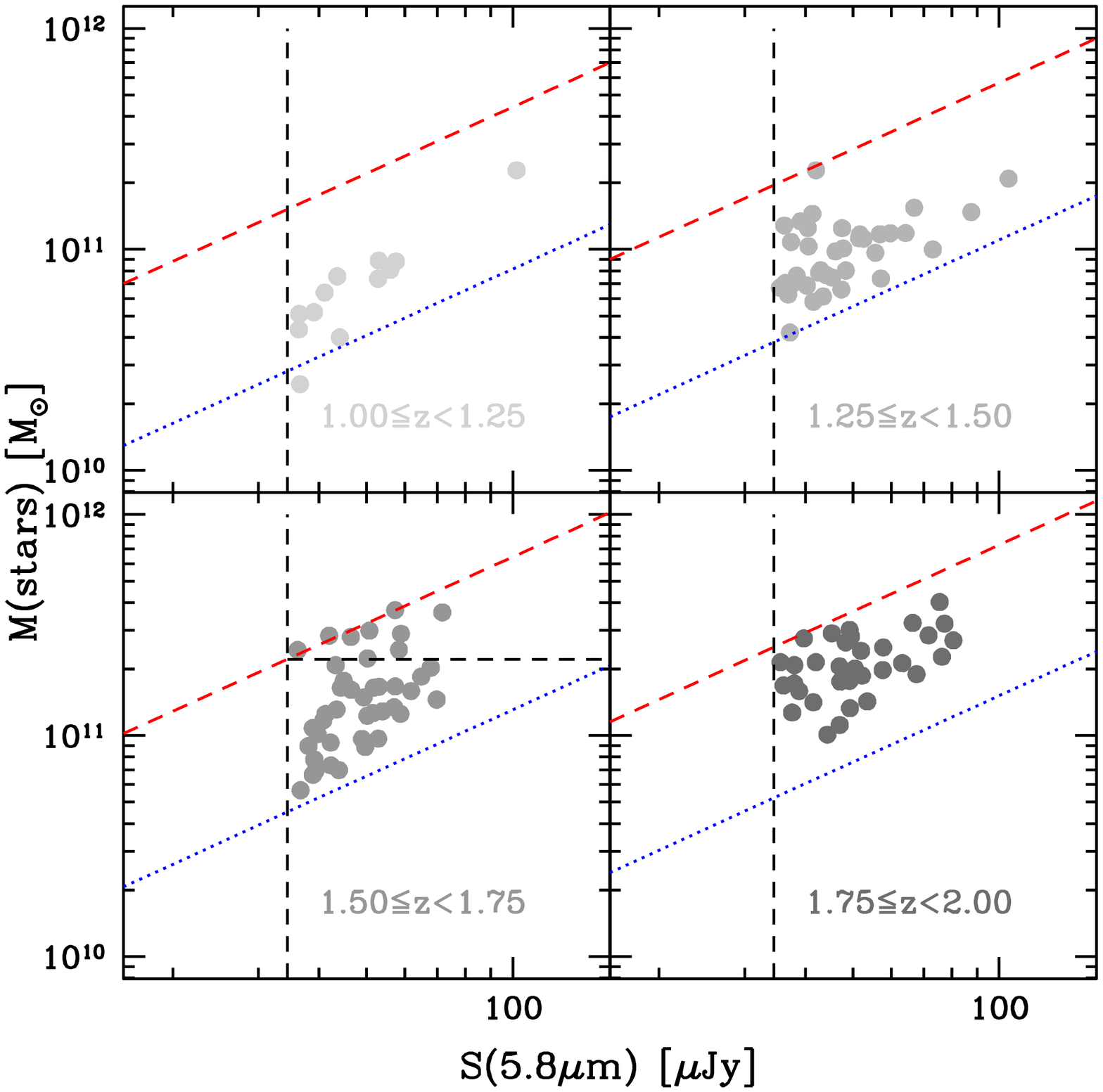}
\includegraphics[height=0.49\textwidth]{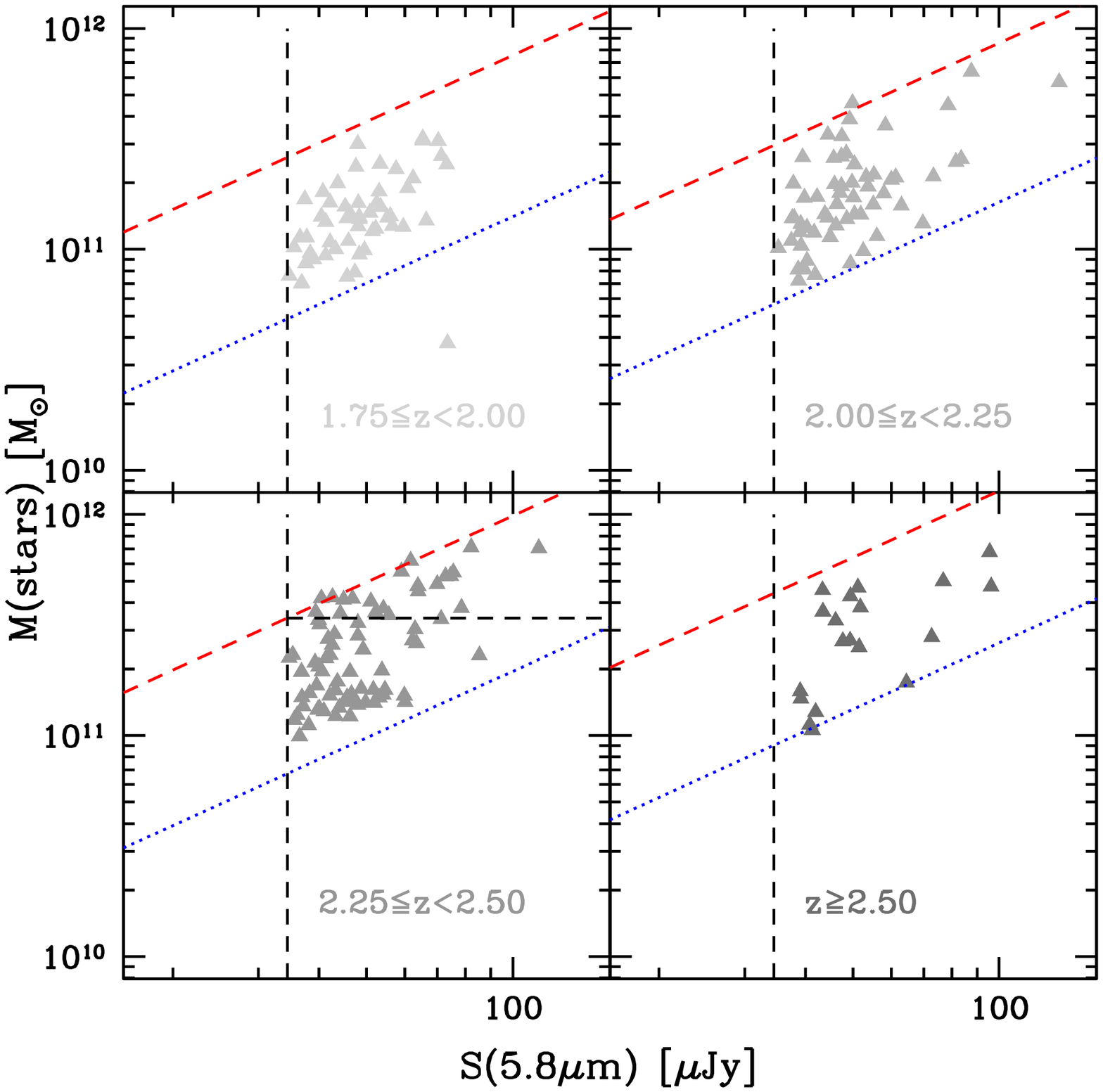}
\caption{Mass incompleteness. {\em Left} and {\em right} panels 
belong to 4.5$\mu$m-peakers and 5.8$\mu$m-peakers, respectively. 
The distribution of stellar mass as a function of observed fluxes is shown for 
different redshift bins. The diagonal lines represent the tracks at the
bin's central redshift described by the minimal (dotted) and maximal (dashed)
$M_\star/L$ ratios in our samples. The vertical dashed lines set the 
adopted $S(5.8\mu\textrm{m})=34.4$ $\mu$Jy flux limit.
The horizontal lines represent the 
mass threshold above which the samples are fully complete. See text 
and \citet{fontana2004} for details on the technique adopted to correct
incompleteness.}
\label{fig:mass_flux}
\end{figure*}

In order to compute the comoving number density of IR-peakers 
 at the given redshift, we 
have adopted the well-known $1/V_a$ method \citep[e.g.][]{schmidt1968}. 
The {\em accessible volume}, $V_a$,
in which each galaxy could be detected in the survey is computed 
by taking into account the selection criteria:
\begin{itemize}
\item the flux cut at $S(5.8\mu\textrm{m})=34.4\ \mu \textrm{Jy}$,
\item the IR-peak color selection.
\end{itemize}
The former provides the classical $V_{max}$ estimate \citep{schmidt1968}, 
related to the maximum redshift $z_{max}$ at which the given galaxy would be
observable in our survey.
The $z_{max}$ is computed by adopting the best fit SED model 
(see Sect. \ref{sect:fit}), redshifted and k-corrected 
until cosmological dimming fades it below the adopted flux limit 
\citep[e.g.][]{hogg1999}:
\begin{equation}\label{eq:cosmo_dimming}
S(\nu)=\frac{L\left(\nu\right)}{4\pi d_L^2}K\left(\nu,z\right)\textrm{,}
\end{equation}
where $d_L$ is the luminosity distance and 
$K(\nu,z)$ is the $k$-correction through the filter $T(\nu)$:
\begin{equation}\label{eq:k_corr}
K\left(\nu,z\right)=\left(1+z\right)\frac{\int_{\nu_1}^{\nu_2}L\left[\nu\left(1+z\right) 
\right]T\left(\nu\right)d\nu}
{\int_{\nu_1}^{\nu_2}L\left(\nu\right)T\left(\nu\right)d\nu}\textrm{.}
\end{equation}

The IR-peak condition defines a spherical shell, delimited by redshifts 
$z_{min}^{peak}$ and $z_{max}^{peak}$, where the IR-peak selection is valid, i.e.
where the 1.6$\mu$m restframe peak is detected in the IRAC 4.5 or 5.8$\mu$m channel.
The effective accessible volume for each source in our color-selected survey 
is thus given by:
\begin{equation}\label{eq:access_V}
V_a = \frac{\Omega}{4\pi}\int_{z_{min}^{peak}}^{min\left(z_{max},z_{max}^{peak}\right)} \frac{dV}{dz}dz \textrm{,}
\end{equation}
where $\Omega$ is the surveyed sky area, and the volume element $dV$ 
depends on the adopted cosmology.

It is worth noting that the $(K_s-3.6)>0$ cut, applied in order to avoid 
low-redshift interlopers, does not affect the accessible volume estimate, 
because the cutoff redshift is always smaller than $z_{min}^{peak}$.

\subsection{Completeness}\label{sect:completeness}

The IR-peaker sample was first selected by applying a flux cut
at 5.8$\mu$m. Despite the fact that we are directly probing the restframe near-IR
emission of these galaxies, which is primarily powered by low-mass stars
dominating their stellar mass budget, it is not possible to define 
a sharp mass limit encompassing the whole sample.

In fact, the mass-to-light ratios ($M_\star/L$) of these galaxies spans a 
relatively wide range, between $\sim$0.03 and $\sim$0.5 at 3.6$\mu$m 
(restframe). At a given redshift 
and flux, these translate into very different mass values, as shown by the 
minimal and maximal galaxy tracks shown in Fig. \ref{fig:mass_z_normal_fit} 
(left panel). As an example, it would thus be more correct to 
say that at $z=2$ the sample is limited to masses $M>7\times10^{10}$ or $M>3.5\times10^{11}$ 
M$_\odot$ for low or high $M_\star/L$ objects, respectively.

This effect has been thoroughly examined by \citet{dickinson2003b}, 
\citet{fontana2006} and \citet{fontana2004}. 
The latter perform a very detailed analysis 
and provide a valuable recipe to partially correct the mass incompleteness 
of flux-limited samples. 

Figure \ref{fig:mass_flux} reports the distribution of stellar masses
as a function of the observed 5.8$\mu$m flux for the galaxies in our sample, 
split into different redshift sub-bins. The left panels refer to 4.5$\mu$m-peakers, 
while on the right side 5.8$\mu$m-peak galaxies are shown.
The vertical dashed line represents the adopted 5.8$\mu$m flux 4 $\sigma$ cut.
The diagonal dashed and dotted lines are the tracks described by the sources with 
minimal and maximal $M_\star/L$ ratios in the sample (see also Fig. \ref{fig:mass_z_normal_fit}),
at the central redshift of each sub-bin. 

\begin{figure*}[!ht]
\centering
\rotatebox{-90}{
\includegraphics[height=0.65\textwidth]{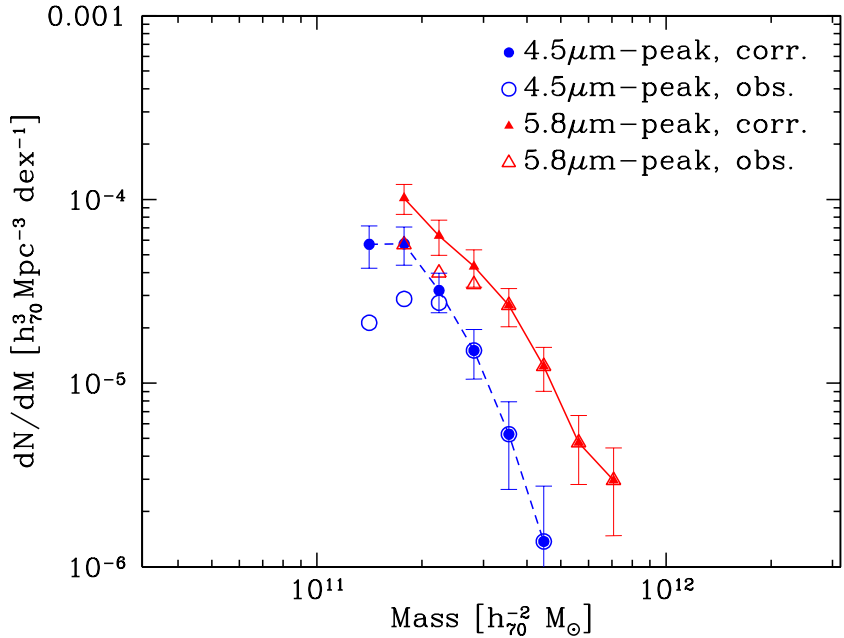}
}
\caption{Mass function of 4.5$\mu$m-peakers (dashed line, circles) and 
5.8$\mu$m-peakers (solid line, triangles) in the ELAIS-S1 area, as obtained 
with the standard spectro-photometric fit and the 
Padova-94 library (see Sect. \ref{sect:fit}). Filled symbols belong to the completeness-corrected mass 
function, while open ones represent the data before any correction was applied.
Error bars account only for Poisson noise statistics.}
\label{fig:mf_standard}
\end{figure*}

Since the majority of 4.5$\mu$m- and 5.8$\mu$m-peak sources 
are in the $1.50\le z<1.75$ and $2.25\le z<2.50$ sub-bins
(see the redshift distribution in Fig. \ref{fig:mass_z_normal_fit}), 
we will use these as reference to derive the threshold completeness mass 
for the two different classes.

The horizontal lines in the two panels focused on these redshift sub-bins
represent the stellar mass levels above which the samples are definitely complete, 
because these thresholds are set by the maximal $M_\star/L$ observed values.
These masses turn out to be $\sim2\times10^{11}$ and $3.5\times10^{11}$ M$_\odot$
for the two populations. Below these values, the sample becomes increasingly
incomplete.

Following \citet{fontana2004}, it is possible to compute the fraction of 
sources lost at each given mass, by exploiting the observed distribution
of $M_\star/L$ ratios and fluxes. 
We defer the reader to their paper for further details on this technique.
We keep those mass bins for which the fraction of missing sources 
turns out to be $\lesssim0.5$.
In this way it was possible to extend our analysis down to 
$1.25\times10^{11}$ and $1.6\times10^{11}$ M$_\odot$ for 
4.5$\mu$m- and 5.8$\mu$m-peakers respectively (see Tab. \ref{tab:obs_mf}).
Note also that below these limits the amplitude of random flux fluctuations 
due to sky-noise is large and the contamination of the IR-peaker
sample by other classes of sources can not be controlled (see Sect. \ref{sect:skynoise}).

\subsection{The observed galaxy comoving number density}\label{sect:obs_MF}

Within the $V_a$ formalism, the comoving number of galaxies per unit volume, 
in each redshift bin and in the mass bin $\Delta M$, is obtained by:
\begin{equation}\label{eq:rho}
\Phi\left(M\right)\Delta M = \sum_i \frac{1}{V_a^i}\Delta M \textrm{,}
\end{equation}
where the sum is made over all galaxies in the given mass bin.

Figure \ref{fig:mf_standard} shows the distribution of 
the comoving number of galaxies as a function of stellar mass (i.e. the 
``observed'' stellar mass function) for 4.5$\mu$m-peakers (dashed line, filled 
circles) and 5.8$\mu$m-peakers (solid line, triangles).
Table \ref{tab:obs_mf} reports the data.
The stellar masses come from the spectro-photometric fit obtained with 
the Padova-94 library and the Salpeter IMF, and accounting for all the available photometric data and 
upper limits.
Filled symbols represent the mass function corrected for incompleteness, while 
open symbols show the data as obtained before applying any correction.

\begin{figure}[!ht]
\centering
\includegraphics[width=0.45\textwidth]{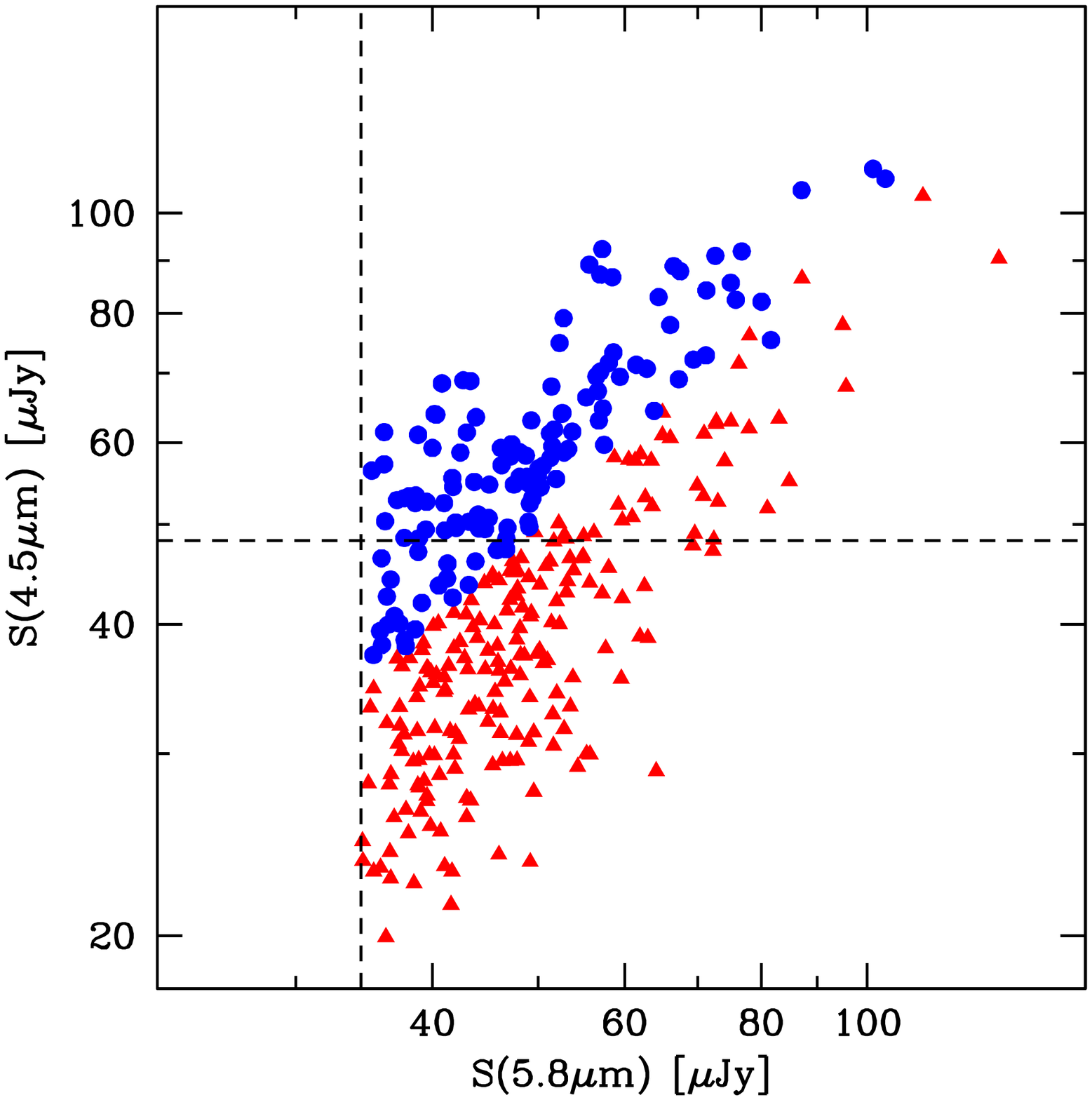}
\caption{Distribution of 4.5$\mu$m and 5.8$\mu$m fluxes for IR-peakers.
Circles represent 4.5$\mu$-peakers, triangles belong to 5.8$\mu$m-peakers.
The vertical dashed line corresponds to the $S(5.8\mu\textrm{m})=34.4$ $\mu$Jy
flux cut. The horizontal dashed line sets the $S(4.5\mu\textrm{m})=48.2$ $\mu$Jy
limit, obtained by assuming a restframe $S(H)/S(K)=1.4$ ratio for typical IR-peak galaxies.}
\label{fig:flux_distr}
\end{figure}

The number density of 4.5$\mu$m-peak population turns out to be 
significantly lower than that provided by 5.8$\mu$m-peakers. It is worth 
noting that the selection based on a 5.8$\mu$m flux cut is optimized 
for the detection of the 1.6$\mu$m stellar peak in IRAC channel 3, while 
in order to have a comparable selection for 4.5$\mu$m-peakers we would 
have needed to apply a 4.5$\mu$m flux cut. 
The 5.8$\mu$m flux cut corresponds to a restframe $H$ band selection for 
5.8$\mu$m-peakers and to a $K$ band selection for 4.5$\mu$m-peakers.
Assuming that the typical $S(H)/S(K)$ flux ratio of a galaxy (e.g. a IR-peaker)
is $\sim1.4$, then the $S(5.8\mu\textrm{m})=34.4$ $\mu$Jy cut 
corresponds to $S(4.5\mu\textrm{m})=48.2$ $\mu$Jy for a 4.5$\mu$m-peak galaxy.
As a confirmation of this effect, Figure \ref{fig:flux_distr} shows the distribution of 4.5$\mu$m and 5.8$\mu$m
fluxes for our sources. 

However, the 5.8$\mu$m-peakers sample includes also all those
galaxies not detected at 8.0$\mu$m, but still consistent  
with the 5.8$\mu$m-peak selection, when using the 8.0$\mu$m
upper limit.
Unfortunately, since the 5.8$\mu$m band is the least sensitive 
among the 3.6, 4.5 and 5.8 $\mu$m SWIRE/IRAC
channels, it is not possible to perform a comparable 4.5$\mu$m selection.
In fact, in order to include similar objects 
to the 4.5$\mu$m-peakers sample, all the $z=1-2$ galaxies with 
$S(4.5\mu\textrm{m})\simeq34.4$ $\mu$Jy, not detected at 5.8um,
but still consistent with the 4.5um-peak selection, should 
be taken into account. 
The resulting sources are detected only in two IRAC channels 
(3.6$\mu$m and 4.5$\mu$m) and are barely detected in J or Ks. Hence 
aliasing by low-$z$ interlopers or contamination by power-law objects 
would be hard to control.

We conclude that our completeness correction is not able to control this 
selection deficit in the 4.5$\mu$m-peakers sub-sample.
Therefore we will limit the parametric derivation of the mass 
function to the 5.8$\mu$m peakers only.

The error bars on the mass function plotted in Fig. \ref{fig:mf_standard}
account only for Poisson statistics. Nevertheless, the uncertainties on 
the stellar mass estimate, caused by degeneracies in the SFH space and 
errors on photometric redshifts, introduce a non negligible contribution to the 
uncertainty on the observed stellar mass function and must be taken into
account.

Due to these uncertainties, an object can move from the mass bin
it formally belongs to and actually contribute to the stellar mass function 
in another mass regime. The probability for each galaxy to contribute to 
the comoving number density in other mass bins is given by the 
$\chi^2$ distribution obtained during SED fitting with the ASA algorithm.

For sake of clarity we do not plot these additional error bars in Fig. 
\ref{fig:mf_standard}, but they 
will be fully taken into account when fitting the observed data 
with a parametric mass function, by using a Bayesian approach (see Sect. 
\ref{sect:MF_fit}).

\begin{table*}[!ht]
\centering
\begin{tabular}{c | c c c | c c}
\hline
\hline
$\log(\textrm{Mass})$ & $\Phi(M)$ 4.5$\mu$m-p. & $\Phi(M)$ 5.8$\mu$m-p.& $\Phi(M)$ 5.8$\mu$m-p (evo) & \multicolumn{2}{|c}{Missing fraction} \\
$[$h$_{70}^{-2}$ M$_\odot]$ & \multicolumn{3}{c|}{$[10^{-5} $h$_{70}^3$ Mpc$^{-3}$ dex$^{-1}]$} & 4.5$\mu$m-p.& 5.8$\mu$m-p.\\
\hline
11.15 $\pm$ 0.05	&	5.70 $\pm$ 1.47 &	  --			& 	--		& 0.53 & -- \\
11.25 $\pm$ 0.05	&	5.73 $\pm$ 1.35 &	 10.26 $\pm$ 1.89	& 7.29 $\pm$ 1.35	& 0.42 & 0.54\\
11.35 $\pm$ 0.05	&	3.19 $\pm$ 0.78 &	 6.34  $\pm$ 1.38	& 5.24 $\pm$ 1.14	& 0.14 & 0.44\\
11.45 $\pm$ 0.05	&	1.51 $\pm$ 0.45 &	 4.32  $\pm$ 0.99	& 3.75 $\pm$ 0.86	& -- & 0.37\\
11.55 $\pm$ 0.05	&	0.53 $\pm$ 0.26 &	 2.65  $\pm$ 0.62	& 2.45 $\pm$ 0.58	& -- & 0.20\\
11.65 $\pm$ 0.05	&	0.14 $\pm$ 0.14 &	 1.23  $\pm$ 0.33	& 1.20 $\pm$ 0.32	& -- & -- \\
11.75 $\pm$ 0.05	&	--		&	 4.74  $\pm$ 0.19	& 4.39 $\pm$ 0.18	& -- & -- \\
11.85 $\pm$ 0.05	&	--		&	 2.95  $\pm$ 0.15	& 2.88 $\pm$ 0.14	& -- & -- \\
\hline
\end{tabular}
\caption{Observed mass function of SWIRE IR-peakers, as obtained after correction
for incompleteness effects, for a Salpeter IMF and Padova-94 stellar tracks. 
We report the measure values of the galaxy comoving number density and the 
fraction of missing sources in each mass bin (see the discussion on completeness in Sect. \ref{sect:completeness}).
For 5.8$\mu$m-peakers the mass function obtained after evolutionary correction is also included (column 4).}
\label{tab:obs_mf}
\end{table*}

\begin{figure}[!ht]
\centering
\includegraphics[width=0.45\textwidth]{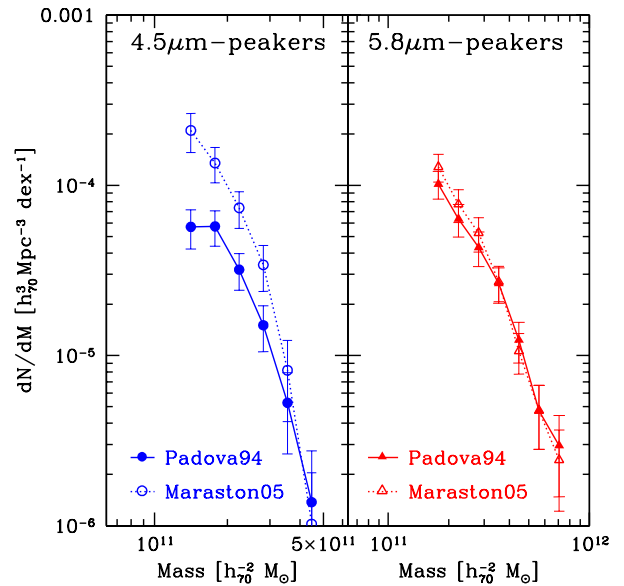}
\caption{Comparison of the mass function as obtained with the Padova-94 (filled symbols, solid lines)
and Maraston (\citeyear{maraston2005}, open symbols, dotted lines) libraries, after correction 
of incompleteness.}
\label{fig:sfr_mf_std_maraston}
\end{figure}

\subsection{Results with the TP-AGB enhanced phase}

The IR-peakers mass function obtained with the Padova-94 library is 
compared to the results from the \citet{maraston2005} approach in 
Fig. \ref{fig:sfr_mf_std_maraston}. Data are corrected for incompleteness, 
in the same way described for the Padova-94 case.

As far as 4.5$\mu$m-peakers are concerned, the two libraries lead to very different mass 
functions, mainly because low-mass objects migrate to higher mass values, when the 
\citet{maraston2005} SSPs are used (see also Fig. \ref{fig:fit_maraston} and Sect. 
\ref{sect:fit_Maraston}). Nevertheless, this mass function is still 
affected by a strong incompleteness effect, which cannot be recovered.

In the case of 5.8$\mu$m-peakers, the difference between the two results 
is less dramatic and the two libraries lead to very similar mass functions.

Note, however, that the comparison of our results to previous estimates of the stellar 
mass function at high redshift \citep[e.g.][ among others]{fontana2006,fontana2004,drory2005,
franceschini2006}, will make use of the results obtained exploiting the 
Padova-94 library.

\subsection{Evolutionary effects}

\begin{figure*}[!ht]
\centering
\includegraphics[width=0.46\textwidth]{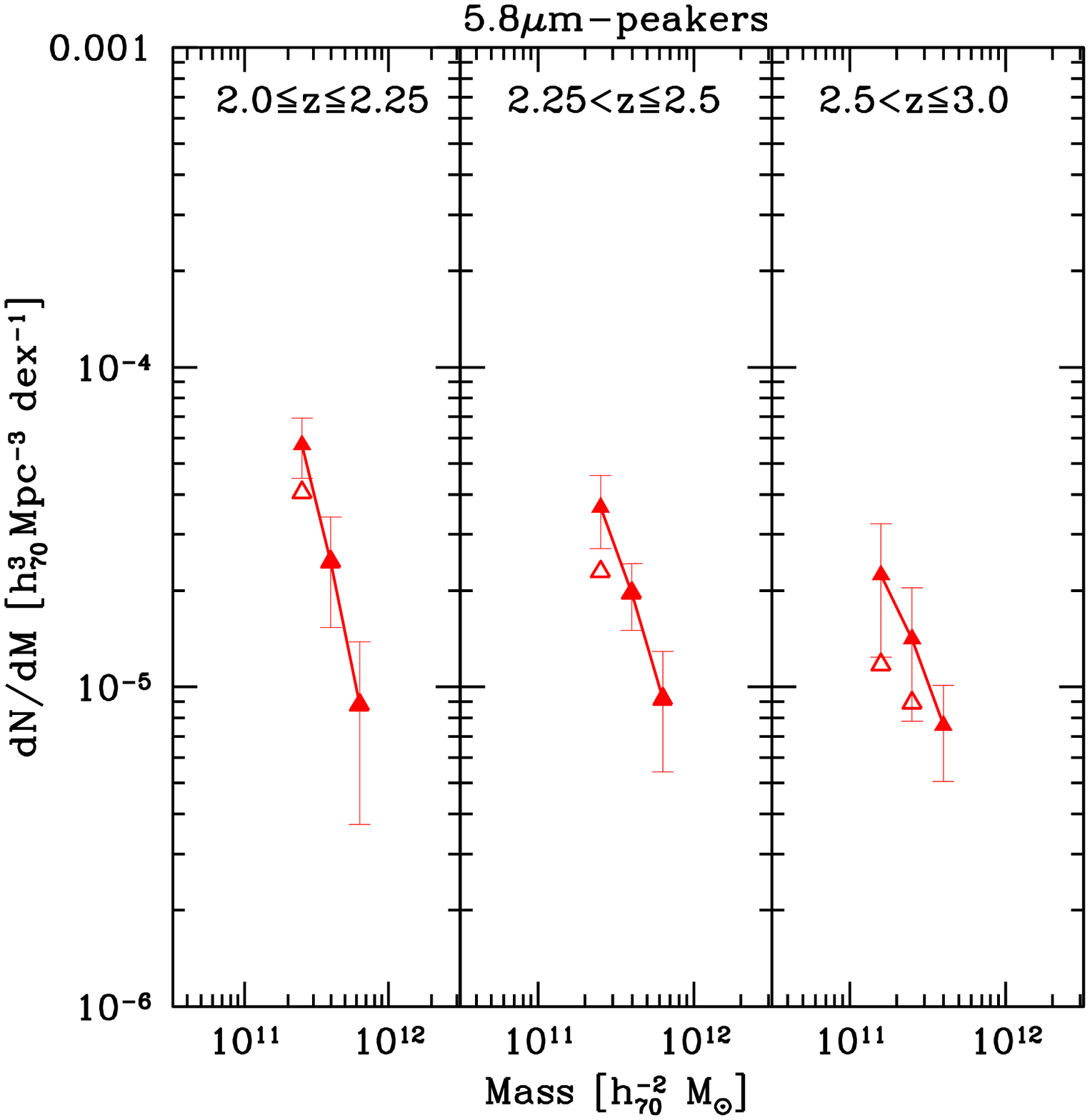}
\includegraphics[width=0.48\textwidth]{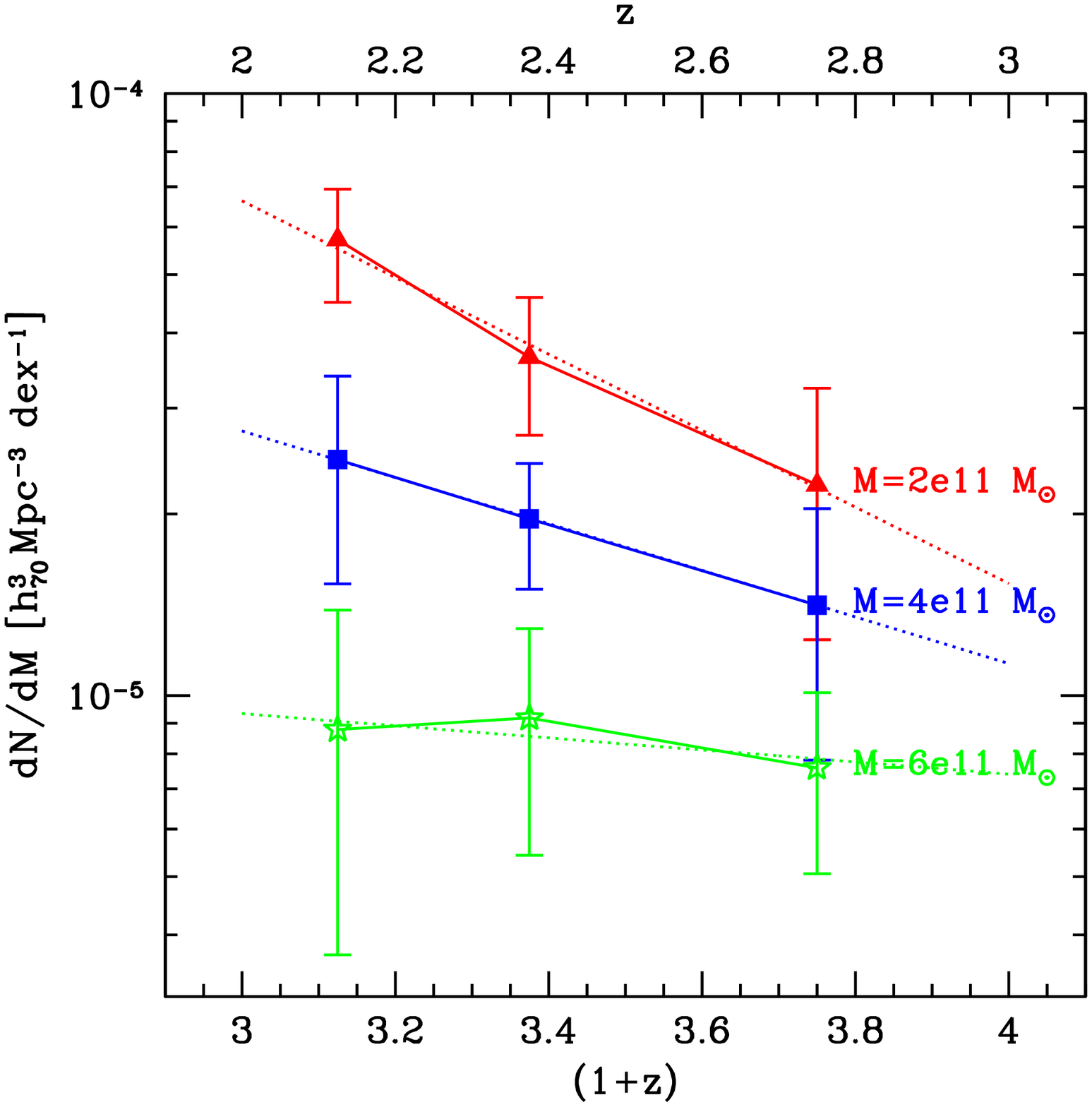}
\caption{{\em Left} panel: mass function of 5.8$\mu$m-peakers split into three 
redshift sub-bins (same symbols as in Fig. \ref{fig:mf_standard}). {\em Right} panel: 
evolutionary trend of the 5.8$\mu$m-peakers stellar mass density as a function of 
redshift for three different mass bins. Solid lines connect the observed data, 
dotted lines represent a power law fit to the evolution.}
\label{fig:mf_evo}
\end{figure*}

\begin{table}
\centering
\begin{tabular}{c c c}
\hline
\hline
Mass range & $a$ & $b$ \\
$[10^{11}$ M$_\odot]$ &  &  \\
\hline
1.6-3.2 & -0.64 $\pm$ 0.06 & -2.27 $\pm$ 0.19\\
3.2-5.0 & -0.39 $\pm$ 0.04 & -3.40 $\pm$ 0.09\\
5.0-7.0 & -0.10 $\pm$ 0.11 & -4.73 $\pm$ 0.28\\
\hline
\end{tabular}
\caption{Evolutionary coefficients describing the dependence of the stellar mass density
on redshift and mass, in the form of a $\left(1+z\right)$ power law (see text for more details).}
\label{tab:mf_evo_param}
\end{table}

\begin{figure}
\centering
\includegraphics[width=0.45\textwidth]{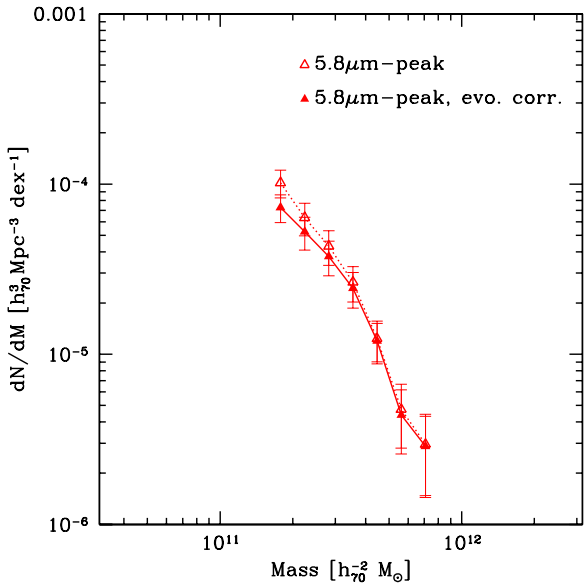}
\caption{Effect of evolutionary correction on the 
stellar mass function of IR-peakers. The open symbols and the dotted line belong to 
the mass function derived without evolutionary correction, while the filled symbols and 
the solid line account for it.}
\label{fig:evo_corr} 
\end{figure}

The mass density of galaxies increases very rapidly between redshift 3 and 2, 
by a factor of $\sim4$, according to Fontana et al. (\citeyear{fontana2006}, 
see also Sect. \ref{sect:integrated_rho}).
It is thus worthwhile studying the evolutionary details of the 
galaxy stellar mass assembly over this redshift range.
The left panel in Fig. \ref{fig:mf_evo} reports the stellar mass 
function of 5.8$\mu$m-peakers split for the first time into three redshift 
sub-bins: $2.0\le z\le2.25$, $2.25<z\le2.5$ and $2.5<z\le3.0$. 
Note that the integration boundaries in Eq. \ref{eq:access_V} are now set by 
the used redshift sub-bins.

The number of massive galaxies decreases in the higher redshift sub-bins, 
as the normalization of the MF becomes smaller by a factor 2-4, depending on mass.
Despite the large error bars, a flattening of the stellar 
mass function seems to be detected, as redshift increases. 
The right panel in Fig. \ref{fig:mf_evo} shows the evolution of the stellar 
mass density as a function of redshift and for three different mass bins,
highlighting that the highest-mass tail ($M\ge4\times10^{11}$ M$_\odot$) of the mass function evolves 
less rapidly than lower-mass bins.

For each mass range considered, we have reproduced the evolutionary trends with a power law:
\begin{equation}\label{eq:mf_evo}
\frac{dN}{dM}\left( z \right) \propto b\left(M\right)\times \left(1+z\right)^{a\left(M\right)}\textrm{.}
\end{equation}
The resulting evolutionary parameters are reported in Tab. \ref{tab:mf_evo_param}. Dotted lines in 
the right panel of Fig. \ref{fig:mf_evo} represent the power law fit.

In any case, it is worth to point out that above $z=2.5$ too few galaxies are included in the 5.8$\mu$m sample, 
and the significance of the derived mass function is rather poor.
Only by exploiting the whole SWIRE area (49 deg$^2$) will it be 
possible to build a catalog of massive galaxies at $z=2-3$ large enough to 
study the evolution of the number density at these epochs.

In order to compare our results to literature data 
\citep[e.g.][]{fontana2006,drory2005,gwyn2005}, the whole $z=2-3$ bin 
will be considered in the following analyses. 
Computing the stellar mass density on this wider redshift range 
allows one to exploit a higher S/N ratio, but on the other hand dilutes the 
information on the evolution of the mass function across redshift. 

As far as a broad redshift range is considered, it is very important to correct 
the derived stellar mass density against evolutionary effects.

Since this is a flux-limited sample, fainter (less massive) objects 
probe smaller volumes than more massive objects. 
Therefore the highest mass tail of the mass function is averaged 
over the full $z$ range considered, but the low-mass bins are contributed by 
galaxies populating only the low-$z$ part of the whole $z=2-3$ 
redshift range. Hence at the faint-end the effect of negative evolution is not balanced 
and tends to steepen the mass function.

If on one hand the volume effect is corrected by the $V_a$ and 
completeness analyses, on the other hand the intrinsic evolution is not accounted for
in this way. Moreover this evolution strongly depends on mass.

We therefore computed the necessary correction factor for each galaxy, 
by comparing the average stellar mass density over the $z=2-3$ range
to that predicted at the given redshift by our power-law fit, as a function of mass.
This additional coefficient is then applied to Eq. \ref{eq:rho} when computing the 
stellar mass function. Figure \ref{fig:evo_corr} compares the stellar mass function 
obtained without and with the evolutionary correction. The low-mass end flattens, as expected.
The fourth column of Tab. \ref{tab:obs_mf} reports the mass function of 5.8$\mu$-peakers corrected for evolution.

\subsection{Parametric stellar mass function}\label{sect:MF_fit}

An independent characterization of the stellar mass function of massive galaxies at 
high redshift is the parametric approach by 
Sandage, Tammann, Yahil (\citeyear{sandage1979}, STY).
We focus the analysis on 5.8$\mu$m-peakers only, 
because the 4.5$\mu$m-peaker sample turned out to be affected by
incompleteness effects that we could not control (see Sect. \ref{sect:obs_MF}).

The results of this analysis will then also be used to estimate the contribution 
of this population to the global stellar mass 
density of galaxies.

As commonly found in the literature, we choose to reproduce the 
observed data with the usual parametric description by \citet{schechter1976}:
\begin{equation}\label{eq:schechter}
\Phi\left(M\right) = \Phi^\ast \ \left(\frac{M}{M^\ast}
\right)^\alpha \ e^{-\frac{M}{M^\ast}}\textrm{,}
\end{equation}
where --- for the sake of clarity --- $M$ here denotes the galaxy stellar 
mass, previously called $M_\star$.

The three free parameters $\Phi^\ast$, $\alpha$ and $M^\ast$ represent the 
normalization, the slope in the low mass regime, and the transition mass 
between a power-law and the exponential drop-off or the e-folding mass of the 
latter.

Since we are sampling the very high mass tail of the galaxy stellar mass 
function, SWIRE IR-peakers are well suited to constrain the $M^\ast$ parameter, 
while we expect to probe $\alpha$ only in a poor way. We will discuss this issue 
later.

While performing this analysis, it is very important to 
account for the main sources of uncertainty on the galaxy number density, 
as described above (see Sect. \ref{sect:obs_MF}). The first source of uncertainty 
is represented by Poisson noise, since the sample consists only of 
203 5.8$\mu$m-peakers.  
The second is that the stellar mass
estimate for each individual galaxy has non-negligible uncertainty, which 
must be accurately included in the parametric fit.

We used a Markov-Chain Monte Carlo (MCMC) sampling of the parameter 
space, to explore the
posterior probability function of the model, with parameters
comprising both the three Schechter parameters and also
additional hyper-parameters to represent each of the individual galaxy
masses.  

According to the Bayes' theorem:
\begin{equation}\label{eq:bayes}
{\cal P}\left(\theta|d\right) = \frac{{\cal P}\left(d|\theta\right) {\cal P}\left(\theta\right)}{{\cal P}\left(d\right)} \textrm{,}
\end{equation}
i.e. the conditional probability of the Schechter parameters, given the data,
is equal to the conditional probability of the data, given the parameters,
times the prior probability of the parameters, divided by a normalization factor.
The theorem can be paraphrased as:
\begin{equation}\label{eq:bayes2}
\textrm{posterior} = \frac{\textrm{likelihood}\times\textrm{prior}}{\textrm{evidence}}\textrm{.}
\end{equation}
The model's posterior
probability function (given the data) is explored using MCMC sampling. 

The likelihood of the Schechter parameters is determined using 
the standard STY method \citep{sandage1979}, derived from the 
luminosity function formalism.
The mass function prior probability is assumed to be a top-hat (i.e. weak) 
prior in the Schechter parameters, 
with boundaries sufficiently wide as to
have negligible effects. The only adopted exceptions are that 
$M^\ast$ and $\Phi^\ast$ are assumed to be non-negative 
(as we are dealing with physical stellar masses), while $\alpha<0$.
The latter assumption is made in order to prevent $\Phi(M)$ deflections 
in the low mass domain, where no constraints are available for SWIRE 
IR-peakers.

The hyper-parameters are constrained solely by the prior
knowledge of the stellar mass probability distribution function (PDF, 
provided by the SED fitting); therefore 
their marginalization automatically accounts for the mass uncertainties.
The prior probability is thus computed as 
\begin{equation}\label{eq:prior}
\textrm{Prior} \propto e^{- \chi^2/2} \textrm{.}
\end{equation}

We use a Metropolis-Hastings algorithm as our MCMC sampler.
In order for this sampler to converge efficiently, we use a combined
strategy for generating proposal steps.  For the Schechter
parameters, we draw new steps from univariate
Gaussian distributions, whose widths were given by 
dummy MCMC runs.
The galaxy mass sampling for each galaxy is made by drawing randomly
an input-sample from the set for each galaxy.  
We use rejection
sampling, based on the $\Delta \chi^2$ of each input sample, relative to
the best fit \citep{vonneumann1951} to improve the efficiency of this part
of the sampling.
We note that this second part of the sampling is independent of the
current position in parameter space; this makes the
Metropolis-Hastings acceptance criterion easy to assess.

Because each sample is quick to evaluate, we are able to
calculate a very
large number ($10^7$) of steps in one hour of 3 GHz CPU time.

\subsection{Results of parametric analysis}\label{sect:res_MF_fit}

\begin{figure*}[!ht]
\centering
\includegraphics[width=0.49\textwidth]{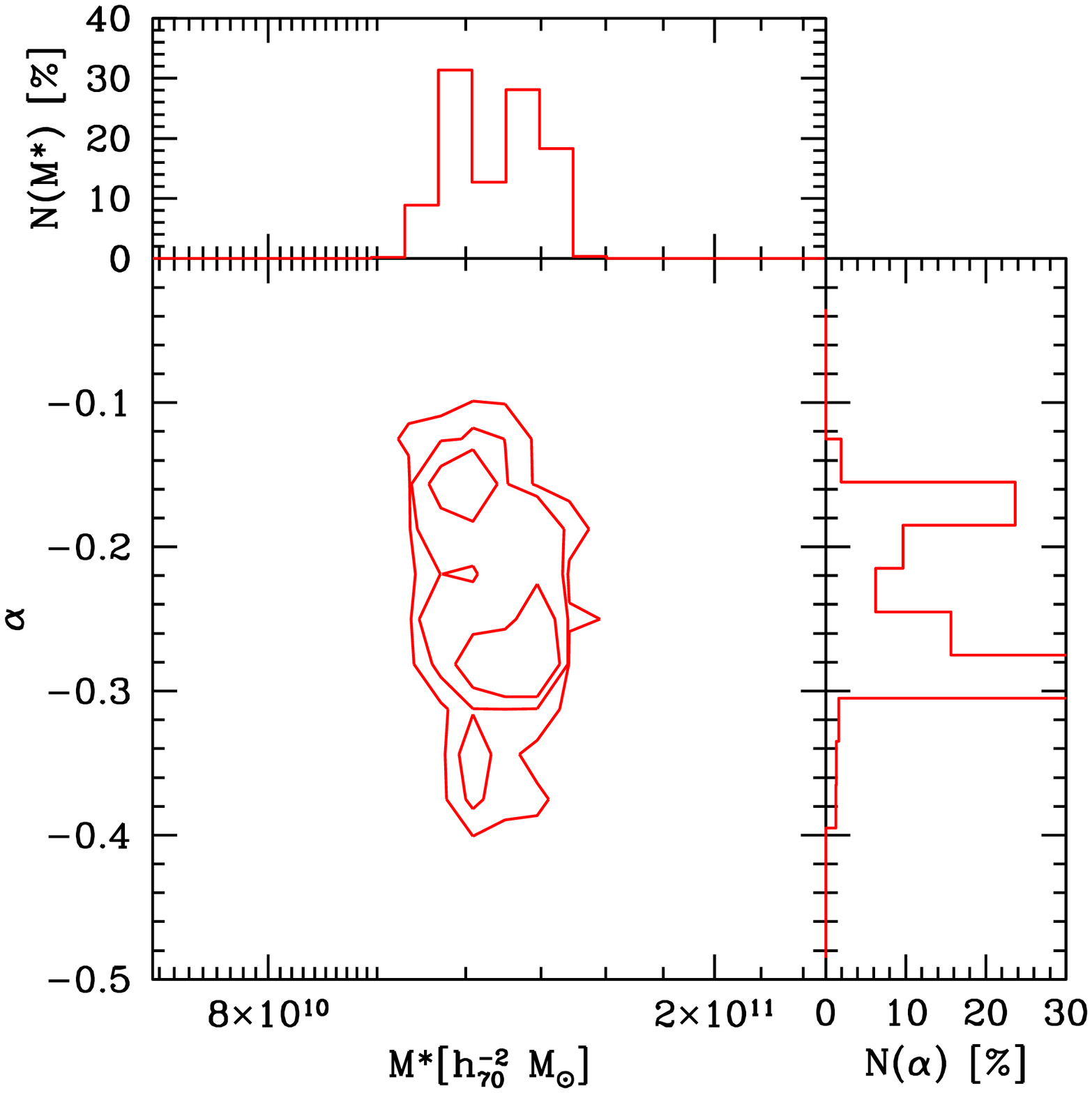}
\includegraphics[width=0.49\textwidth]{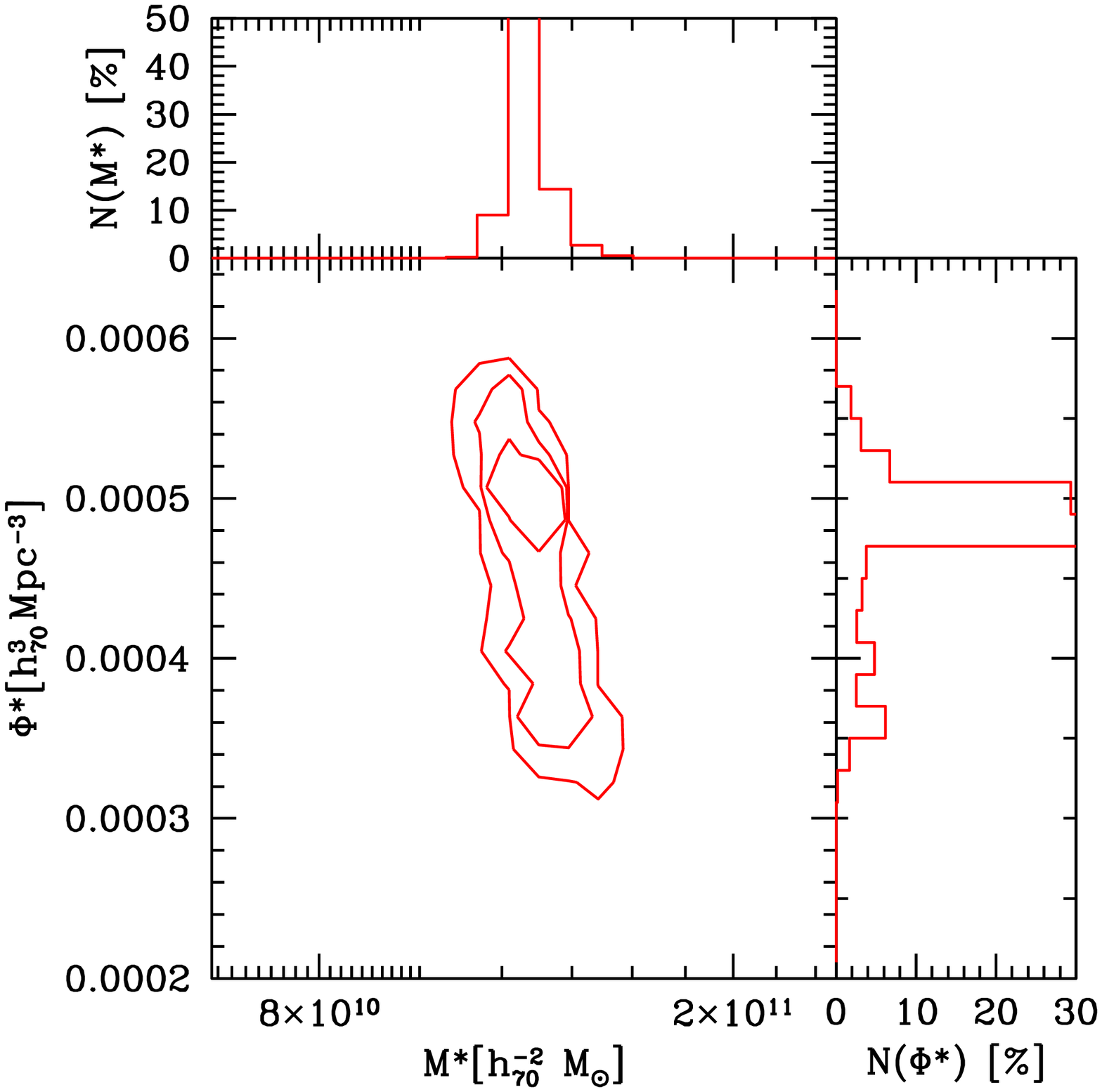}\\
\rotatebox{-90}{
\includegraphics[height=0.49\textwidth]{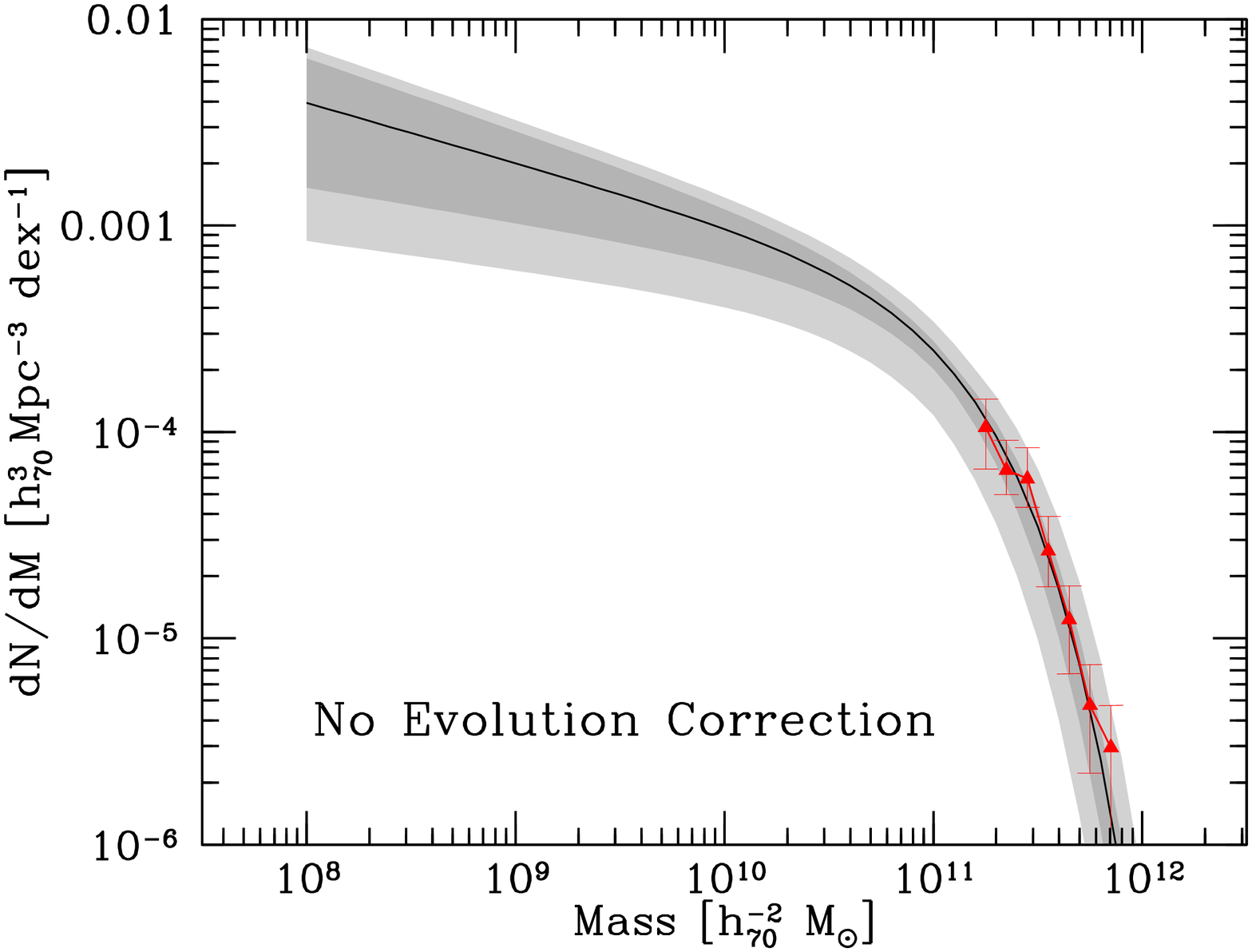}
}
\rotatebox{-90}{
\includegraphics[height=0.49\textwidth]{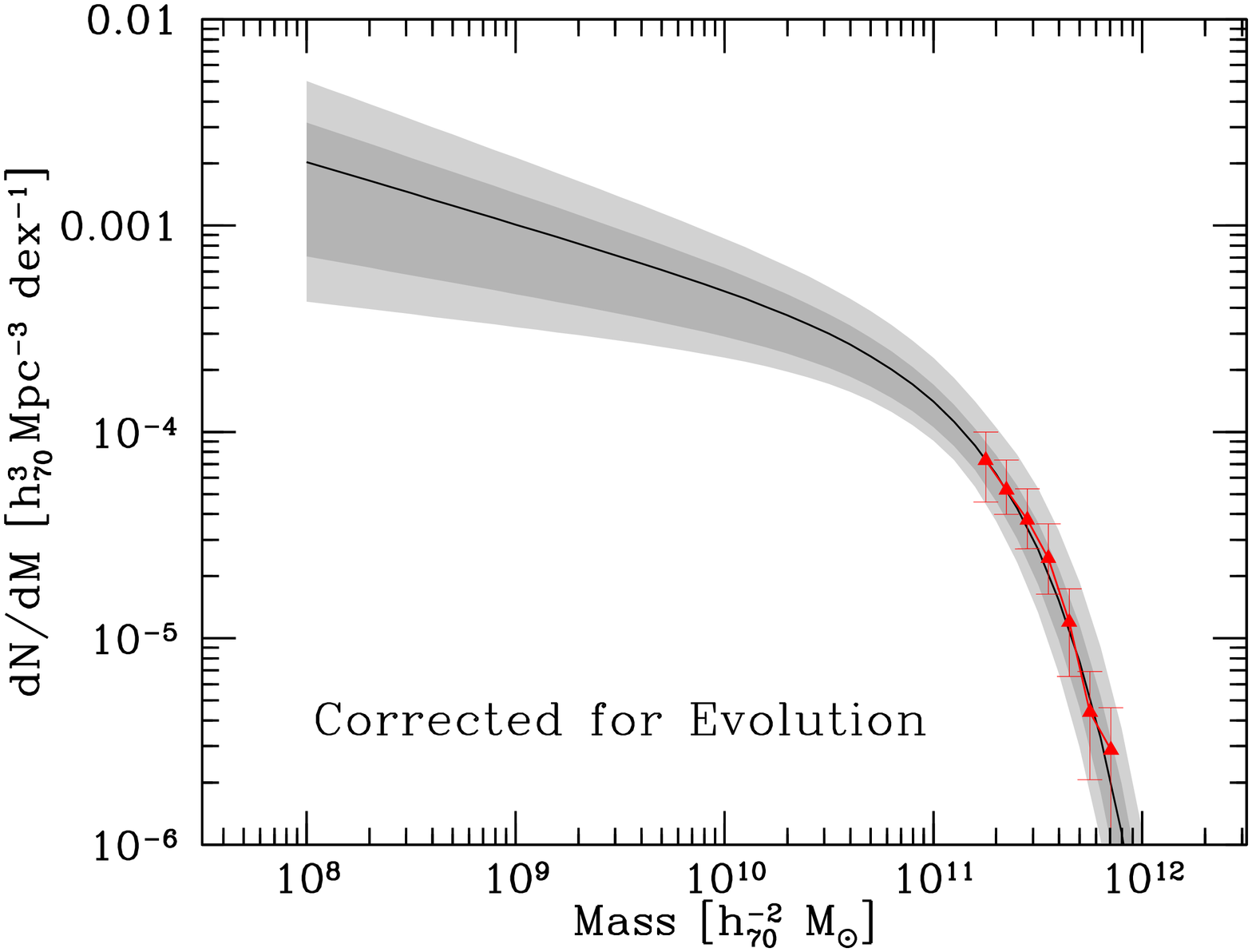}
}
\caption{Results of Schechter parametric analysis of the observed comoving 
number density of 5.8$\mu$m-peakers (Padova-94 case). The two {\em top} panels report the 
distribution of models in the $M^\ast,\ \alpha,\ \Phi^\ast$ space explored
by MCMC sampling, in the case of no evolutionary correction. Contours belong to 68.3, 95.5 and 99.7\% confidence levels.
The {\em bottom} plots show the possible Schechter mass functions overlaid on 
the observed comoving number density of 5.8$\mu$m-peakers, as obtained without ({\em
left}) and with ({\em right}) correction for evolutionary effects. The shaded areas represent 
the 1$\sigma$ and 3$\sigma$ ranges obtained in the fit. Error bars include the probability
of galaxies to be shifted from one mass bin to another, due to uncertainties in
the stellar mass estimate.}
\label{fig:res_MCMC}
\end{figure*}

The results of the Schechter parametric analysis of the observed comoving 
number density of 5.8$\mu$m-peakers are shown in Fig. \ref{fig:res_MCMC}, 
 based on the Padova-94 library.

By considering the Schechter sub-space using a Bayesian approach,
we automatically marginalize over the hyper-parameters, 
producing the posterior probability distributions
for the mass function parameters.

The top left panel in Fig. \ref{fig:res_MCMC} reports the 1,~2,~3 $\sigma$ contours 
in the $(M^\ast,\ \alpha)$ 
space, as computed to include the 68.3, 95.5 and 99.7 \% of the total volume 
occupied by the explored samples.

The median values obtained for the Schechter characteristic mass $M^\ast$ 
are $1.66\times10^{11}$ and $1.32\times10^{11}$ $[$ h$_{70}^{-2}$ M$_\odot]$, 
in case the evolutionary correction has or has not been considered.
The 3$\sigma$ uncertainty range is $\sim1$ dex in both cases.

As mentioned above, we are now probing only the very-massive tail of the galaxy
stellar mass function, hence a very loose constraint can be set on the $\alpha$ 
parameter. We find a median value of $-0.30$, which in fact is reasonably 
similar to the value $\alpha = -0.35$ obtained by \citet{fontana2006} on a
GOODS-MUSIC sample ranging from $\sim10^9$ to $\sim4\times 10^{11}$ M$_\odot$.
If no evolutionary correction is applied, the MCMC space exploration highlights the presence of 
a secondary solution at $\alpha=-0.16$ (see top left panel in Fig. \ref{fig:res_MCMC}).
The range of suitable models extends between $\alpha=-0.12$ and $-0.40$.

We therefore decided to fix the value of $\alpha$ to $-0.35$ and proceed to
explore the $(M^\ast,\ \Phi^\ast)$ space 
(upper right panel in Fig. \ref{fig:res_MCMC}). 
The parameter $\Phi^\ast$ has a most probable median value of
$0.00022$ $[$ h$_{70}^3$ Mpc$^{-3}]$, with a 1$\sigma$ range smaller than
$0.1$ dex, but doubles if no evolutionary correction is considered. 
Finally, $M^\ast$ is similarly distributed as in the former case 
with $\alpha$ as a free parameter.

The bottom panels in Fig. \ref{fig:res_MCMC} overlay the Schechter 
results on the actual observed 5.8$\mu$m-peaker comoving number density, 
with and without the correction for evolutionary effects in the $z=2-3$ redshift range.
The shaded areas belong to the models obtained varying the 
mass function parameters within the 1$\sigma$ (dark) and 3$\sigma$ (light) 
ranges, as derived with our Bayesian approach.
The difference between the two cases is very significant, particularly 
as far as $M^\ast$ and $\Phi^\ast$ are concerned.

The error bars shown in this Figure account also for the possible shifts
of galaxies from one mass bin to another, caused by the uncertainty on 
stellar mass, as derived from SED fitting. The error bars are computed 
taking into account the probability of a galaxy 
to be shifted, which is given by the actual $\chi^2$ distribution of 
all $\sim10^5$ solutions in the exploration of the SED parameter space.

The same MCMC Schechter analysis has been performed also on the galaxy mass function obtained 
by adopting the \citet{maraston2005} SSP library. The median values 
of the Schechter parameters are very similar to those obtained with the Padova-94
library.

Table \ref{tab:res_MCMC} summarizes the results, including the median and the 
1$\sigma$, 3$\sigma$ ranges for the Schechter parameters, in both the Padova-94 and the 
\citet{maraston2005} cases, and with or without evolutionary correction.

\begin{table}[!ht]
\centering
\begin{tabular}{l c c c}
\hline
\hline
\multicolumn{4}{c}{Padova-94 (no evo.)}\\
\hline
  & median & 1$\sigma$ & 3$\sigma$ \\
\hline
$\log(M^\ast)$	& 11.12 & 11.07 to 11.13 & 11.01 to 11.19 \\
$\Phi^\ast$	& 0.00048 & 0.00046 to 0.00052 & 0.00032 to 0.00056 \\
$\alpha$	& $-0.29$ & $-$0.36 to $-$0.17 & $-$0.40 to $-$0.12 \\
\hline
\multicolumn{4}{c}{}\\
\multicolumn{4}{c}{}\\
\hline
\hline
\multicolumn{4}{c}{Padova-94 (evo.)}\\
\hline
  & median & 1$\sigma$ & 3$\sigma$ \\
\hline
$\log(M^\ast)$	& 11.22 & 11.18 to 11.24 & 11.14 to 11.27 \\
$\Phi^\ast$	& 0.00022 & 0.00019 to 0.00025 & 0.00018 to 0.00031 \\
$\alpha$	& $-0.30$ & $-$0.34 to $-$0.18 & $-$0.37 to $-$0.12 \\
\hline
\multicolumn{4}{c}{}\\
\multicolumn{4}{c}{}\\
\hline
\hline
\multicolumn{4}{c}{\citet[][ no evo.]{maraston2005}}\\
\hline
  & median & 1$\sigma$ & 3$\sigma$ \\
\hline
$\log(M^\ast)$	& 11.10 & 11.04 to 11.12 & 10.97 to 11.15 \\
$\Phi^\ast$	& 0.00050 & 0.00046 to 0.00053 & 0.00035 to 0.00057 \\
$\alpha$	& $-0.34$ & $-$0.37 to $-$0.14 & $-$0.41 to $-$0.11 \\
\hline
\multicolumn{4}{c}{}\\
\multicolumn{4}{c}{}\\
\hline
\hline
\multicolumn{4}{c}{\citet[][ evo.]{maraston2005}}\\
\hline
  & median & 1$\sigma$ & 3$\sigma$ \\
\hline
$\log(M^\ast)$	& 11.18 & 11.16 to 11.23 & 11.12 to 11.31 \\
$\Phi^\ast$	& 0.00025 & 0.00019 to 0.00027 & 0.00017 to 0.00034 \\
$\alpha$	& $-0.34$ & $-$0.39 to $-$0.19 & $-$0.42 to $-$0.15 \\
\hline
\end{tabular}
\caption{Results of Schechter STY analysis of the comoving number density of 
5.8$\mu$m-peakers.}
\label{tab:res_MCMC}
\end{table}


\section{Discussion}

\begin{figure*}[!ht]
\centering
\rotatebox{-90}{
\includegraphics[height=0.85\textwidth]{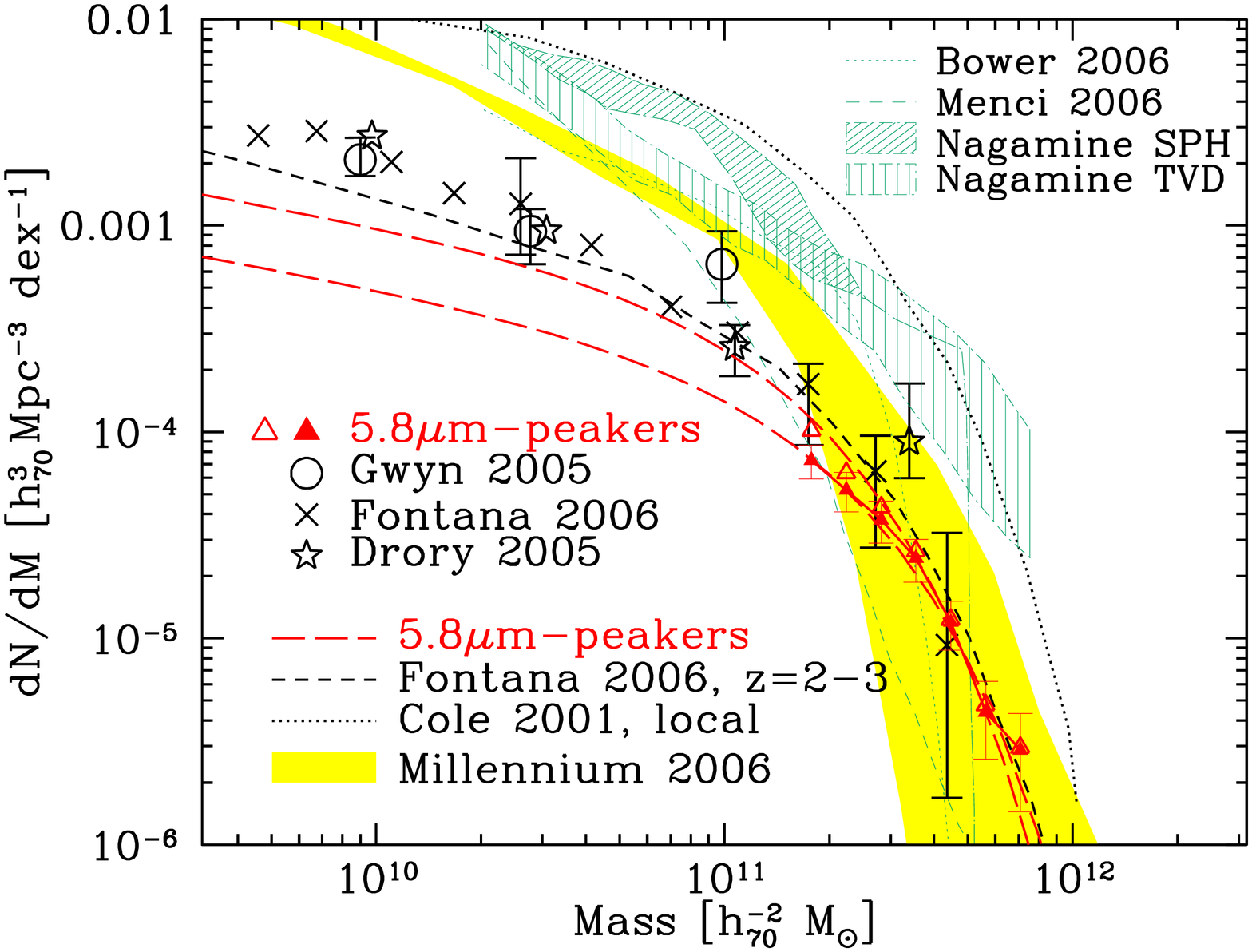}
}
\caption{Comparison between the observed $z=2-3$ stellar mass function of 
5.8$\mu$m-peakers and literature data, transformed to a Salpeter IMF 
(in the range $0.1-100$ M$_\odot$). Symbols represent the observed comoving
galaxy number densities from \citet{fontana2006}, \citet{gwyn2005} and 
\citet{drory2005}. 
The two thick, long dashed curves belong to the Schechter STY fit to the
5.8$\mu$m-peakers, as obtained with (lower line) and without (upper line) the evolutionary
correction.
The thick, sort-dashed line represents the Schechter fit to $z=2-3$ galaxies in
\citet{fontana2006}.
The thick dotted line is the local galaxy stellar mass function by \citet{cole2001}. 
Thin lines represents the predictions from semi-analytic \citep{bower2006,menci2006}
and hydro-dynamical \citep{nagamine2005b,nagamine2005a} models; the solid-shaded area
belongs to the Millennium Simulation in the concordance $\Lambda$-CDM cosmogony 
\citep{kitzbichler2006}.}
\label{fig:mf_literature}
\end{figure*}

We have taken advantage of the wide area offered by the SWIRE survey and the 
extensive multiwavelength coverage in the ELAIS-S1 field to identify 
very massive ($M>10^{11}$ M$_\odot$) galaxies at high redshift.
Thus the very massive tail of the $z=2-3$ galaxy stellar mass function has been
sampled with unprecedented detail.

\subsection{The stellar mass function at $z=2-3$}

Figure \ref{fig:mf_literature} compares the observed comoving number  
of 5.8$\mu$m-peakers to $z=2-3$ data drawn from the literature 
\citep{fontana2006,drory2005,gwyn2005} and transformed to a Salpeter IMF ($0.1-100$ M$_\odot$).
Only Poisson error bars are shown 
here, in order to directly compare our results to literature estimates. 

Our data are quite consistent with previous derivations of the stellar 
mass function, at the bright end, but deviate in the lower mass
bins. This is mainly due to fact that no evolutionary correction is usually applied in
the literature. For comparison, we show the IR-peakers data
as obtained both with (filled triangles) and without (open triangles) this correction.
The two thick long-dashed lines represent the two corresponding Schechter best fits.
The \citet{fontana2006} estimate (short dashes) is very similar to ours, when no
evolution is taken into account.

SWIRE has the advantage of probing the highest masses in an enormous 
cosmic volume, approaching
$10^{12}$ M$_\odot$, a regime where no previous studies have ever succeeded 
because of the rarity of sources. 
On the other hand, SWIRE is a shallow survey and not enough information is available 
in the low mass regime. As a consequence, below $M^\ast$ the SWIRE IR-peakers
Schechter function is poorly constrained.

A significant evolution with respect to the local galaxy stellar mass function 
is confirmed in the highest mass bins ($M>5\times10^{11}$ M$_\odot$).
This evolution is of the same order to that derived at lower masses from literature data:
the number of galaxies at $z=2-3$ has decreased by a factor of $\gtrsim 10$
at all mass regimes, with respect to the current epoch.
On the other hand, at lower redshift only a weak evolution is detected 
for massive galaxies ($M>10^{11}$ M$_\odot$), at least to $z\sim1.5$ 
\citep[e.g.][ among others]{franceschini2006,bundy2005,fontana2006}, while 
the number of lower mass objects is significantly lower at $z\sim1$ than 
the local value.

The solid-shaded area in Fig. \ref{fig:mf_literature} represents the prediction
by the Millennium $\Lambda$-CDM Simulation, and is taken from \citet{kitzbichler2006}.
The wide area covered by the model is meant to represent measurement
errors in $\log \left(M_\star\right)$. The results of this simulation are 
fairly consistent with our 5.8$\mu$m-peaker data, above $2\times10^{11}$
M$_\odot$. Nevertheless, the slope of their prediction appears to be too 
steep, because the number of $z=2-3$ 
galaxies between $M=1-2\times10^{11}$ M$_\odot$ is overpredicted.
Note also that below $10^{11}$ M$_\odot$, the 
galaxy stellar mass function is significantly overestimated by the Millennium 
model, with respect to the observed literature data.

Thin lines and other shaded areas belong to semi-analytic \citep[SAM,][]{bower2006,menci2006}
and hydro-dynamical \citep{nagamine2005b,nagamine2005a} models.
Both the two SAMs include AGN feedback on star formation, although in 
two different ways: in the Durham simulation \citep{bower2006} the AGN feedback is 
related to the gas smooth accretion onto the central black hole, 
continuing into the AGN quiescent phase;
while in the \citet{menci2006}
case it is produced by shock waves originated during the short active AGN phase only.
Further details on the two models and their differences are found in \citet{menci2006}.

The \citet{nagamine2005b,nagamine2005a} hydro-dynamical simulations 
include radiative heating and cooling, supernovae feedback, and standard 
star formation recipes.
The two models have been obtained with two different approaches. In 
\citet{nagamine2005b}
a Smoothed Particle Hydrodynamics (SPH) entropy formulation was adopted, while 
the other case is based on a Eulerian mesh code with a Total Variation 
Diminishing (TVD) scheme. 
The SPH simulation takes into account also feedback by galactic winds and 
a multi-component interstellar medium influencing the star formation process.

The agreement of these four models with the data is --- unfortunately --- 
rather poor.
The two hydro-dynamical simulations systematically overpredict the 
number density of galaxies at $z=2-3$, 
at $10^{10}<M<10^{11}$ M$_\odot$, as shown by comparison to literature 
data. Moreover the SPH run applies a cutoff at $M=5\times10^{11}$ M$_\odot$, 
while our data also show a non-null number density at these very high 
masses.

The SAM predictions are closer to the actual data, but merely intersect 
the observed $\Phi(M)$, under-predicting the comoving number density 
of very massive objects and overpredicting that of lower-mass sources.
This happens at $M\simeq 10^{11}$ M$_\odot$ for the \citet{menci2006}
model and at $M\simeq 3\times 10^{11}$ M$_\odot$ for \citet{bower2006}.
The latter shows a very steep slope above $M^\ast$, too steep to be consistent
with the observational evidence.

\subsection{The integrated stellar mass density}\label{sect:integrated_rho}

\begin{figure*}[!ht]
\centering
\begin{minipage}{0.74\textwidth}
\rotatebox{-90}{
\includegraphics[height=\textwidth]{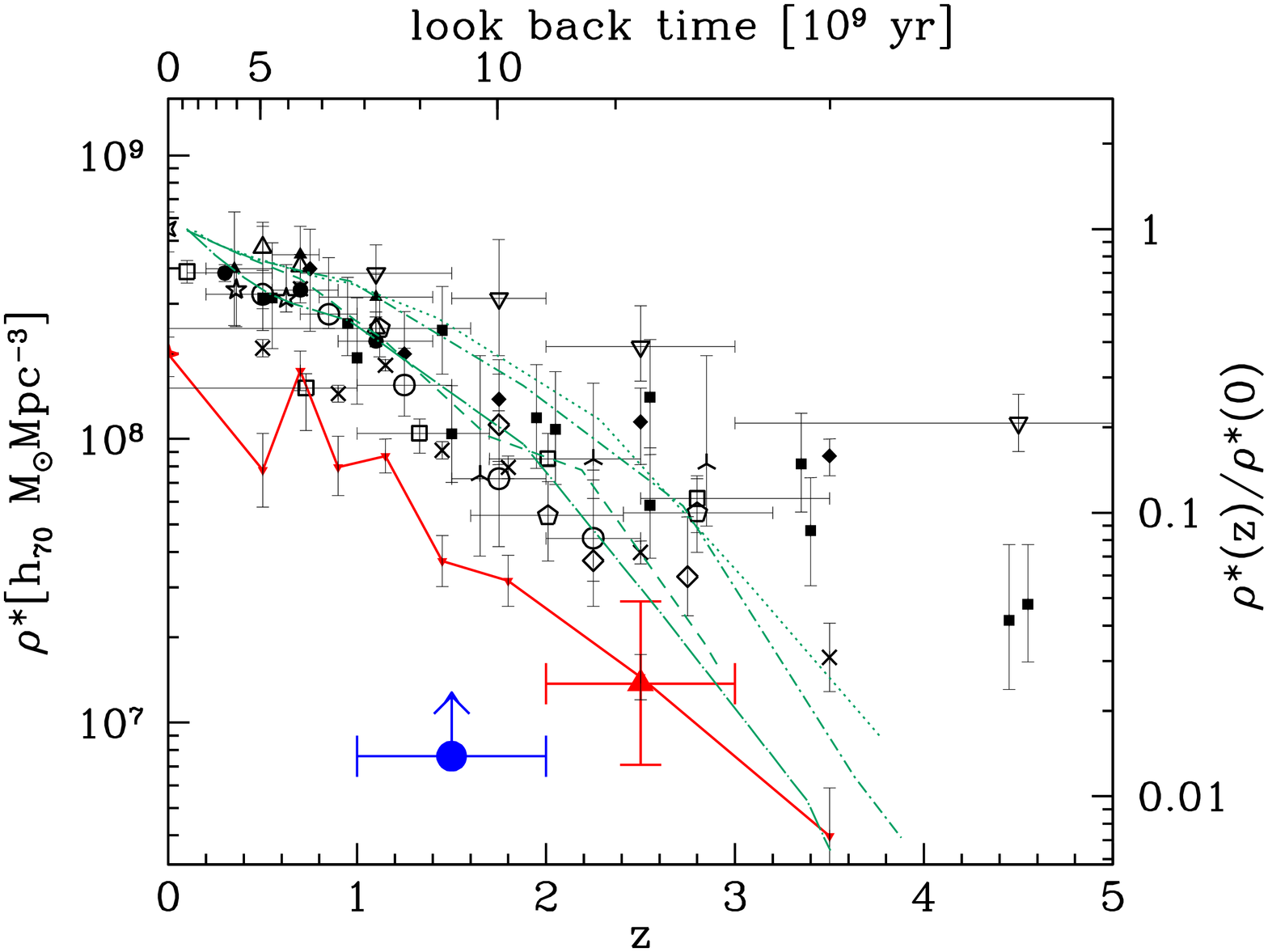}
}
\end{minipage}
\begin{minipage}{0.25\textwidth}
\includegraphics[width=\textwidth]{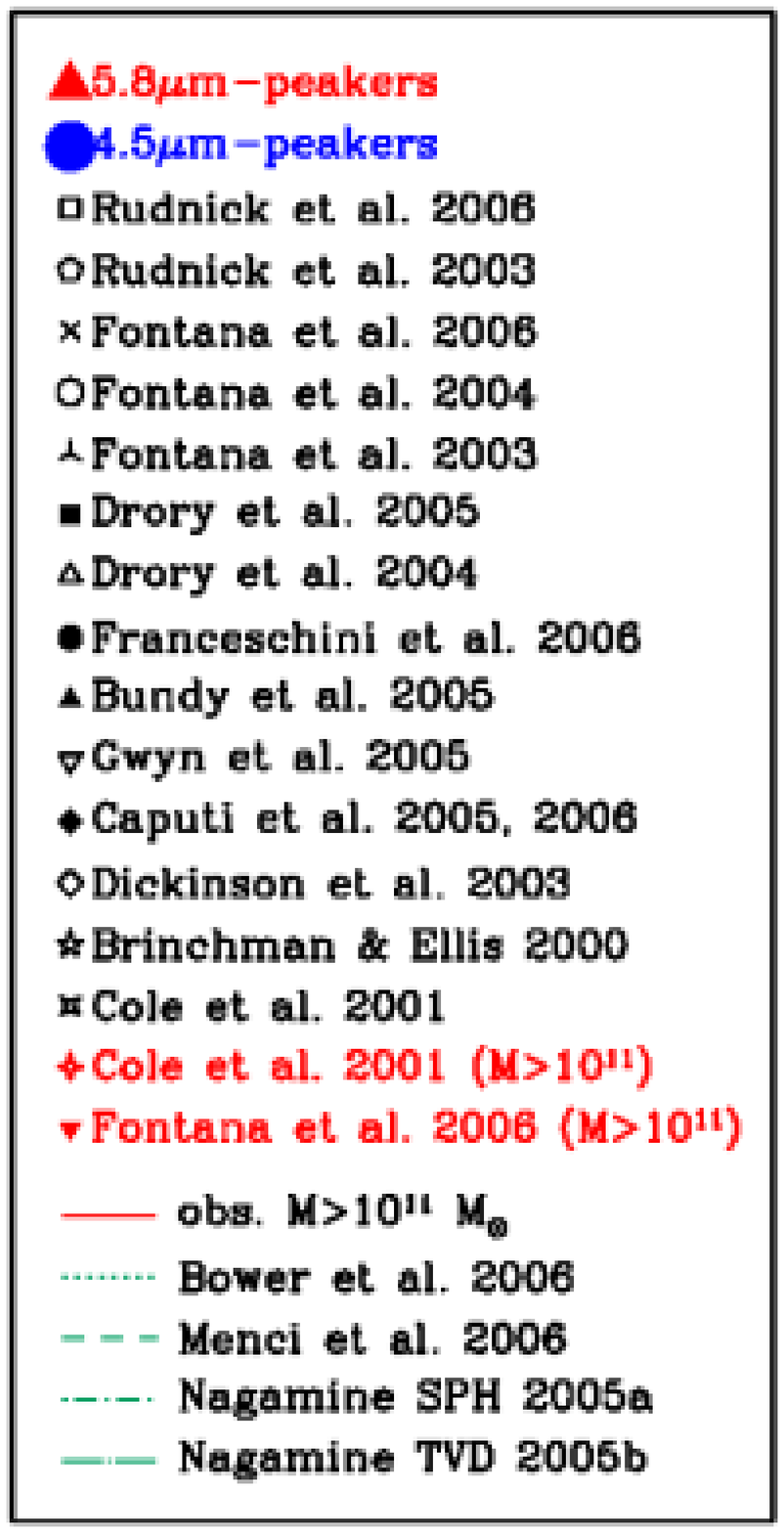}
\end{minipage}\\
\caption{Integrated stellar mass density as a function of redshift. The 
two large symbols represent the values of $\rho_\star$ derived from our data.
The filled triangle represents the contribution to the stellar mass density by 
5.8$\mu$m-peakers above $1.6\times10^{11}$ M$_\odot$, as obtained by 
integrating the Schechter solutions of our MCMC fit. The filled circle 
sets a lower limit to the stellar mass density of 4.5$\mu$m-peakers above 
$1.25\times10^{11}$ M$_\odot$, obtained by integrating their observed number density.
Literature data from numerous works are shown (see {\em right} panel 
for references), integrated down to $10^8$ M$_\odot$. Thin lines belong to different models, while the thick solid 
line simply connects the $\rho_\star$ obtained for galaxies heavier than $10^{11}$ 
M$_\odot$ by \citet{fontana2006} and \citet{cole2001}. 
All data and models have been transformed to a Salpeter IMF, extended to the mass
range $0.1-100$ M$_\odot$. 
}
\label{fig:rho}
\end{figure*}

A complete view on the contribution of high-redshift massive 
galaxies to the mass budget in the Universe is given by their 
global stellar mass density, as obtained by integrating 
the stellar mass function.

As far as 5.8$\mu$m-peakers are concerned, we have derived this quantity by 
exploiting the properties of the Schechter function. Its integral  
can be expressed as:
\begin{eqnarray}
\nonumber  \rho_\star \left(M>M_{inf}\right) & = & \int_{M_{inf}}^\infty \Phi(M)dM \\
\label{eq:schechter_int} & = &  M^\ast \Phi^\ast \times\Gamma\left(\alpha+1,\frac{M_{inf}}{M^\ast} \right)
\end{eqnarray}
where $\Gamma(a,x)$ is the incomplete gamma function, estimated from $a$ to infinity,
and $M_{inf}$ is the lower-mass integration cutoff (set by the completeness limit, in 
our case).
We integrate the stellar mass function over the mass range 
for which we were able to apply a reliable completeness 
correction, i.e.
$M>1.6\times10^{11}$ M$_\odot$, taking into account the large uncertainties 
in the Schechter parameters, due to poor sampling in the low mass regime.
The resulting stellar mass density (accounting for evolutionary correction) is $\rho_\star = 1.18 \times 10^7$ 
$[$h$_{70}$ M$_\odot$ Mpc$^{-3}]$, with a 3$\sigma$ range between 
$6.0\times 10^6$ and $2.3\times 10^7$ $[$h$_{70}$ M$_\odot$ Mpc$^{-3}]$.

In the case of 4.5$\mu$m-peakers, instead, we can set only a lower limit to the actual 
stellar mass density built in $M\ge10^{11}$ M$_\odot$. By simply 
integrating the observed data, we obtain $\rho_\star \ge 6.55 \times 10^6$  
$[$h$_{70}$ M$_\odot$ Mpc$^{-3}]$.
Table \ref{tab:rho} reports the stellar mass densities thus obtained.

\begin{table}[!ht]
\centering
\begin{tabular}{l | c c}
\hline
\hline
 & 4.5$\mu$m-peak &  5.8$\mu$m-peak\\ 
 & \multicolumn{2}{c}{$\rho_\star$ $[$h$_{70}$ M$_\odot$ Mpc$^{-3}]$}\\
\hline 
Observed ($V_a$, no evo)& $>6.55 \times 10^6$ & $1.45 \pm 0.3 \times 10^7$ \\
Schechter STY fit (no evo) & -- &  $1.56 \times 10^7$\\
Schechter 3$\sigma$ (no evo) & -- & $0.5-2.9 \times 10^7$ \\ 
\hline
Observed ($V_a$, evo)& -- & $1.23 \pm 0.3 \times 10^7$ \\
Schechter STY fit (evo) & -- &  $1.18 \times 10^7$\\
Schechter 3$\sigma$ (evo) & -- & $0.6-2.3 \times 10^7$ \\ 
\hline
\end{tabular}
\caption{Stellar mass density for 4.5$\mu$m-peakers ($z=1-2$, $M\ge1.25\times10^{11}$ M$_\odot$)
and 5.8$\mu$m-peakers ($z=2-3$, $M\ge1.6\times10^{11}$ M$_\odot$),  
as obtained with the $V_a$ and parametric analyses (Padova-94 library).}
\label{tab:rho}
\end{table}

Figure \ref{fig:rho} compares the $\rho_\star$ based on IR-peakers to 
a collection of data from the literature 
\citep{rudnick2006,rudnick2003,fontana2006,fontana2004,fontana2003,drory2005,
drory2004,franceschini2006,bundy2005,gwyn2005,caputi2006,caputi2005,dickinson2003b,brinchmann2000,
cole2001}, all transformed to a Salpeter IMF extended between 0.1 and 100 M$_\odot$.
This graph highlights a dramatic scatter in the 
current estimate of the stellar mass density, with significant 
discrepancies between the various authors. 

While the literature data were obtained by integrating the mass function down to $10^8$ M$_\odot$,
the IR-peaker data points belong
to sources above $\sim10^{11}$ M$_\odot$ only.
The solid thick line simply connects the $\rho_\star$ estimate for galaxies more massive
than $10^{11}$ M$_\odot$ by \citet{fontana2006}. Despite the large uncertainty
on the actual value, due to degeneracies in the Schechter fit, our measured 
$\rho_\star$ for the 5.8$\mu$m-peakers is fully consistent with 
\citet{fontana2006} data. On average between 30\% and 50\% of the total stellar 
mass in galaxies at $z=2-3$ is stored in our population of massive 
($M>1.6\times10^{11}$ M$_\odot$) 5.8$\mu$m-peakers.
Although very unlikely \citep[on the basis of data by][]{fontana2006},
this value could in fact grow 
to a higher fraction if we account for the uncertainties and/or compare to the lowest 
estimate of the overall stellar mass density from the literature.

The thin lines in Fig. \ref{fig:rho} represent the evolution of 
the global stellar mass density, as predicted by the same semi-analytical 
and hydro-dynamical models shown in Fig. \ref{fig:mf_literature} 
\citep{bower2006,menci2006,nagamine2005b,nagamine2005a}.
For all models the evolution of the stellar mass density becomes steeper
at $z\ge1.5-2.0$.
The values of $\rho_\star$ predicted by 
different models gradually diverge at redshift $z>1$, therefore
the observational data could in principle constrain 
the actual scenario, but the current wide scatter and 
large error bars in the stellar mass density 
makes it very hard to disentangle the various models.
However, we provide the first tight constraint at the largest masses,
for $z=2-3$.

\begin{figure}[!ht]
\centering
\rotatebox{-90}{\includegraphics[height=0.48\textwidth]{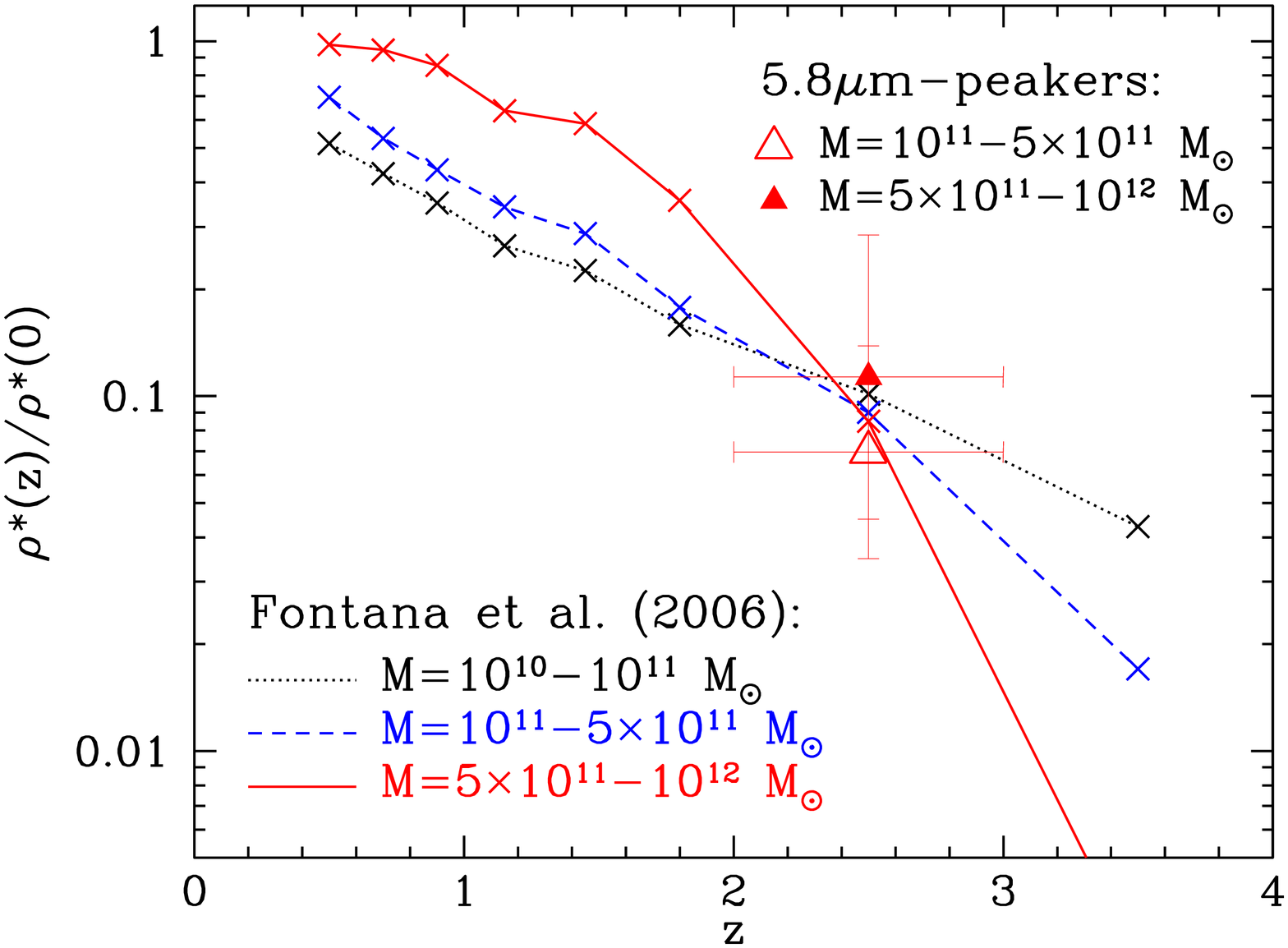}}
\rotatebox{-90}{\includegraphics[height=0.48\textwidth]{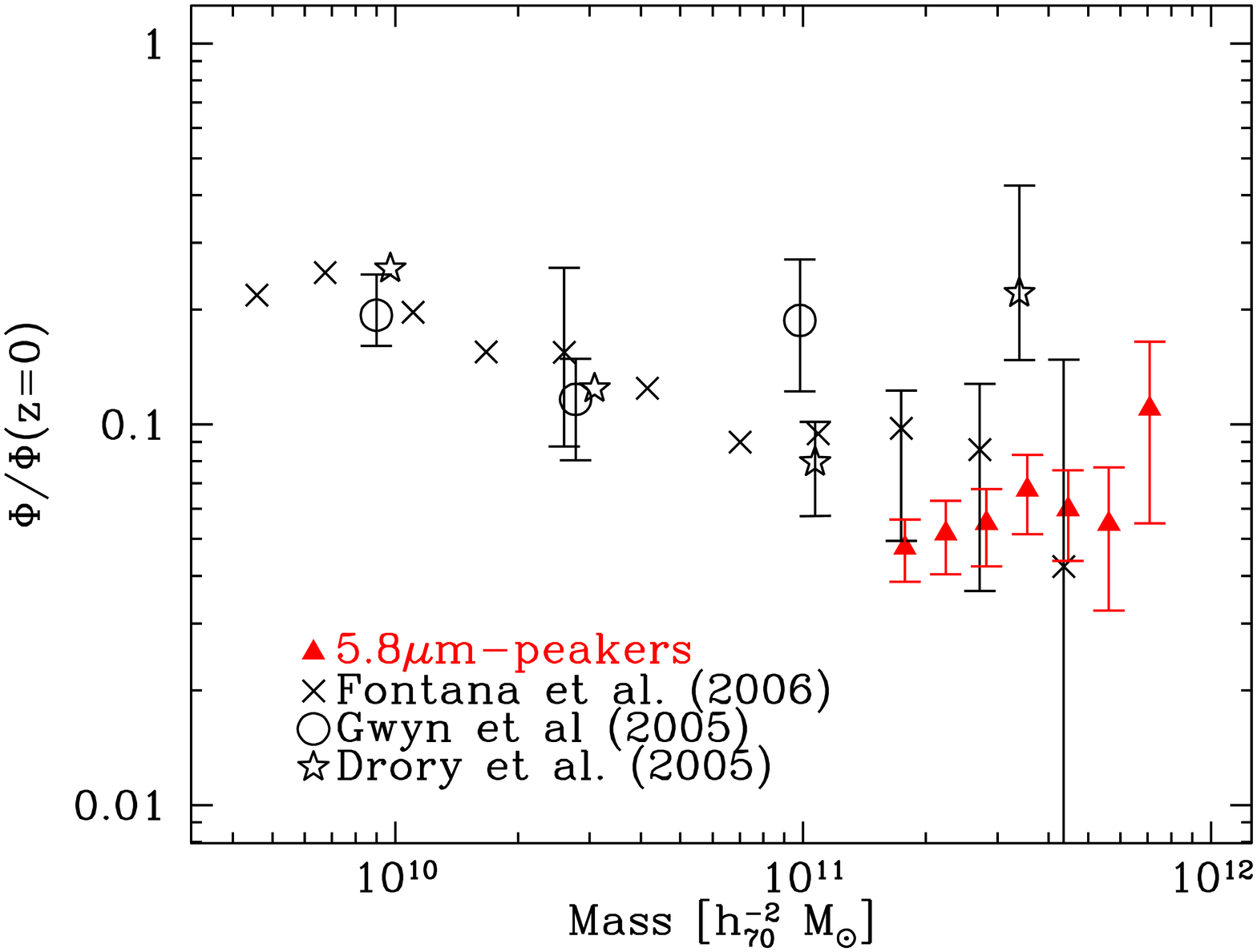}}
\caption{{\em Top} panel: ratio of the stellar mass density at a given
redshift and in the local Universe, as a function of redshift and for different 
mass bins. Datapoints with crosses were computed by integrating the \citet{fontana2006} Schechter fit
to the stellar mass function. {\em Bottom} panel: ratio of the mass function at 
$z=2-3$ to the local one, as derived for 
5.8$\mu$m-peakers (triangles) and from the literature (other symbols). 
The \citet{cole2001} local mass function was adopted.}
\label{fig:mf_ratio}
\end{figure}

The top panel of Fig. \ref{fig:mf_ratio} shows the ratio of the stellar mass density 
at a given
redshift and in the local Universe, for different 
mass bins. Crosses represent the data obtained by integrating the Schechter fit to the 
mass function described by Fontana et al. (\citeyear{fontana2006}, see also their Fig.9). 
We have chosen these authors, because 
their analysis extends to the redshift bin covered by 5.8$\mu$m-peakers and 
because they provide Schechter parameters in all redshift bins.  
The dotted, dashed and solid lines represent three different mass bins. 
Triangles represent 5.8$\mu$m peakers, and are obtained by integrating 
our Schechter fit in the  
$10^{11}\le M<5\times10^{11}$ and $5\times10^{11}\le M<10^{12}$ M$_\odot$
mass bins.

Despite the poor statistics and the high masses and the large ``noise'' previously 
pointed out in Fig. \ref{fig:rho}, \citet{fontana2006} data clearly show
that very massive galaxies evolve more rapidly than lower mass objects and reach 
the local density at earlier epochs. It is also interesting to note that 5.8$\mu$m-peakers
sample a redshift range where $\rho_\star/\rho_\star\left(0\right)$ assumes similar values
in all the three mass bins considered. This effect explains why the ratio
of the stellar mass function at $z\sim2.5$ to the local one is relatively flat (bottom 
panel of Fig. \ref{fig:mf_ratio}). The latter plot shows that $\phi/\phi\left(0\right)$ 
spans values within a factor of $3-4$ in the whole mass range between $5\times 10^9$ 
and $10^{12}$ M$_\odot$.

At least 50-60\% of the present-day stellar mass in very massive systems 
seems to have assembled between $z\simeq2.5$ and $z=1.5$, i.e. when the 
Universe was between $\sim2.5$ and $\sim4.5$ Gyr old.
This process was roughly complete at $z\sim1$, when more than 80\% of 
massive galaxies were already in place.
SWIRE 5.8$\mu$m-peakers ($z=2-3$) represent less than $\sim$10\% of the stellar mass 
in the local population of massive galaxies ($M=10^{11}-10^{12}$ M$_\odot$).

According to theoretical models, the formation of galaxies and large scale 
structure occurs in the frame of some variant of `biased' hierarchical buildup
within a $\Lambda$-CDM cosmology \citep[e.g.][]{cole2000,hatton2003,granato2004}.
In these scenarios the most massive objects (e.g. $M_\star > \textrm{ several }
10^{11}$ M$_\odot$) are predicted to assemble earlier, more quickly and
in richer environments than less massive ones 
\citep[e.g.][]{somerville2001,nagamine2005a}.

It is worth noting that no information is currently available 
on the environment hosting these very massive galaxies at high-redshift.
The study of the clustering properties of mid-IR powerful-emitting IR-peakers 
has shown that their spatial correlation length is
$r_0=14.40\pm1.99$ $[$h$^{-1}$ Mpc$]$ at $z=2-3$ \citep{farrah2006}.
This value is consistent with a populations residing in $M\simeq10^{13}$
M$_\odot$ dark matter haloes, thus they are excellent candidates for being  
part of protoclusters at high redshifts.
If this is the case, and if less active IR-peakers (i.e. with lower mid-IR
luminosities or not detected by MIPS at all) showed analogous properties, 
they could trace quiescent stages of similar systems.
Broadly speaking, the studied 5.8$\mu$m-peakers, sampling the very massive 
tail of the stellar mass function at $z=2-3$, 
might represent a different class with respect to the 
field general population. They could, in fact, be at the center of large 
dark matter haloes, their companions being too faint to 
be detected by SWIRE. Different environmental conditions would 
rule their evolution and partially explain 
the mass ``downsizing'' effect exemplified by a 
faster high-$z$ evolution of massive systems.


\section{Summary and conclusions}

Sampling the very massive tail of the stellar mass function at high redshift
and estimating its contribution to the global stellar mass density 
is a critical task in modern cosmology, motivated by the recent evidence 
that a ``downsizing'' effect exists in the evolution of stellar mass across
cosmic time.

We have exploited SWIRE/Spitzer and ancillary data in the ELAIS-S1 
area to perform a systematic search for high redshift ($z\gtrsim1.0$)
massive ($M>10^{11}$ M$_\odot$) galaxies.
High redshift systems have been isolated by identifying the 1.6$\mu$m
restframe stellar peak shifted to IRAC wavelengths (3.6$-$8.0 $\mu$m).
The availability of near-IR ($J$ and $K_s$ band) data allowed us to 
avoid low-redshift interlopers and selection aliasing due to bright 3.3 $\mu$m
PAH features at $z\simeq0.4$.
A total of 203 5.8$\mu$m-peakers and 123 4.5$\mu$m-peakers have been 
identified over one square degree in the ELAIS-S1 field.

We have performed an extensive SED analysis, based on mixed stellar
population synthesis, focused on deriving the stellar masses
of the selected sample. The advantage of near-IR and mid-IR constraints, 
as well as the dependence of results on the choice of the IMF and SSP library 
have been explored in detail. The main results are:
\begin{itemize}
\item because of the shallow 5.8$\mu$m flux cut adopted, 
the SWIRE IR-peaker sample consists of very massive galaxies, the
majority of sources having $M_\star>10^{11}$ M$_\odot$. Objects in the 
range $z=1-2$ peak at $M_\star\simeq10^{11}$ M$_\odot$, while the distribution
of 5.8$\mu$m-peakers ($z=2-3$) is centered at 
$M_\star\simeq2\times 10^{11}$ M$_\odot$.
Typical uncertainties in stellar mass estimate (due to degeneracies in the SFH
space) range between 0.1 and 0.3 dex, depending on multiwavelength coverage. 
The emission of $\sim$30\% of the sources turns out to be dominated by stars 
older than 1 Gyr, and also in the majority of the remaining cases old 
stellar populations do contribute to the observed SEDs.
\item the availability of mid-IR data provides a valuable constraint on the
recent star formation history of individual galaxies. If no 24$\mu$m flux (nor
upper limit) were available, 
the resulting stellar masses could be underestimated and the spread in
mass would be wider.
\item despite being very useful in the selection process, near-IR ($J$ and $K_s$) data 
turned out to be effective in constraining the D4000 break
only in $\sim$15\% of cases, and preferentially for
4.5$\mu$m-peakers. 
\item the choice of a \citet{chabrier2003} IMF, instead of a \citet{salpeter1955}
one, leads to systematically lower stellar masses, resulting 
in a rigid shift of the stellar mass distribution by $\sim0.3$ dex.
\item when including thermally-pulsing AGB stars in the SSP library
\citep{maraston2005}, the fraction of
objects dominated by old stellar populations increases, but the overall
stellar mass distribution does not change significantly, because of the
different $M_\star/L$ ratios of SSPs.
\end{itemize}
The stellar mass estimates have been used to compute the comoving number
density of galaxies as a function of stellar mass (i.e. the observed stellar
mass function), adopting the accessible volume formalism.
Following \citet{fontana2004}, we have corrected the samples for 
mass incompleteness and recovered the mass function down to $1.25\times10^{11}$ and 
$1.6\times10^{11}$ M$_\odot$ for 4.5$\mu$m- and 5.8$\mu$m-peakers
respectively. 

Unfortunately, the selection, based on a 5.8$\mu$m flux cut turned out to be 
only partially sensitive to 4.5$\mu$m peakers, therefore only a lower limit on
the mass function of $z=1-2$ sub-sample could be set.

The observed stellar mass function of 5.8$\mu$m-peakers was reproduced with a 
parametric function, using the STY \citep{sandage1979} approach. 
The uncertainties in the stellar masses of individual 
sources were automatically included in the analysis by using a Bayesian
formalism, and the best fit was obtained with a MCMC sampling of the parameter
space. The stellar mass function was finally integrated to derive the stellar
mass density locked in 5.8$\mu$m-peakers with $M>1.6\times10^{11}$ M$_\odot$,
at $z=2-3$. The results of the mass function analysis are:
\begin{itemize}
\item the wide area surveyed by SWIRE allows the very massive tail of the 
stellar mass function to be probed with unprecedented detail at $z=2-3$, 
extending previous analyses up to $M_\star=7\times10^{11}$ 
$[$h$_{70}^{-2}$ M$_\odot]$.
\item at $M<5\times10^{11}$ M$_\odot$, a significant intrinsic evolution has been detected across the redshift range
$z=2-3$, strongly dependent on mass. 
The dependence or the 5.8$\mu$m-peakers number density on $\left(1+z\right)$
has powers of $\sim-0.4$ and $\sim-0.65$ for $M\sim2\times10^{11}$ and
$4\times10^{11}$ M$_\odot$ respectively.
In the highest mass bins ($M\ge5\times10^{11}$
M$_\odot$) the number density of IR-peakers keeps nearly constant. 
\item comparison to literature data for the $z=2-3$ mass function shows an
overall agreement of the 5.8$\mu$m-peaker comoving number density to that of the K20,
MUNICS, and GOODS-MUSIC surveys, in the higher mass regime. At lower masses a
significant evolutionary correction should be applied, when the mass function is averaged over
a wide redshift range.
\item a significant evolution of the stellar mass function of 
$M\gtrsim10^{11}$ M$_\odot$ galaxies
with respect to the local estimate was detected: 
$\Phi(z=2-3)\le0.1\times\Phi(z=0)$. Combining 5.8$\mu$m-peakers and literature data 
\citep[e.g.][]{fontana2006}, this implies that the bulk of massive galaxies 
was not yet in place by the time the Universe was $\sim3$ Gyr old, 
but must have been assembled in the following $\sim1.5$ Gyr of evolution.
\item current hydro-dynamical models significantly
overestimate the number density of massive galaxies, while the semi-analytic 
approach underestimates it.
\item since SWIRE 5.8$\mu$m-peakers sample only the very massive tail of the
mass function, the Schechter slope $\alpha$ cannot be constrained. Despite its very low
significance, the best fit value $\alpha=-0.30$ is similar to that found in the
literature. The other best fit parameters are: $M^\ast = 1.66 \times10^{11}$ 
$[$h$_{70}^{-2}$ M$_\odot]$ $\pm$1 dex; $\Phi^\ast=0.00022^{+0.00004}_{-0.00009}$ 
$[$ h$_{70}^3$ Mpc$^{-3}]$.
\item the integrated stellar mass density of 5.8$\mu$m peakers is 
$\rho_\star = 1.18 \times 10^7$ $[$h$_{70}$ M$_\odot$ Mpc$^{-3}]$, with a 
3$\sigma$ range of $\pm 0.3$ dex. Only a lower limit could be set for
4.5$\mu$m-peak galaxies: $\rho_\star\ge6.55\times10^6$ 
$[$h$_{70}$ M$_\odot$ Mpc$^{-3}]$.
\item on average SWIRE massive 5.8$\mu$m-peakers provide 30$-$50\% of 
the total stellar mass density in galaxies at $z=2-3$.
\item 5.8$\mu$m-peakers provide less than $\sim10$\% of the stellar 
mass locked in massive galaxies ($M=10^{11}-10^{12}$ M$_\odot$) in 
the local Universe.
\end{itemize}
The analysis carried out on SWIRE massive galaxies at $z>1.5$ 
over one square degree, highlighted the complementarity of 
wide-shallow and deep pencil-beam surveys. 

On one hand, sampling the faint end of the luminosity function, i.e. the
low-mass end of the mass function, is needed in order to constrain the shape 
of the mass function and the total stellar mass density in galaxies at high
redshift. 

On the other hand, very massive galaxies 
are rare objects on the sky, with a number density $1.2\times10^{-4}$ 
$[$h$_{70}^3$ Mpc$^{-3}]$ and it is necessary to explore large volumes
of Universe in order to fully characterize them. 
The natural extension of this analysis is to build the stellar mass function 
of high-$z$ galaxies over the whole SWIRE 49 deg$^2$ area, taking advantage of  
what we have learned thanks to the full multi-wavelength coverage in ELAIS-S1.

At the time being, not much information is known about the 
environment hosting these massive high-$z$ objects.
Further analyses of this population should be carried out 
probing also the environmental frame they belong to, in order to correctly 
interpret their role in the ``downsizing'' scenario.


\begin{acknowledgements}

We also would like to thank the whole SWIRE team for preparing the ES1 
Spitzer data. 
We are grateful to Stephan Charlot for very useful discussions about
SED fitting, IMF, and AGB stars, and
Paolo Cassata on
the stellar mass function and units. Finally, SB wishes to thank 
people at the Sussex Astronomy Center for their warm hospitality.

This work made use of 
observations collected at the European Southern
Observatory, Chile: ESO projects No. 168.A-0322, 170.A-0143, 
073.A-0446, 075.A-0428.

The Spitzer Space Telescope
is operated by the Jet Propulsion Laboratory, California Institute of Technology, under
contract with NASA. 
SWIRE was supported by NASA through the SIRTF Legacy
Program under contract 1407 with the Jet Propulsion Laboratory.

\end{acknowledgements}


\bibliographystyle{aa}
\bibliography{7491bib}   

\end{document}